\documentclass[aps,prd,twoside,singlecolumn,superscriptaddress,floatfix,nofootinbib]{revtex4}

\usepackage{graphicx,natbib,amssymb,amsmath,stmaryrd,makecell}
\usepackage[utf8]{inputenc}
\UseRawInputEncoding
\usepackage{amsmath}
\usepackage{bm}
\usepackage{xcolor}
\usepackage{natbib}
\usepackage{graphicx}
\usepackage{hyperref}
\hypersetup{
    colorlinks=blue,
    linkcolor=blue,
    filecolor=magenta,      
    urlcolor=blue,
}

\newcommand\beq{\begin{equation}}
\newcommand\eeq{\end{equation}}
\newcommand\beqn{\begin{eqnarray}}
\newcommand\eeqn{\end{eqnarray}}

\newcommand{\ba}{\begin{eqnarray}}
\newcommand{\ea}{\end{eqnarray}}
\newcommand{\be}{\begin{equation}}
\newcommand{\ee}{\end{equation}}

\begin{document}

\defcitealias{krolewski_2020}{K20}
\defcitealias{krolewski2021cosmological}{K21}

\title{Constraining the galaxy-halo connection of infrared-selected \emph{unWISE} galaxies with galaxy clustering and galaxy-CMB lensing power spectra}

\author{Aleksandra Kusiak}
\affiliation{Department of Physics, Columbia University, New York, NY, USA 10027}
\email{akk2175@columbia.edu}

\author{Boris Bolliet}
\affiliation{Department of Physics, Columbia University, New York, NY, USA 10027}

\author{Alex Krolewski}
\affiliation{AMTD Fellow, Waterloo Centre for Astrophysics, University of Waterloo, Waterloo ON N2L 3G1, Canada}
\affiliation{Perimeter Institute for Theoretical Physics, 31 Caroline St. North, Waterloo, ON NL2 2Y5, Canada}

\author{J.~Colin Hill}
\affiliation{Department of Physics, Columbia University, New York, NY, USA 10027}
\affiliation{Center for Computational Astrophysics, Flatiron Institute, New York, NY, USA 10010}

\begin{abstract}

We present the first detailed analysis of the connection between galaxies and their dark matter halos for the \emph{unWISE} galaxy catalog --- a full-sky, infrared-selected sample built from \emph{WISE} data, containing over 500 million galaxies.  Using \emph{unWISE} galaxy-galaxy auto-correlation and \emph{Planck} CMB lensing-galaxy cross-correlation measurements down to 10 arcmin angular scales, we constrain the halo occupation distribution (HOD), a model describing how central and satellite galaxies are distributed within dark matter halos, for three \emph{unWISE} galaxy samples at mean redshifts $\bar{z} \approx 0.6$, $1.1$, and $1.5$, assuming a fixed cosmology at the best-fit \emph{Planck} $\Lambda$CDM values. 
We constrain the characteristic minimum halo mass to host a central galaxy, 
$M_\mathrm{min}^\mathrm{HOD} = 
1.83^{+0.41}_{-1.63} \times 10^{12} M_\odot/h$, $5.22^{+0.34}_{-4.80} \times 10^{12} M_\odot/h$, $6.60 ^{+0.30}_{-1.11} \times 10^{13} M_\odot/h$ 
and the mass scale at which one satellite galaxy per halo is found, 
$M^{\prime}_1 = 1.13 ^{+0.32}_{-0.70} \times 10^{13} M_\odot/h$,  $1.18 ^{+0.30}_{-1.11} \times 10^{13} M_\odot/h$,  $1.23 ^{+0.14}_{-1.17} \times 10^{14} M_\odot/h$ for the \emph{unWISE} samples at $\bar{z}\approx 0.6$, $1.1$, and $1.5$, respectively.  We find that all three samples are dominated by central galaxies, rather than satellites.
Using our constrained HOD models, we infer the effective linear galaxy bias for each \emph{unWISE} sample, and find that it does not evolve as steeply with redshift as found in previous perturbation-theory-based analyses of these galaxies.  We discuss possible sources of systematic uncertainty in our results, the most significant of which is the uncertainty on the galaxy redshift distribution.  Our HOD constraints provide a detailed, quantitative understanding of how the \emph{unWISE} galaxies populate the underlying dark matter halo distribution.  These constraints will have a direct impact on future studies employing the \emph{unWISE} galaxies as a cosmological and astrophysical probe, including measurements of ionized gas thermodynamics and dark matter profiles via Sunyaev-Zel'dovich and lensing cross-correlations.

\end{abstract}

\maketitle

\section{Introduction}
\label{sec:intro}
The connection between galaxies and their host dark matter halos plays a crucial role in both cosmology and astrophysical models of galaxy formation.  To maximize the cosmological constraining power of current galaxy surveys, the modeling of large-scale structure requires understanding and treatment of the galaxy-halo connection. On the other hand, since galaxies form within dark matter halos, understanding the link between them is crucial for improving our theoretical understanding of galaxy formation (see, e.g.,~\cite{2018ARA&A..56..435W} for a review).

The goal of this work is to constrain a leading model for the galaxy-halo connection, the halo occupation distribution, for the \emph{unWISE} galaxies~\cite{krolewski_2020,Schlafly19}.  The halo occupation distribution (HOD) is a description of galaxy clustering in a larger halo model framework, which describes the spatial fluctuations of cosmological observables in terms of the contributions from dark matter halos \cite{Cooray:2002dia, seljak2000, peacock2000}. It is based on the assumption that each dark matter particle belongs to one dark matter halo. The standard HOD model from Zheng \textit{et al.} \cite{Zheng_2007}, which characterized the Sloan Digital Sky Survey \cite{zehavi2005_sdss} and DEEP2 Galaxy Redshift Survey \cite{coil2006_deep2} galaxies in the HOD framework, and which we adopt, assumes that each halo contains central and satellite galaxies.  Central galaxies are located in the center of a halo, and satellites are distributed according to a specified radial profile. With this empirical approach, it is possible to constrain several physical characteristics of a given galaxy sample, such as the mean number of centrals and satellites for a given halo mass, or the minimum halo mass to host a central galaxy, as done in, e.g., the Dark Energy Survey (DES) Year 3 analysis \cite{Zacharegkas2021} or for the infrared \emph{Herschel} galaxies \cite{Hermes2010}. 

In this paper, we use HOD modeling to constrain the galaxy-halo connection for \emph{unWISE} galaxies. The \emph{unWISE} catalog is constructed from data from the \emph{Wide-field Infrared Survey Explorer} (\emph{WISE}) and \emph{NEOWISE} missions, covering the full sky and containing over 500 million objects. It is divided into three subsamples using infrared color and magnitude cuts (Table~\ref{table:unwise_cuts}), denoted \emph{blue}, \emph{green}, and \emph{red}. These subsamples have mean redshifts $\bar{z} \approx 0.6, 1.1, 1.5$. The \emph{unWISE} samples were constructed, validated, and characterized in Krolewski \textit{et al.} \cite{krolewski_2020} (\citetalias{krolewski_2020} hereafter), where the authors measured tomographic cross-correlations of \emph{unWISE} galaxies with \emph{Planck} CMB lensing maps with combined $S/N \approx 80$ over the multipole range $100 < \ell < 1000$. 
The measurements were further used in a companion paper~\cite{krolewski2021cosmological} (\citetalias{krolewski2021cosmological} hereafter) to constrain the cosmological parameters $\sigma_8$ (the amplitude of low-redshift density fluctuations) and $\Omega_m$ (the matter density fraction). The combined \emph{unWISE} samples yielded a value for the combination of these parameters $S_8 = \sigma_8(\Omega_m /0.3 )^{0.5}$ of $S_8=0.784 \pm 0.015$ (68\% confidence interval), consistent with low-redshift lensing measurements~\cite{DESY3cosmology,KiDS1000cosmology,Hikage2019}, yet in moderate tension with \emph{Planck} CMB results for this parameter~\cite{Planck2018}.  The \emph{unWISE} sample was also used to measure the kinematic Sunyaev-Zel'dovich effect with the ``projected-field'' estimator \cite{Dore2004, DeDeo, Hill2016, Ferraro2016} in Ref.~\cite{Kusiak_2021}, where the product of the baryon fraction $f_b$ and free electron fraction $f_{\rm free}$ was constrained to be $(f_b / 0.158)(f_{\rm free} / 1.0) = 0.65 \pm 0.24$, $2.24 \pm 0.25$, and $2.87 \pm 0.57$ for \emph{unWISE} blue, green, and red, respectively. 

In this work, we analyze measurements of the \emph{unWISE} galaxy-galaxy auto-correlation and \emph{unWISE} galaxy $\times$ \emph{Planck} CMB lensing cross-correlation, which are slightly updated from those in \citetalias{krolewski_2020, krolewski2021cosmological}, to constrain the HOD parameters describing the three \emph{unWISE} galaxy samples, such as the minimum mass of a halo to host a central galaxy. The results are obtained by fitting a theoretical halo model of the galaxy-galaxy and galaxy-CMB lensing cross-correlations to the updated measurements from \citetalias{krolewski_2020}. The best-fit model describes the data well, with $\chi^2=$ 11.8, 7.9, 15.3 for a joint fit with galaxy-galaxy and galaxy-CMB lensing measurements (19 data points in total), separately for each \emph{unWISE} sample, here for the blue ($\bar{z}\approx 0.6$), green ($\bar{z} \approx 1.1$), and red ($\bar{z} \approx 1.5$) sample, respectively. We constrain the characteristic minimum halo mass to host a central galaxy, $M_\mathrm{min}^\mathrm{HOD}$, to be $M_\mathrm{min}^\mathrm{HOD} = 
1.83^{+0.41}_{-1.63} \times 10^{12} M_\odot/h$, $5.22^{+0.34}_{-4.80} \times 10^{12} M_\odot/h$, $6.60 ^{+0.30}_{-1.11} \times 10^{13} M_\odot/h$ and the mass scale at which one satellite galaxy per halo is found, $M^{\prime}_1$, to be $M^{\prime}_1 = 1.13 ^{+0.32}_{-0.70} \times 10^{13} M_\odot/h$,  $1.18 ^{+0.30}_{-1.11} \times 10^{13} M_\odot/h$,  $1.23 ^{+0.14}_{-1.17} \times 10^{14} M_\odot/h$ for \emph{unWISE} blue, green, and red, respectively. We also derive other quantities from our best-fit HOD model, such as the effective linear bias and the number of central and satellite galaxies per halo for each sample.  We find that the effective linear bias does not evolve as steeply with redshift as found in \citetalias{krolewski_2020} and \citetalias{krolewski2021cosmological}, yet it is in rough agreement with the bias measurements from \citetalias{krolewski_2020} and \citetalias{krolewski2021cosmological}, obtained by cross-matching the \emph{unWISE} galaxies with spectroscopic
quasars from BOSS DR12 \cite{SDSSquasars_eBOSS_DR122017} and eBOSS DR14 \cite{eBOSS_dr14_2017} and galaxies from BOSS CMASS and LOWZ \cite{BOSSgal_2015}.  Future work to further constrain the redshift distributions of these samples, e.g., using DESI~\cite{2013arXiv1308.0847L}, will be extremely useful.

\citetalias{krolewski_2020} and \citetalias{krolewski2021cosmological} used HOD-populated $N$-body mocks to assess the redshift evolution of the bias within each sample, and to test the cosmology modeling pipeline. Their models were adjusted to match the observed galaxy auto-correlation, CMB lensing cross-correlation, and bias evolution (as measured from cross-correlations with spectroscopic samples in narrow redshift bins),
aiming for approximate ($\sim 10\%$ level) agreement.
In contrast, our analysis provides a more systematic and quantitative fit
to the angular power spectra, and thus supersedes
the HOD approach taken in \citetalias{krolewski_2020}
and \citetalias{krolewski2021cosmological}.  To our knowledge, this analysis is the first high-precision HOD model fit to the clustering and lensing measurements for the \emph{unWISE} galaxies.

The HOD constraints obtained in this analysis can be further used to study ionized gas residing in the \emph{unWISE} galaxies, e.g., to probe its pressure profile through the thermal Sunyaev-Zel'dovich effect in the halo model (e.g.,~\cite{Vikram2017,Hill2018,pandey2020,Koukoufilippas:2019ilu, pandey2021crosscorrelation}).  When combined with the results for the \emph{unWISE} gas density profile obtained with kinematic Sunyaev-Zel'dovich effect measurements \cite{Kusiak_2021}, it is possible to constrain the thermodynamics of gas in \emph{unWISE} galaxies \cite{Battaglia2017}, as done in, e.g., Ref.~\cite{schaan2020act, amodeo2021, vavagiakis_2021} for BOSS CMASS galaxies.  The HOD approach also opens the doors to study other cross-correlations involving \emph{unWISE} galaxies in the halo model framework, enabling more detailed characterization of the galaxies, dark matter, ionized gas, neutral gas, thermal dust, and other components associated with the galaxies in these enormous samples.

Throughout this analysis, we assume a flat $\Lambda$CDM cosmology with \emph{Planck} 2018 best-fit parameter values (last column of Table~II of Ref.~\cite{Planck2018}): $\omega_{cdm}= 0.11933$, $\omega_b=0.02242$, $H_0 = 67.66$ km/s/Mpc, $\ln(10^{10}A_s)=3.047$ and $n_s=0.9665$ with $k_\mathrm{pivot}=0.05\,\mathrm{Mpc}^{-1}$, and $\tau_\mathrm{reio}=0.0561 $.  All error bars, unless stated otherwise, are 1$\sigma$ and represent the 68\% confidence intervals. In our analysis, we work in units of $M_\odot/h$ for masses and we adopt the $M_{200c}$ halo mass definition everywhere, i.e., the mass enclosed within the spherical region whose density is 200 times the critical density of the universe, and the corresponding mass-dependent radius $r_{200c}$, which encloses mass $M_{200c}$. 

The paper is organized as follows. We start by describing the two theoretical building blocks for this analysis, the halo occupation distribution in Section~\ref{sec:HOD} and the halo model in Section~\ref{sec:HM}. In the remainder of Section~\ref{sec:theory}, we give detailed prescriptions for the angular power spectra used in this work in the halo model. Then in Section~\ref{sec:data} we present the data: the \emph{unWISE} galaxy catalog and \emph{Planck} CMB lensing map, along with the pipeline to obtain the desired auto- and cross-correlation measurements.  Section~\ref{sec:HODfitting} discusses the HOD model and parameter fitting. In Section~\ref{sec:results}, we present the results of fitting the auto- and cross-correlations to the HOD model, and in Section~\ref{sec:discussion} we discuss the results and how the obtained constraints can be further utilized.

\section{Theory}
\label{sec:theory}
In this section we describe the formalism that we use to model the observables of interest, namely, the galaxy-galaxy angular power spectra, $C_\ell^{gg}$, and the CMB lensing-galaxy cross-power spectra, $C_\ell^{g \kappa_{\rm{cmb}}}$. 
The crucial ingredient is the HOD model described in subsection~\ref{sec:HOD}. We then give the full halo model expressions for the angular power spectra in subsection~\ref{sec:HM}. 

\subsection{Galaxy Halo Occupation Distribution}
\label{sec:HOD}
The galaxy HOD is a statistical framework that describes how galaxies populate the underlying dark matter halo distribution. In this approach, dark matter halos contain two types of galaxies: satellites and centrals. Each halo can host either one central galaxy that is located in the center of a halo or no centrals at all. Satellite galaxies, on the other hand, are distributed within the host dark matter halo according to a specified profile. 
The number of satellites per halo is not limited in this approach. Following the DES Year 3 (DES-Y3) galaxy halo model analysis~\cite{Zacharegkas2021} and other previous works, we adopt the HOD model introduced in Zheng \textit{et al.} \cite{Zheng_2007}, and developed in Zehavi \textit{et al.} \cite{Zehavi_11}, parametrized by a number of HOD parameters, which we describe below. 

In this model, the expectation value for the number of central galaxies  $N_c$ in a halo of mass $M$ is given by
\begin{equation}
    N_c (M) =\frac{1}{2}\left[1+\mathrm{erf}\left(\frac{\log M - \log M_\mathrm{min}^\mathrm{HOD}}{\sigma_{\log M}}\right)\right] \,,
    \label{eq:N_c}
\end{equation}
where $M_\mathrm{min}^\mathrm{HOD}$ is the characteristic minimum mass of halos that can host a central galaxy,  $\sigma_{\mathrm{log} M}$, is the width of the cutoff profile \cite{Zheng_2007}, and \emph{erf} denotes the error function. 

The expectation value for the number of satellite galaxies $N_s$ in a halo is given by a power law and coupled to $N_c$ in the following way:
\begin{equation}
    N_s(M) = N_c(M) \left[\frac{M-M_0}{M_1^\prime}\right]^{\alpha_s},
    \label{eq:N_s}
\end{equation}
where $\alpha_s$ is the index of the power law of the satellite profile, $M_0$ is the mass scale above which the number of satellites grows, and $M_1^\prime$ sets the amplitude.  

In total, this standard HOD prescription consists of five free parameters; two for the central galaxies ($M_\mathrm{min}^\mathrm{HOD}$,  $\sigma_{\mathrm{log} M}$,), and three for the satellite profile ($\alpha_s$, $M_0$, and $M_1^{\prime}$). Following the DES-Y3 HOD modeling in Ref.~\cite{Zacharegkas2021}, in our work we constrain  $\sigma_{\rm{log}M}$, $\alpha_s$,  $M_\mathrm{min}^\mathrm{HOD}$, and $M_1^{\prime}$, and set $M_0=0$. In this case the $M_1^{\prime}$ parameter denotes the mass scale at which one satellite galaxy per halo is found. Typical values of these parameters for the DES-Y3 galaxies can be found in Ref.~\cite{Zacharegkas2021} (note that the masses there are in units of $M_\odot$). In Table~\ref{table:priors} we present the priors on these parameters used in our analysis, largely motivated by the DES-Y3 priors, but broadened in some cases to encompass the range preferred by the \emph{unWISE} galaxy samples, as determined by initial, exploratory runs, where the posterior distributions were hitting the edges of some of the HOD priors.

      \begin{table}[t!]
        \setlength{\tabcolsep}{16pt}
        \renewcommand{\arraystretch}{1.5}
        \begin{tabular}{ |c|c|c|c|} 
        \hline
         \textbf{Parameter}  & \textbf{Prior Blue} & \textbf{Prior Green} &  \textbf{Prior Red} \\ 
        \hline\hline
        $\sigma_{\mathrm{log} M}$& 0.01--1.20   &  0.01--2.0 &  0.01--2.00 \\ 
        $\alpha_s$ & 0.10--2.50  & 0.10--2.50 &  0.10--2.50\\ 
        $\mathrm{log}(M_\mathrm{min}^\mathrm{HOD})$  & 10.85--12.85 & 10.85--14.35   & 10.85--15.85\\ 
        $\mathrm{log}(M_{1}^`)$  & 11.35--13.95   & 11.35--14.85  & 11.35--15.85\\ 
        $\lambda$ & 0.10--1.80   & 0.10--3.00 & 0.10--3.00\\
        $10^{7}A_{\mathrm{SN}}$ & -2.00--2.00 &  -2.00--2.00 & -3.00--3.00 \\
        \hline
        \end{tabular}
        \caption{Prior ranges for the six model parameters, $\{ \alpha_{\mathrm{s}}$, $\sigma_{\mathrm{log} M}$, $M_\mathrm{min}^\mathrm{HOD}$, $M^{\prime}_1$, $\lambda$, $A_{\mathrm{SN}} \}$, varied in the joint fit for each of the \emph{unWISE} samples. All priors are uniform in the quantities given in the first column. Details of the fitting procedure are presented in Section~\ref{sec:HODfitting}. } 
        \label{table:priors}
        \end{table}

\subsection{Angular Power Spectra in the Halo Model Formalism}
\label{sec:HM}
In this section we describe the halo model and present its predictions for the cross- and auto-correlation power spectra used in our analysis, namely galaxy-CMB lensing  $C_{\ell}^{g \kappa_{\rm{cmb}} }$, galaxy-galaxy $C_{\ell}^{gg}$, CMB lensing-lensing magnification $C_{\ell}^{\kappa_{\rm{cmb}} \mu_g}$, galaxy-lensing magnification $C_{\ell}^{g \mu_g}$, and lensing magnification-lensing magnification $C_{\ell}^{\mu_g \mu_g}$, which are defined below.  Here and in all of the following, we use ``galaxy'' to refer to ``galaxy number overdensity''.  For the numerical implementation  we use the publicly available code \verb|class_sz|, version \verb|v1.0|\footnote{\href{https://github.com/borisbolliet/class_sz}{https://github.com/borisbolliet/class\_sz}} \cite{Bolliet:2017lha}, an extension of \verb|CLASS| \cite{CLASS} version \verb|v2.9.4|, which enables halo model computations of various cosmological observables.  

\subsubsection{General Formalism}
\label{subsec:HMframework}

The halo model is a formalism that uses dark matter halos to build an analytic model for the nonlinear matter density field and other cosmological fields~\cite[see, e.g.,][and references therein]{seljak2000,peacock2000,Cooray:2002dia}. Its main yet very simple assumption is that each particle can be part of only one dark matter halo. With a further assumption that all matter is enclosed in halos, it allows us to construct the entire density field or other cosmological fields, in a fully non-perturbative framework \cite{dodelson2020modern}. The halo model formalism enables computations of power spectra, bispectra, and higher moments of the matter density field. Here we present the halo model predictions for various cross- and auto-correlation angular power spectra relevant to this work. 

In the halo model, power spectra are computed as the sum of a 1-halo and a 2-halo term. The 1-halo term accounts for correlations between mass elements located within the same halo, while in the 2-halo term the mass elements are located in two distinct halos. Formally, the angular power spectrum $C_\ell^{ij}$ between two tracers $i$ and $j$ is defined as 
\begin{equation}
    C_\ell^{ij}=C_\ell^{ij,\mathrm{1h}} +  C_\ell^{ij,\mathrm{2h}} \label{eq:cl_ij}
\end{equation}
where $ C_\ell^{ij,\mathrm{1h}}$ is the 1-halo term of the correlation between $i$ and $j$ and $ C_\ell^{ij,\mathrm{2h}}$ the 2-halo term. 

The 1-halo term of the power spectrum between tracers $i$ and $j$ is an integral over halo mass, $M$, and redshift, $z$, given by  
\begin{equation}
    C_\ell^{ij,\mathrm{1h}}=\int_{z_\mathrm{min}}^{z_\mathrm{max}}  \mathrm{d} z \frac{\mathrm{d}^2 V}{\mathrm{d} z \mathrm{d} \Omega} \int_{M_\mathrm{min}}^{M_\mathrm{max}}  \mathrm{d}M \frac{\mathrm{d}n}{\mathrm{d}M} u_\ell^i(M,z) u_\ell^j(M,z), \label{eq:cl1h_ij}
\end{equation}
where $\mathrm{d} V $ is the cosmological volume element, defined in terms of redshift $z$ and comoving distance $\chi(z)$ to redshift $z$ as $\mathrm{d}V= \chi^2\mathrm{d}\chi=\frac{c\chi^2}{H}(1+z)\mathrm{d}\ln(1+z)$, with $H$ the Hubble parameter, $\mathrm{d} \Omega $ is the solid angle of this volume element,  $\mathrm{d}n/(\mathrm{d}M$) is the differential number of halos per unit mass and volume, defined by the halo mass function (HMF), where in our analysis we use the Tinker \textit{et al.} analytical fitting fuction~\cite{Tinker_2008}, and the quantities $u_\ell^i(M,z)$ and $u_\ell^j(M,z)$ are the multipole-space kernels of the various large-scale structure tracers of interest, e.g., CMB weak lensing or galaxy overdensity, which we define below. In \verb|class_sz|, we set the mass bounds of the integral to $M_\mathrm{min} = 7\times10^8 \, M_\odot/h$ and ${M_\mathrm{max}} = 3.5\times10^{15} \, M_\odot/h$ and the redshift bounds to $z_\mathrm{min}=0.005$ and $z_\mathrm{max} =4$, the latter dictated by the upper redshift limit of the \emph{unWISE} galaxy samples that we analyze.  We verify that all calculations are converged with these choices.  Further discussion of our modeling choices for the HMF and the satellite galaxy profile parametrization can be found in Appendices~\ref{app:HMF} and~\ref{app:profile}, respectively.

The 2-halo term of the power spectrum of tracers $i$ and $j$ is given by
\begin{equation}
C_\ell^{ij, \mathrm{2h}}=\int_{z_\mathrm{min}}^{z_\mathrm{max}}  \mathrm{d} z \frac{\mathrm{d}^2 V}{\mathrm{d} z \mathrm{d} \Omega}  \left| \int_{M_\mathrm{min}}^{M_\mathrm{max}}  \mathrm{d}M_i \frac{\mathrm{d}n}{\mathrm{d}M_i} b(M_i, z) u_\ell^i(M_i, z)  \right| \left| \int_{M_\mathrm{min}}^{M_\mathrm{max}} \mathrm{d}M_j \frac{\mathrm{d}n}{\mathrm{d}M_j}  b(M_j, z) u_\ell^j(M_j, z)  \right| P_{\mathrm{lin}}\left(\frac{\ell+\tfrac{1}{2}}{\chi},z\right),
\label{eq:cl2h_ij}
\end{equation}
where $P_{\mathrm{lin}}$ is the linear matter power spectrum (computed with \verb|CLASS| within  \verb|class_sz|) and $b(M, z)$ is the linear bias describing the clustering of the two tracers (e.g.,~\cite{gatti2021crosscorrelation, pandey2021crosscorrelation}). We model the linear halo bias using the Tinker \textit{et al.} (2010) \cite{Tinker_2010} fitting function.

\subsubsection{CMB lensing-galaxy cross-correlation}
\label{subsec:kg}

For the CMB weak lensing field, in the Limber approximation \cite{Limber, LimberExtended} the multipole-space kernel $u_\ell^{\kappa_{\rm cmb}}$ is 
\begin{equation}
     u_\ell^{\kappa_{\rm cmb}} (M, z) = W_{\kappa_{\rm cmb}}(z) u^m_\ell(M, z), 
    \label{eq:ulkappacmb}\\    
\end{equation}
where $u_{\ell}^m$ is the Fourier transform of the dark matter density profile (defined below), and the CMB lensing kernel $W_{\kappa_{\rm cmb}}(z)$ is 
\begin{equation}
    W_{\kappa_{\rm cmb}}(z) = \frac{3}{2}\Omega_m \frac{H_0^2}{\chi^2(z)} \frac{(1+z)}{H(z)} \frac{\chi(z)}{c}\frac{\chi(z_\star)-\chi(z)}{\chi(z_\star)},
    \label{eq:wkappacmb}
\end{equation}
where $\Omega_{m}$ is the matter density as a fraction of the critical density at $z = 0$, $z_\star \approx 1090$ is the redshift of the surface of last scattering, and $H_0$ is the present-day value of the Hubble parameter. 
For the Fourier transform of the dark matter density profile $u_{\ell}^m$, we model it using the usual truncated Navarro, Frenk, and White (NFW) dark matter profile \cite{Navarro_1997}, with truncation at $r_{\mathrm{out}} = \lambda r_{200c}$, which is given by an analytical formula \cite{Scoccimarro:2000gm} 
\begin{equation}
    u_\ell^m (M,z)= \frac{M}{\rho_{m,0}}\left(\cos(q)[\mathrm{Ci}((1+\lambda c_{200c})q)-\mathrm{Ci}(q)]+\sin(q) [\mathrm{Si}((1+ \lambda c_{200c})q)-\mathrm{Si}(q)]- \frac{\sin(\lambda c_{200c}q)}{(1+ \lambda c_{200c})q)} \right) f_{_\mathrm{NFW}}(\lambda c_{200c})
    \label{eq:ulm}
\end{equation}
where $\rho_{m,0}$ is the mean matter density az $z=0$, $\mathrm{Ci}(x)=\int_x^{\infty}\mathrm{d}t\cos(t)/t$ and $\mathrm{Si}(x)=\int_0^x\mathrm{d}t\sin (t)/t$ are the cosine and sine integrals, the $f_{_\mathrm{NFW}}$ function is given by
\begin{equation}
f_{_\mathrm{NFW}}(x)=[\ln(1+x)-x/(1+x)]^{-1},
\end{equation} 
and the argument $q$ is defined as 
\begin{equation}
    q=k \frac{r_{200c}}{c_{200c}} 
\end{equation}
where $k=(\ell +1/2)/\chi$ is the wavenumber 
and $c_{200c}$ is the concentration parameter computed with the concentration-mass relation defined in Ref.~\cite{Bhattacharya_2013}.

The galaxy overdensity multipole-space kernel $u_\ell^g(M,z)$ is 
\begin{equation}
u_\ell^g(M,z)=W_g(z)\bar{n}_g^{-1}\left[N_c+ N_s u_\ell^m(M,z)\right],
\label{eq:ulg}
\end{equation}
where $u_{\ell}^m$ is the Fourier transform of the dark matter density profile defined in Eq.~\ref{eq:ulm}, $N_c$ and $N_s$ are the expectation value for the number of centrals and satellites, given in Eq.~\ref{eq:N_c} and \ref{eq:N_s}, $\bar{n}_g$ is the mean number density of galaxies given by 
\begin{equation}
    \bar{n}_g(z) =\int_{M_\mathrm{min}}^{M_\mathrm{max}}\mathrm{d}M \frac{\mathrm{d}n}{\mathrm{d}M}\left(N_c+N_s\right),
    \label{eq:n_g}
\end{equation}
and $W_g(z)$ is the galaxy kernel defined as 
\begin{equation}
    W_g(z)= \frac{H(z)}{c}\frac{\varphi_g^\prime(z)}{\chi^2},
\label{eq:wgz}
\end{equation}
where $\varphi_g^\prime(z)$ is the normalized galaxy distribution of the given galaxy catalog
\begin{equation}
    \varphi_g^\prime(z) = \frac{1}{N_g^\mathrm{tot}}\frac{\mathrm{d}N_g}{\mathrm{d}z},\quad \mathrm{with}\quad N_g^\mathrm{tot}=\int \mathrm{d}z\frac{\mathrm{d}N_g}{\mathrm{d}z}.  \label{eq:varphig}
\end{equation}
We show the normalized galaxy distributions for the \emph{unWISE} samples in Section~\ref{sec:data} in Fig.~\ref{fig:dndz_COSMOS}, which were obtained by cross-matching the \emph{unWISE} objects with the \emph{COSMOS} catalog objects \citep{Laigle_2016}, as will be explained later in Section~\ref{sec:data}. 


\subsubsection{Galaxy-galaxy auto-power spectrum}
\label{subsec:gg}
The second correlation we consider is galaxy clustering. As described in Section~\ref{subsec:HMframework}, the 1-halo term for the galaxy-galaxy power spectrum is given by

\begin{equation}
    C_\ell^{gg,\mathrm{1h}}=\int_{z_\mathrm{min}}^{z_\mathrm{max}}  \mathrm{d} z \frac{\mathrm{d}^2 V}{\mathrm{d} z \mathrm{d} \Omega} \int_{M_\mathrm{min}}^{M_\mathrm{max}}  \mathrm{d}M \frac{\mathrm{d}n}{\mathrm{d}M} \langle |u_\ell^g(M,z)|^2 \rangle , 
\label{eq:clgg1h}
\end{equation}
where we cannot simply use the form of the galaxy multipole space kernel $u_\ell^g(M,z)$ squared, but rather require its second moment (see Section~2.2 in Ref.~\cite{vandenBosch2013}), which is given by (see Eqs. 15 and 16 in Ref.~\cite{Koukoufilippas:2019ilu})

\begin{equation}
\langle |u_\ell^g(M,z)|^2 \rangle=W_g(z)\bar{n}_g^{-2}\left[N_s^2 u_\ell^m(M,z)^2 + 2N_s u_\ell^m(M,z)\right],
\label{eq:ulg^2}
\end{equation}
where $N_s$ is the expectation value for the number of satellites, given in Eq.~\ref{eq:N_s}, and $\bar{n}_g$ is the mean number density of galaxies (Eq.~\ref{eq:n_g}). 

The 2-halo term of the galaxy-galaxy power spectrum is given by 
\begin{equation}
    C_\ell^{gg,\mathrm{2h}}=\int_{z_\mathrm{min}}^{z_\mathrm{max}}  \mathrm{d} z \frac{\mathrm{d}^2 V}{\mathrm{d} z \mathrm{d} \Omega} \left|\int_{M_\mathrm{min}}^{M_\mathrm{max}}  \mathrm{d}M \frac{\mathrm{d}n}{\mathrm{d}M} b(M,z) u_\ell^g(M,z) \right|^2 P_{\mathrm{lin}}\left(\frac{\ell+\tfrac{1}{2}}{\chi},z\right),\label{eq:clgg2h}
\end{equation}
where $b(M,z)$ is the Tinker \textit{et al.} (2010) \cite{Tinker_2010} bias, and $u_\ell^g(M,z)$ is the first moment of the galaxy multipole kernel given in Eq.~\ref{eq:ulg}. Note that we do not consider cross-correlations between different galaxy samples in this work, but only the auto-correlation of each sample.

\subsubsection{CMB lensing-galaxy lensing magnification cross-correlation}
\label{subsec:km}
An additional quantity that must be taken into account in our model is galaxy lensing magnification. The magnification bias contribution arises from the fact that the luminosity function of a galaxy sample is steep at the faint end, near the threshold for detection. Magnification bias is characterized by the logarithmic slope of the galaxy number counts as a function of apparent magnitude $m$ near the magnitude limit of the survey defined as $s = \frac{\mathrm{d log_{10}} N}{\mathrm{d}m}$.

The observed galaxy number density fluctuation $\delta_g^{\rm{obs}}$ is the sum of the intrinsic galaxy overdensity $\delta_g$ and the magnification bias contribution $\mu_g$:
\begin{equation}
    \delta_g^{\rm obs} = \delta_g + \mu_g.
\end{equation}
The galaxy magnification bias gives a non-zero contribution to each correlation that includes the galaxy overdensity field $\delta_g$, i.e., $C_{\ell}^{i g, \rm{ obs}} = C_{\ell}^{i g} +  C_{\ell}^{i \mu_g}$, where $i$ is a tracer. As we show below when computing our predictions, for the low-redshift (blue) \emph{unWISE} galaxies the lensing magnification bias is negligible, but for the higher redshift samples (\emph{unWISE} green and red), it is usually non-negligible \cite{Kusiak_2021, krolewski_2020}.

Therefore, the observed cross-correlation of the CMB lensing and galaxy overdensity fields includes a contribution from the lensing magnification field $\mu_g$:

\begin{equation}
    C_{\ell}^{\kappa_{\rm{cmb}} g, \rm{ obs}} = C_{\ell}^{ \kappa_{\rm{cmb}} g} +  C_{\ell}^{\kappa_{\rm{cmb}} \mu_g}, 
\end{equation}
where $C_{\ell}^{ \kappa_{\rm{cmb}} g}$ is --- as defined in Eq.~\ref{eq:cl_ij} --- the sum of the 1- and 2-halo terms, and the exact prescription for this cross-correlation is given in Section~\ref{subsec:kg}. 
The CMB lensing-lensing magnification term $C_{\ell}^{\kappa_{\rm{cmb}} \mu_g}$ can be similarly written down in the halo model as 
\begin{equation}
     C_{\ell}^{\kappa_{\rm{cmb}} \mu_g} =  C_{\ell}^{\kappa_{\rm{cmb}} \mu_g, \mathrm{1h}} + C_{\ell}^{\kappa_{\rm{cmb}} \mu_g, \mathrm{2h}}. 
\end{equation}
where the 1- and 2-halo terms  can be computed according to the prescription in Section~\ref{subsec:HMframework}. 

The lensing magnification multipole-space kernel $u_\ell^{\mu_g}$ is given by 
\begin{equation}
    u_\ell^{\mu_g}(M,z) = (5s-2) W_{\mu_g}(z) u^m_\ell(M,z),
    \label{eq:ulmu}
\end{equation}
where $u_{\ell}^m$ is defined in Eq.~\ref{eq:ulm} and the lensing magnification bias kernel $W_\mathrm{\mu_{_{g}}}$ is

\begin{equation}
 W_\mathrm{\mu_{_{g}}}(z)=\frac{3}{2}\frac{\Omega_\mathrm{m}(H_0/c)^2}{\chi^2(z)}(1+z) \chi(z) I_g(z)\quad\mathrm{with}\quad I_g(z)=\int_z^{z_\mathrm{max}}\mathrm{d}z_g \varphi^\prime(z_g)\frac{\chi(z_g)-\chi(z)}{\chi(z_g)}.\label{eq:wkg}
\end{equation}
where $\chi(z_g)$ is the comoving distance to galaxies at redshift $z_g$ and $\varphi^\prime$ is the normalized galaxy distribution from Eq.~\ref{eq:varphig}.


\subsubsection{Galaxy-galaxy lensing magnification cross-correlation}
\label{subsec:gm}

Similarly, the observed auto-correlation of a galaxy overdensity map includes contributions from the lensing magnification field $\mu_g$,
\begin{equation}
    C_{\ell}^{gg, \rm{ obs}} = C_{\ell}^{gg} + 2 C_{\ell}^{g \mu_g} + C_{\ell}^{\mu_g \mu_g},
    \label{eq:clgg_obs}
\end{equation}
where $ C_{\ell}^{gg}$ is defined above in Section~\ref{subsec:gg}, $C_{\ell}^{\mu_g \mu_g}$ and $C_{\ell}^{g \mu_g}$ can analogously be written as a sum of 1-halo and 2-halo terms, and computed according the prescription presented in this Section, with the multipole-space kernels $u^g_{\ell}$ and $u_{\ell}^{\mu_g}$ defined in Eq.~\ref{eq:ulg} and \ref{eq:ulmu}. The 1-halo and 2-halo terms of $C_{\ell}^{\mu_g \mu_g}$  are:

\begin{equation}
    C_\ell^{\mu_g \mu g,\mathrm{1h}}=\int_{z_\mathrm{min}}^{z_\mathrm{max}}  \mathrm{d} z \frac{\mathrm{d}^2 V}{\mathrm{d} z \mathrm{d} \Omega} \int_{M_\mathrm{min}}^{M_\mathrm{max}}  \mathrm{d}M \frac{\mathrm{d}n}{\mathrm{d}M} |u_\ell^{\mu_g}(M,z)|^2  ,
\label{eq:clmm1h}
\end{equation}

\begin{equation}
    C_\ell^{\mu_g \mu_g,\mathrm{2h}}=\int_{z_\mathrm{min}}^{z_\mathrm{max}}  \mathrm{d} z \frac{\mathrm{d}^2 V}{\mathrm{d} z \mathrm{d} \Omega} \left|\int_{M_\mathrm{min}}^{M_\mathrm{max}}  \mathrm{d}M \frac{\mathrm{d}n}{\mathrm{d}M} b(M,z) u_\ell^{\mu_g}(M,z) \right|^2 P_{\mathrm{lin}}\left(\frac{\ell+\tfrac{1}{2}}{\chi},z\right),\label{eq:clmm2h}. 
\end{equation}

From now on we denote the observed galaxy field (i.e., including the lensing magnification contributions) generally as $``g"$, unless confusion could arise. 

\subsection{Parameter Dependence}
Out of all the parameters presented in this section, following the standard HOD implementation \cite{Zheng_2007} and the DES-Y3 analysis \cite{Zacharegkas2021}, we consider four varying HOD parameters {$\alpha_{\mathrm{s}}$, $\sigma_{\mathrm{log} M}$, $M_\mathrm{min}^\mathrm{HOD}$, $M^{\prime}_1$}, as well as the parameter $\lambda$ which quantifies the NFW truncation radius $r_{\mathrm{out}}$ (Eq.~\ref{eq:ulm}). Appendix~\ref{app:profile} discusses the subtle difference between the parameter $\lambda$ considered in this analysis, and the parametrization between the satellite galaxies' radial distribution and the matter density profile $a\equiv c_\mathrm{sat}/c_\mathrm{dm}$ considered in the DES-Y3 \cite{Zacharegkas2021} analysis.

\begin{figure}
    \centering
    \includegraphics[scale=0.5]{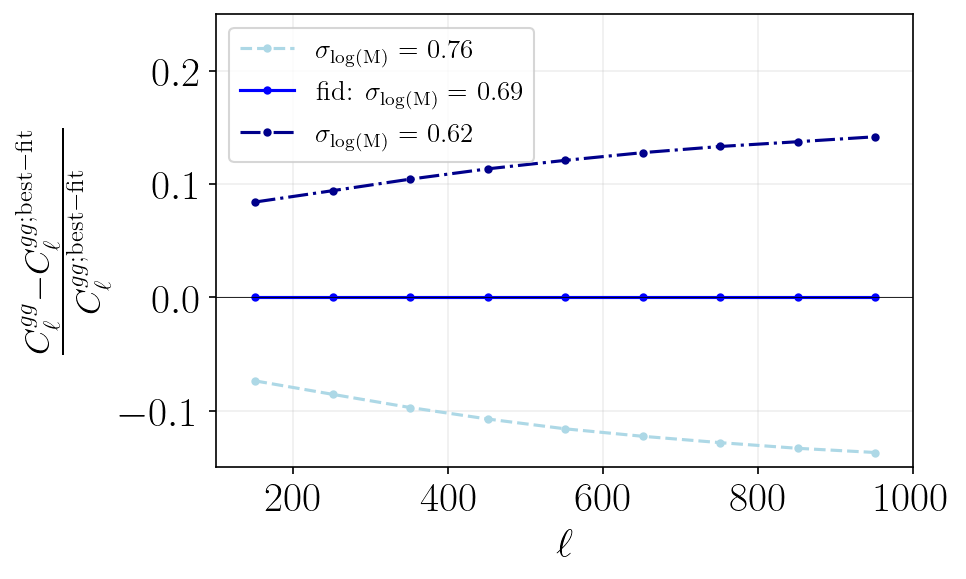}
    \includegraphics[scale=0.5]{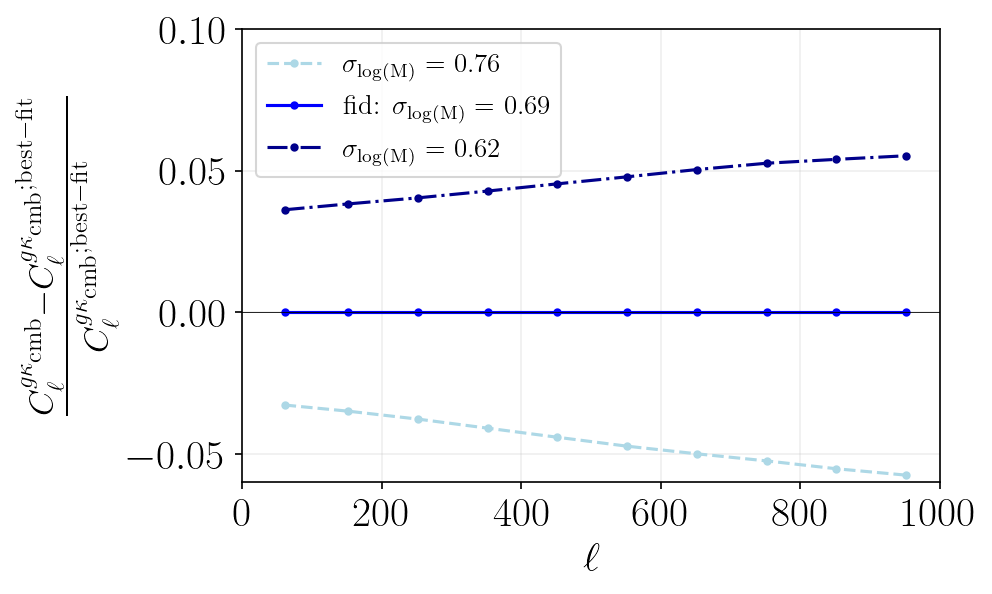}
    \includegraphics[scale=0.5]{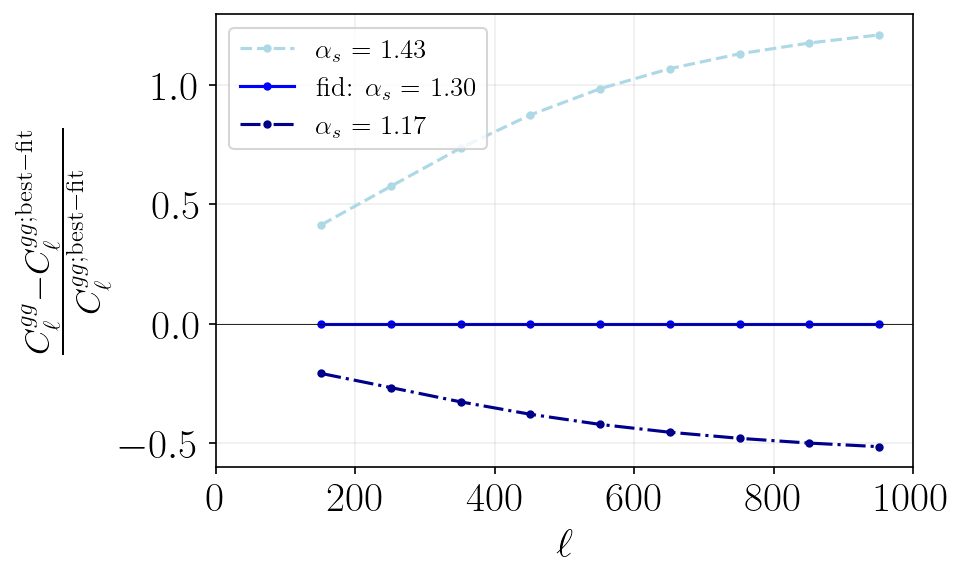}
    \includegraphics[scale=0.5]{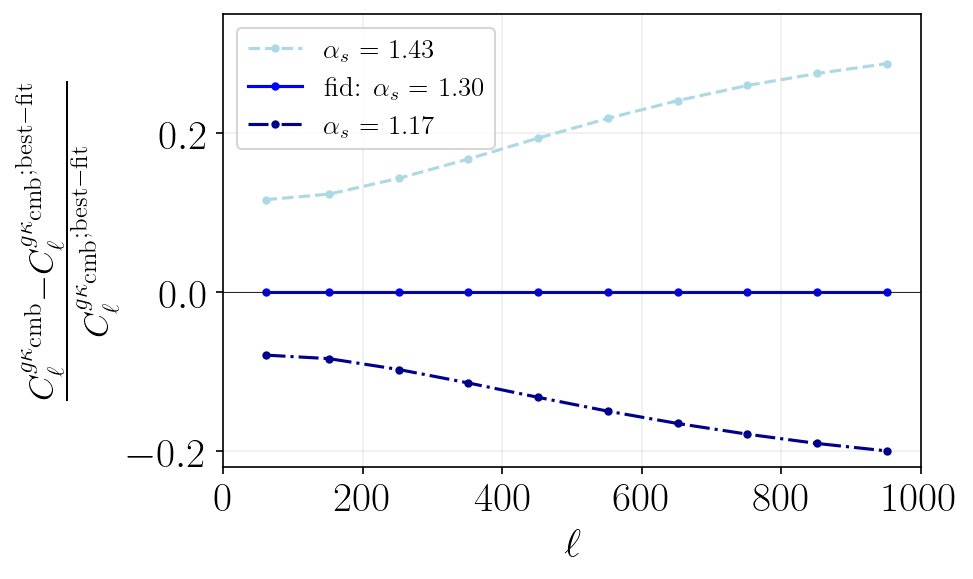}
    \caption{Impact of varying selected HOD parameters in our model. Here we show the fractional change in galaxy auto power spectrum $(C_\ell^{gg}- C_{\ell}^{gg;\mathrm{best-fit}})/C_{\ell}^{gg;\mathrm{best-fit}}$ (left) and in CMB lensing cross power spectrum $(C_\ell^{g\kappa_{\rm{cmb}}}-C_{\ell}^{g\kappa_{\rm{cmb}};\mathrm{best-fit}})/C_{\ell}^{g\kappa_{\rm{cmb}};\mathrm{best-fit}}$ (right) for the \emph{unWISE} blue sample, where $C_{\ell}^{gg;\mathrm{best-fit}}$ is the prediction computed for the best-fit values of the HOD parameters (Table~\ref{table:hod_results}), and $C_{\ell}^{gg}$ is a prediction computed with varying the best-fit value of the $\alpha_{\mathrm{s}}$ and $\sigma_{\mathrm{log} M}$ parameters by $\pm 10$\% (and similarly for galaxy-CMB lensing cross power spectra). The legends of each plot give the corresponding exact values for $\alpha_{\mathrm{s}}$ or $\sigma_{\mathrm{log} M}$. Top left: fractional change in $C_\ell^{gg}$ for $\sigma_{\mathrm{log} M}$. Top right: fractional change in $C_\ell^{g \kappa_{\rm{cmb}}}$ for $\sigma_{\mathrm{log} M}$. Bottom left: fractional change in $C_\ell^{gg}$ for $\alpha_{\mathrm{s}}$. Bottom right: fractional change in $C_\ell^{g \kappa_{\rm{cmb}}}$ for $\alpha_{\mathrm{s}}$. }
    \label{fig:varyHOD}
\end{figure}

In Fig.~\ref{fig:varyHOD} we show the impact of varying selected HOD parameters on our halo model prediction (Section~\ref{sec:HM}) computed with \verb|class_sz|. In Fig.~\ref{fig:varyHOD}, we present the fractional change in the galaxy-galaxy auto power spectrum $(C_\ell^{gg}- C_{\ell}^{gg;\mathrm{best-fit}})/C_{\ell}^{gg;\mathrm{best-fit}}$ and the galaxy-CMB lensing cross-power spectrum $(C_\ell^{g\kappa_{\rm{cmb}}}-C_{\ell}^{g\kappa_{\rm{cmb}};\mathrm{best-fit}})/C_{\ell}^{g\kappa_{\rm{cmb}};\mathrm{best-fit}}$ for the \emph{unWISE} blue sample, where $C_{\ell}^{gg;\mathrm{best-fit}}$ and $C_{\ell}^{g\kappa_{\rm{cmb}};\mathrm{best-fit}}$ denote the prediction computed for the best-fit values of the HOD parameters (see Table ~\ref{table:hod_results} and Section~\ref{sec:results}, where we discuss the final results), and $C_\ell^{gg}$ and $C_\ell^{g \kappa_{\rm{cmb}}}$ denote the predictions computed when varying $\sigma_{\mathrm{log} M}$ and  $\alpha_{\mathrm{s}}$ by 10\%.

From Fig.~\ref{fig:varyHOD}, we note that a 10\% variation in $\sigma_{\mathrm{log} M}$ around its best-fit value for the blue sample changes the computed galaxy-galaxy auto-power spectrum by up to $\approx 15$\%, while for the galaxy-CMB lensing cross-power spectrum, the change is smaller, $\lesssim 5$-6\%. In the case of increasing $\alpha_{\mathrm{s}}$ by 10\%, the increase in the galaxy-galaxy power spectrum is significant, exceeding 100\%, while when decreasing $\alpha_{\mathrm{s}}$ by 10\%, the decrease is only around 50\%. For the CMB lensing cross-correlation prediction, varying $\alpha_{\mathrm{s}}$ has an impact of changing the prediction by 20-30\%. Some of these changes might appear quite large, yet we note that the computed predictions depend on specific values of the other parameters (and their combination) where we adopted our model (see Section~\ref{sec:intro} and Section~\ref{sec:theory}) and the best-fit model values (Table~\ref{table:hod_results}). We chose parameters that quantify the central ($\sigma_{\mathrm{log} M}$) and satellite ($\alpha_{\mathrm{s}}$) contributions to the HOD. The fractional changes for these parameters are similar for the green and red samples.
The analysis is performed at fixed cosmology as noted in Sec.~\ref{sec:intro}. We discuss the impact of varying selected cosmological parameters in Appendix~\ref{app:cosmo}.

\section{Data}
\label{sec:data}
In this section, we describe the \emph{unWISE} galaxy catalog and the \emph{Planck} CMB lensing map used to measure the auto- and cross-correlation $C_\ell^{gg}$ and $C_\ell^{g \kappa_{\rm{cmb}} }$.  We measure the angular power spectra using the pipeline built in \citetalias{krolewski_2020} and \citetalias{krolewski2021cosmological}, with a couple of minor modifications, reviewed hereafter. 

\subsection{\emph{unWISE} galaxy catalog}
\label{subsec:unwise}
The \emph{unWISE} galaxy catalog \cite{Schlafly_2019, krolewski_2020, krolewski2021cosmological} is constructed from the \emph{Wide-Field Infrared Survey Explorer} (\emph{WISE}) satellite mission, including the post-hibernation \emph{NEOWISE} data. \emph{unWISE} contains over 500 million galaxies over the full sky, spanning redshifts $0 < z \lesssim 2$. It is divided into three subsamples: blue, green, and red, which we describe in more detail below. 
    
    \begin{table}[t!]
    \centering
    \setlength{\tabcolsep}{10pt}
    \renewcommand{\arraystretch}{1.3}
    \begin{tabular}{| c| c |c|c|c| } 
    \hline
   \emph{unWISE} & $W2 <$ & $W2 >$ & $W1 - W2 > x$ & $W1 - W2 < x$  \\ 
    \hline\hline
    blue & 15.5 & 16.7 & & $\frac{(17 - W2)}{4}+0.3$ \\
    green & 15.5 & 16.7 & $\frac{(17 - W2)}{4}+0.3$ & $\frac{(17 - W2)}{4}+0.8$\\ 
    red & 15.5 & 16.2 & $\frac{(17 - W2)}{4}+0.8$ &  \\ 
    \hline
    \end{tabular}
    \caption{Cuts made on infrared color and magnitude in the W1 (3.4 $\mu$m) and W2 (4.6 $\mu$m) bands in the \emph{WISE} data to construct the \emph{unWISE} catalogs (see \cite{Schlafly19}, \citetalias{krolewski_2020}, and \citetalias{krolewski2021cosmological} for further details).}
    \label{table:unwise_cuts}
    \end{table}
    
The \emph{WISE} satellite mapped the entire sky at 3.4, 4.6, 12, and 22 $\mu$m (W1, W2, W3, and W4) with angular resolution of $6.1''$, $6.4''$, $6.5''$, and $12''$, respectively \cite{Wright_2010}. \emph{unWISE} galaxies are selected from the \emph{WISE} objects based on cuts on infrared galaxy color and magnitude in W1 and W2, which are summarized in Table~\ref{table:unwise_cuts}. Stars are removed from the catalog by cross-matching with \emph{Gaia} catalogs. More details on the construction of the \emph{unWISE} catalog are given in \citetalias{krolewski_2020} and \citetalias{krolewski2021cosmological}, and summarized in \cite{Kusiak_2021}.

\begin{figure}
    \centering
    \includegraphics[scale=0.5]{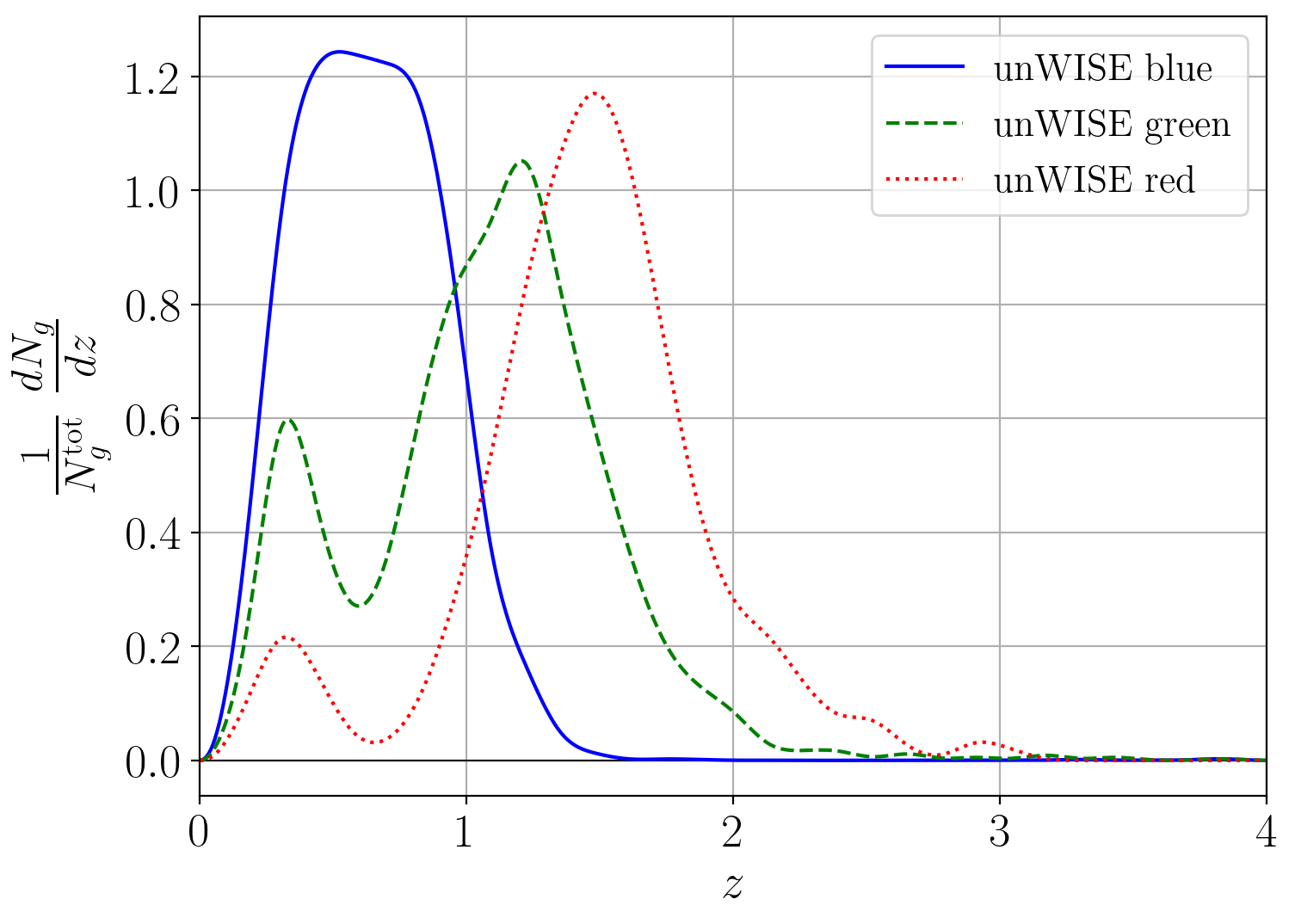}
    \caption{Normalized redshift distributions $\frac{1}{N_g^{\rm{tot}}}\mathrm{d}N_g/\mathrm{d}z$ for each of the \emph{unWISE} galaxy samples: blue (solid), green (dashed), and red (dotted), obtained by cross-matching the \emph{unWISE} objects with the \emph{COSMOS} catalog. Other important characteristics of the \emph{unWISE} samples are presented in Table~\ref{table:unwise}. \label{fig:dndz_COSMOS}}
\end{figure}

Based on W1 and W2 cuts, \emph{unWISE} is further divided into three subsamples (blue, green, and red) of mean redshifts $\bar{z} = 0.6$, 1.1, and 1.5, respectively. The redshift distribution of each of the subsamples, as described in \citetalias{krolewski_2020} and \citetalias{krolewski2021cosmological}, can be obtained by either 1) cross-correlating \emph{unWISE} galaxies with spectroscopic \emph{BOSS} galaxies and \emph{eBOSS} quasars or 2) by direct cross-matching where \emph{unWISE} galaxies are
directly matched to \emph{COSMOS} galaxies, which have precise 30-band photometric redshifts. The first method allows a direct measurement of $b(z) \mathrm{d}N_g/\mathrm{d}z$, the product of the galaxy bias and redshift distribution, while the \emph{COSMOS} cross-matching measures $\mathrm{d}N_g/\mathrm{d}z$ only. Since the two methods are fully consistent (\citetalias{krolewski_2020} and \citetalias{krolewski2021cosmological}), and our halo model approach requires only $\mathrm{d}N_g/\mathrm{d}z$ (see Eq.~\ref{eq:varphig}), we use the COSMOS cross-matched redshift distributions of the \emph{unWISE} galaxies. These normalized distributions are presented in Fig.~\ref{fig:dndz_COSMOS}. As shown on this figure, the blue sample peaks at redshift $z \approx 0.6$, the green one peaks at $z \approx 1.2$, with a second smaller bump at $z \approx 0.3$, and the red one at $z \approx 1.5$, with a smaller peak also at $z \approx 0.3$.
Other important characteristics of each sample are presented in Table~\ref{table:unwise}: the mean redshift $\bar{z}$ and the approximate width of the redshift distribution $\delta_z$, both measured by matching to objects with high-precision photometric redshifts in the \emph{COSMOS} field \citep{Laigle_2016}; the number density per deg$^2$ $\bar{n}$,; and the response of the number density to galaxy magnification $s$ defined as $s = \mathrm{d} \log_{10} N_g/ \mathrm{d}m$, needed to compute the lensing magnification terms.  

Given knowledge of typical galaxy SEDs (e.g., Ref.~\cite{Silva_1998}), we can qualitatively assess the regions of the spectrum that are responsible for the emission in the \emph{unWISE} bands (Table~\ref{table:unwise_cuts}). The W1 band covers the range $3-3.8 \, \mu {\rm m}$ and the W2 band about $4-5 \, \mu {\rm m}$. From the plots in Ref.~\cite{Silva_1998}, we note that the turning point from stellar-dominated to thermal-dust-dominated emission happens at about $2-3 \, \mu {\rm m}$ for starbust galaxies and at about $4-6 \, \mu {\rm m}$ for star-forming galaxies. The polycyclic aromatic hydrocarbons (PAH) lines located at 3.3, 6.25, 7.6, 8.6, 11.3, 12.7 $\, \mu {\rm m}$ (at $z=0$) also contribute to the emission. 
Looking at the redshift distribution of the three \emph{unWISE} samples (Fig.~\ref{fig:dndz_COSMOS}) and redshifting the W1/W2 bands accordingly, we find that the red sample's emission is stellar-dominated (except for the low-$z$ bump). For the green and blue sample (lower redshift), the emission is stellar-dominated as well, unless there are starburst galaxies in the sample (in which case there is a contribution from a mix of thermal dust and stellar emission). Comparing the star formation rates (SFR) for the \emph{unWISE} galaxies obtained with the \emph{COSMOS} SFRs (Appendix A of \citetalias{krolewski_2020}) with those in Ref.~\cite{Silva_1998}, we estimate that the blue sample has 10-30\% starburst galaxies, while the green one consists of 30-50\% starbursts. Since the turning points between stellar-dominated and dust-dominated emission for the starbust galaxies will fall within the redshifted W1 and W2 bands for the green and blue samples, these samples will have a mix of stellar-dominated and dust-dominated emission at the quoted level.  To summarize, the emission in the \emph{unWISE} samples is approximately as follows: 70-90\% stellar-dominated emission and 10-30\% a mixture of stellar and thermal dust emission, with a contribution from the $3.3 \, \mu {\rm m}$ PAH emission for the blue sample; 50-70\% stellar-dominated and 30-50\% mixture, with a small contribution from the $3.3 \, \mu {\rm m}$ PAH emission for the green sample; and stellar-dominated for red.

\begin{figure}
    \centering
    \includegraphics[scale=0.4]{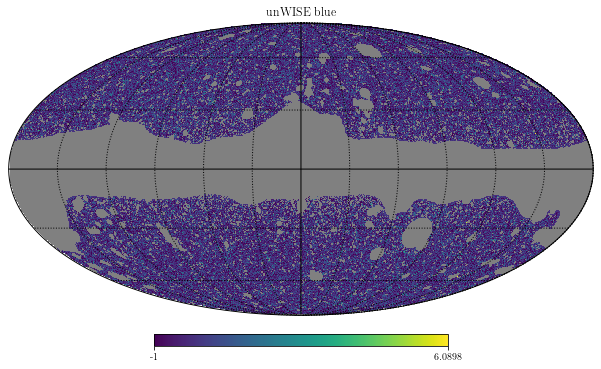}
    \includegraphics[scale=0.4]{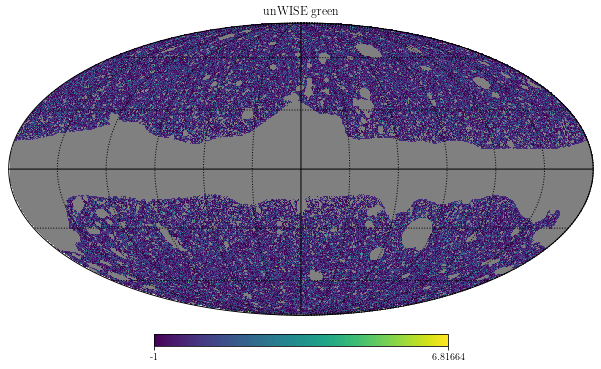}
    \includegraphics[scale=0.4]{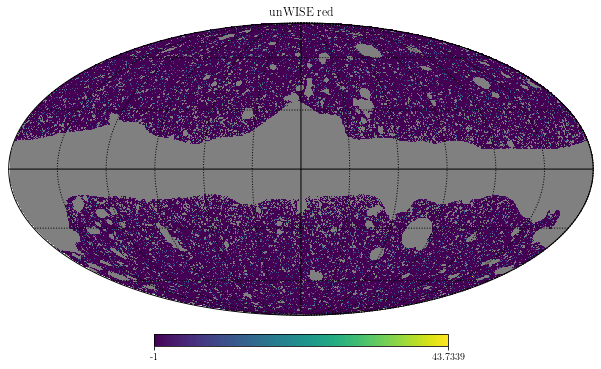}
    \caption{ Galaxy overdensity $\delta_g$ maps for each of the \emph{unWISE} samples, namely:  blue (top left), green (top right), and  red (bottom panel), with the mask applied. The masked regions have a value of 0 and are shown in grey. The galaxy samples are discussed in Section~\ref{sec:data} in more details. \label{fig:unwise_maps}}
\end{figure}

The galaxies in each \emph{unWISE} sample are populated into a \verb|HealPix| map of resolution $N_{\rm side}=2048$, and a galaxy overdensity $\delta_g \equiv (n-\bar{n})/\bar{n}$ map is constructed, where $n$ denotes the number of galaxies in each pixel, and $\bar{n}$ the mean number of galaxies in the map. We show the final \emph{unWISE} overdensity maps in Fig.~\ref{fig:unwise_maps}. 

    \begin{table}[t!]
    \setlength{\tabcolsep}{10pt}
    \renewcommand{\arraystretch}{1.7}
    \begin{tabular}{| c| c |c|c|c| } 
    \hline
    \emph{unWISE} & $\bar{z}$ & $\delta_{z}$ & $\bar{n}$ & $s$ \\ 
    \hline\hline
    blue & 0.6 & 0.3 & 3409 & 0.455\\
    green & 1.1& 0.4 & 1846 & 0.648\\ 
    red &  1.5 & 0.4 & 144 & 0.842\\ 
    \hline
    \end{tabular}
    \caption{Important properties of each \emph{unWISE} sample: $\bar{z}$, mean redshift; $\delta_z$, approximate width of the redshift distribution, both obtained from $\mathrm{d}N_g/\mathrm{d}z$ as measured by matching to objects with precise photometric redshifts in the \emph{COSMOS} field \citep{Laigle_2016} (see Section~\ref{sec:data}); $\bar{n}$, the number density per deg$^2$; and $s$, the response of the number density to lensing magnification $s = \mathrm{d} \log_{10} N_g/ \mathrm{d}m$. See \citepalias{krolewski_2020, krolewski2021cosmological}
    and \cite{Schlafly19} for further details.}
    \label{table:unwise}
    \end{table}

\begin{figure}
    \centering
    \includegraphics[scale=0.4]{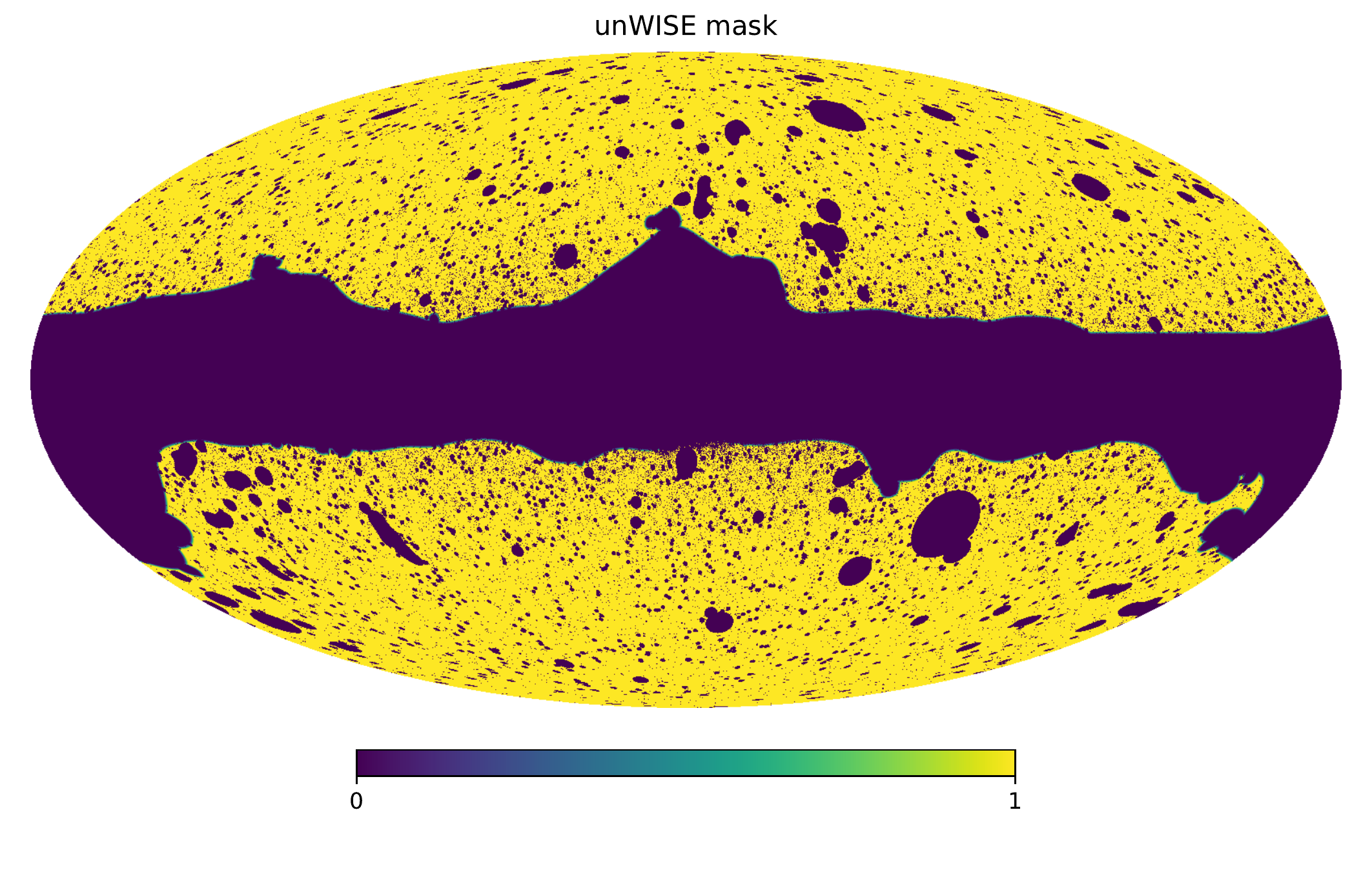}
    \includegraphics[scale=0.4]{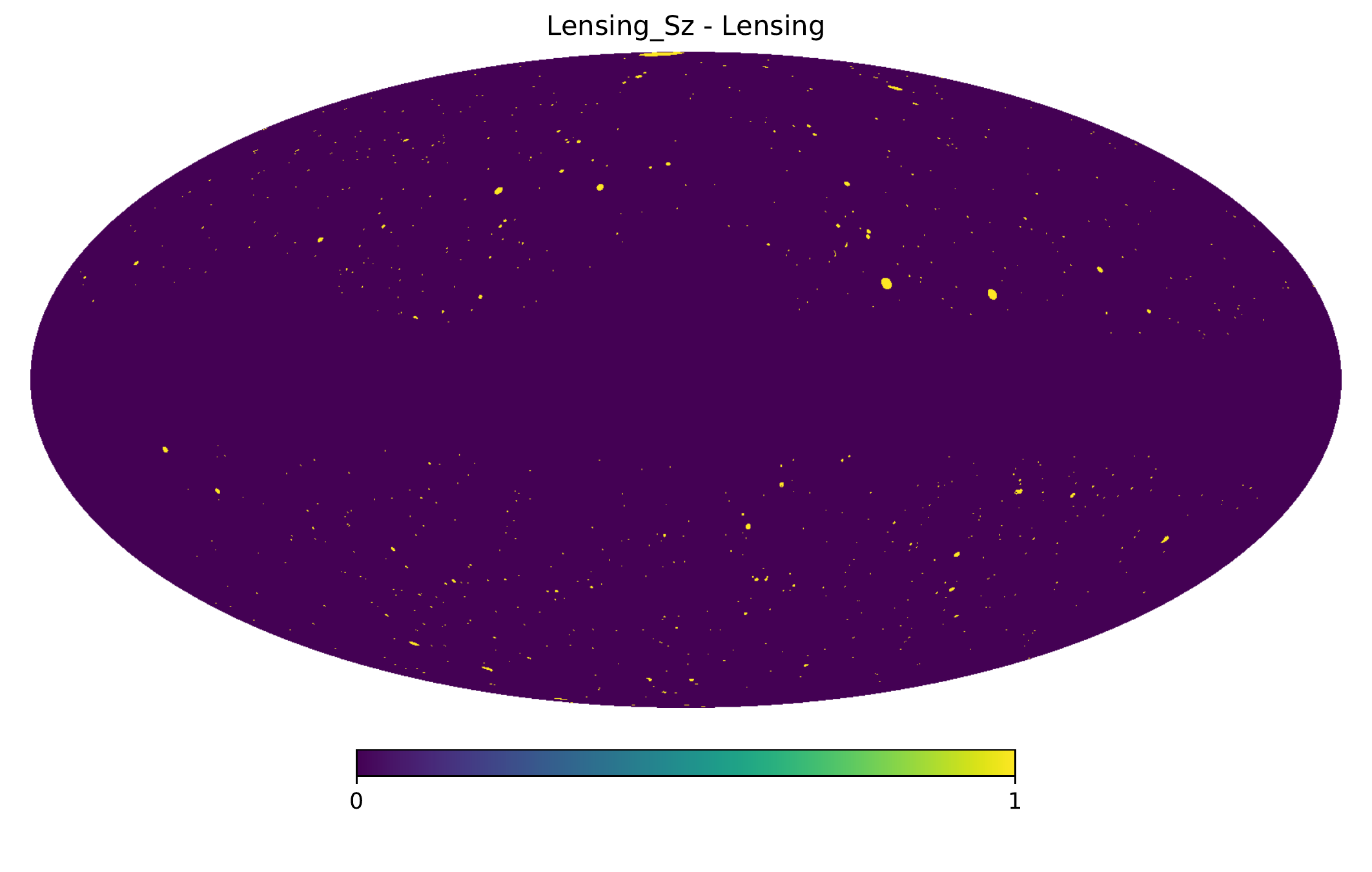}
    \includegraphics[scale=0.4]{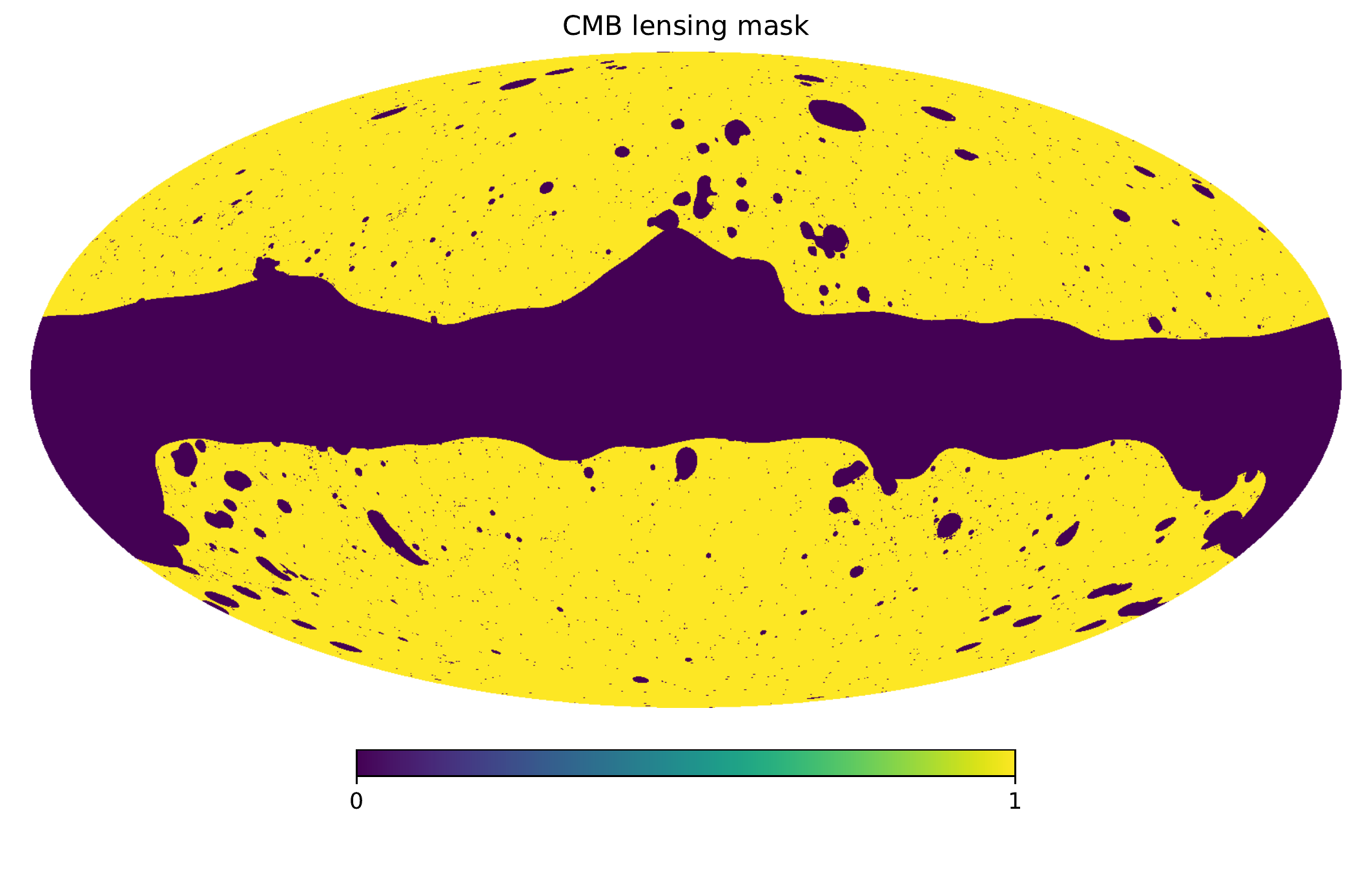}
    \includegraphics[scale=0.4]{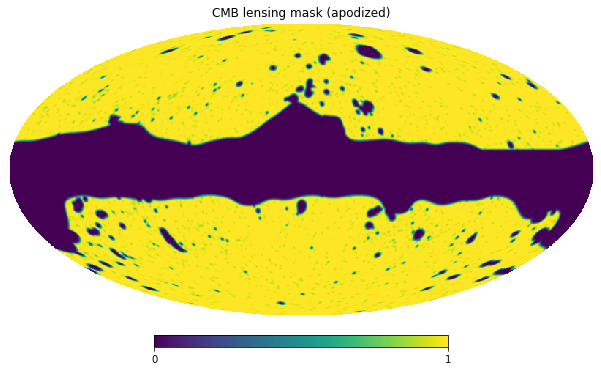}
    \caption{Masks used in the analysis. The masked pixels have value 0 and are shown in purple on the plots, while the unmasked pixels have value of 1 and are shown in yellow. \textit{Top left:} \emph{unWISE} mask used in this analysis with only the Galactic part of the mask apodized, using the Namaster C1 apodization with scale 1\textdegree. \textit{Top right:} Difference between ``Lensing\_Sz'' mask and ``Lensing'' mask, i.e., tSZ-selected clusters that are unmasked in our lensing mask. \textit{Bottom left:}  Unapodized CMB lensing mask. \textit{Bottom right:} CMB lensing mask apodized by smoothing with a 1\textdegree\ FWHM Gaussian smoothing. See Subsection~\ref{subsec:planckCMBmap} for more details. }
    \label{fig:masks_SZdeproj_compare}
\end{figure}

The \emph{unWISE} mask is constructed based on the \emph{Planck} 2018 lensing map as an effective Galactic mask \cite{planck2018_lens}. Furthermore, other bright objects are masked by cross-matching with external catalogs: stars with CatWISE \cite{CatWISE_paper}, bright galaxies with LSLGA\footnote{https://github.com/moustakas/LSLGA}, and planetary nebulae. 
Removal of \emph{Gaia} stars reduces the effective area in a \verb|HEALpix| pixel, as we cut out 2.75'' (i.e., the size of a WISE pixel) around each star.  Therefore, we also mask pixels where more than 20\% of the area is lost to stars, and correct the density in the remainder by dividing by the fractional area covered.
We split the masked areas into a contiguous part around the Galactic plane (the ``Galactic part'') and disconnected sections around bright stars, galaxies, planetary nebulae, and 143 and 217 GHz point sources (from the CMB lensing mask).  We apodize only the Galactic part, with a C1 apodization kernel in \verb|Namaster| \cite{Namaster_github, namaster} with apodization scale 1\textdegree. We leave the rest of the mask unapodized (top left panel of Fig.~\ref{fig:masks_SZdeproj_compare}), in order to preserve as much sky for the measurements as possible.  The validation of this choice is performed using simulations, as described below and in \citetalias{krolewski_2020,krolewski2021cosmological}.  This leaves a total unmasked sky fraction of $f_{\mathrm{sky}}=0.575$ when applied to the \emph{unWISE} maps. 

\subsection{\emph{Planck} CMB lensing maps}
\label{subsec:planckCMBmap}
For the CMB lensing map, \citetalias{krolewski_2020} and \citetalias{krolewski2021cosmological} used the \emph{Planck} 2018 lensing convergence $\kappa$ map with its associated mask \cite{planck2018_lens}. In this analysis we use a slightly different map and mask. For the lensing map, we use the \emph{Planck} ``Lensing-Szdeproj'' map downloaded from the Planck Legacy Archive\footnote{\href{https://pla.esac.esa.int}{https://pla.esac.esa.int}}. This lensing map is built from the \emph{Planck} \emph{SMICA-noSZ} map (temperature only), which has the thermal Sunyaev-Zel'dovich (tSZ) effect deprojected (using its known frequency dependence) prior to the lensing reconstruction operation \cite{planck2018_lens}. In the previous work in \citetalias{krolewski_2020, krolewski2021cosmological}, tSZ clusters were masked. Here, we wish to ensure that tSZ-selected clusters are not masked, so we can avoid having to introduce a selection function in the theoretical modeling.  Because the \emph{SMICA-noSZ} temperature map \cite{Planck2015_comp_sepa} is used for the lensing reconstruction here, we do not need to mask the tSZ-selected clusters \cite{planck_tSZcluster2021}. However, the associated ``Lensing-Szdeproj'' mask (also downloaded from the Planck Legacy Archive) nevertheless still has a value of zero at the tSZ cluster locations. Therefore, we add the signal from these clusters back into the map (i.e., set the value of the mask to one) by adding to the ``Lensing-Szdeproj'' mask the difference between the mask without cluster masking (``Lensing\_Sz'' mask) and the default mask (``Lensing'') (see top right plot of Fig.~\ref{fig:masks_SZdeproj_compare}).  In short, we use a CMB lensing map that includes signal at the location of tSZ clusters, to avoid biasing our interpretation of the cross-correlation measurements (see, e.g.,~\cite{Lembo2021} for further investigation of the effects of cluster masking in CMB lensing maps).

The trade-off for avoiding this potential bias is that we must use a CMB lensing map that has been reconstructed from a component-separated CMB temperature map with the tSZ signal explicitly deprojected using its known frequency dependence.  This is necessary to avoid a different bias, namely, the bias in the CMB lensing reconstruction itself due to the non-Gaussianity of the tSZ signal (and its non-zero correlation with the CMB lensing potential field)~\cite{vanEngelen2014,Osborne2014}.  Indeed, avoiding this bias is the original motivation for masking clusters in CMB lensing reconstruction.  Significant progress has been made in recent years in formulating CMB lensing estimators that use both the frequency dependence of the tSZ effect (and other contaminants) and the geometric structure of lensing to mitigate such foreground biases (e.g.,~\cite{MH2018,chen_2018,Schaan-Ferraro2019,Sailer2020,Darwish2021,Abylkairov2021,Sailer2021,Chen_2022}), ideally without the need for additional masking of individual clusters or sources.  The penalty for using a tSZ-deprojected temperature map in CMB lensing reconstruction is that the noise in the map is higher than that in a pure minimum-variance temperature map.  Thus, our \emph{unWISE} -- \emph{Planck} CMB lensing cross-correlation measurements are slightly noisier than those analyzed in \citetalias{krolewski_2020} and \citetalias{krolewski2021cosmological} (we compare our auto- and cross-correlation measurements to those from \citetalias{krolewski_2020, krolewski2021cosmological} in Fig.~\ref{fig:data_SZdeproj_compare}). 

Following \citetalias{krolewski_2020} and \citetalias{krolewski2021cosmological}, we also mask a small region of the sky with $|b|<10$\textdegree, which leaves $f_{\mathrm{sky}}=0.665$ after apodization. We apodize the lensing mask by smoothing the entire mask with a Gaussian with FWHM = 1\textdegree. The final apodized \emph{Planck} lensing mask used in this analysis is shown in the bottom right panel of Fig.~\ref{fig:masks_SZdeproj_compare}.  

\subsection{Measurements}

The $C_{\ell}^{gg}$ and $C_{\ell}^{g\kappa_{\rm{cmb}}}$ measurements that we use in the work are obtained with the pipeline from \citetalias{krolewski_2020, krolewski2021cosmological}, which we briefly summarize below.  We make two updates with respect to these earlier works: (i) we do not mask the \emph{Planck} tSZ clusters in the CMB lensing map (see previous subsection) and (ii) we use slightly different, more optimal mask apodization settings\footnote{We switch from a Gaussian apodization to a $C^1$ apodization as discussed in \citep{Grain09}.} (see  subsection~\ref{subsec:planckCMBmap} for the description of the \emph{Planck} lensing mask and subsection~\ref{subsec:unwise} for the \emph{unWISE} mask). 

As described in \citetalias{krolewski_2020, krolewski2021cosmological} the pseudo-power spectra are calculated from the masked maps for each sample using the \verb|Namaster| code \cite{Namaster_github, namaster}. Firstly, the lensing mask described above is applied to the CMB lensing map, and the \emph{unWISE} mask is applied to each of the respective galaxy maps. The pixel window function is corrected for in the measurements: no pixel window correction is applied for the CMB lensing map, and one power of the pixel window function correction is applied for each power of the galaxy field, except that the shot noise is not corrected for the pixel window function. The power spectra are calculated from $\ell_{min} = 20$ up to $\ell_{max}=6000$, but we use only measurements at $\ell < 1000$ in the analysis. Moreover, note that \citetalias{krolewski_2020, krolewski2021cosmological} do not use the galaxy auto-correlation data at $\ell < 100$, because these large-scale modes in the \emph{unWISE} galaxy samples are found to be contaminated by residual systematics (for the galaxy-CMB lensing cross-correlation, the bandpowers at $\ell > 20$ are found to be sufficiently free of systematics), and we follow the same approach in our analysis. The measurements are binned into multipole bins, with $\Delta_{\ell}=80$ for the first bin and $\Delta_{\ell}=100$ for the remaining bins, resulting in nine binned $C_{\ell}^{gg}$ and ten binned $C_{\ell}^{g\kappa_{\rm{cmb}}}$ data points.  

To ensure unbiased results, the pipeline to compute $C_{\ell}^{gg}$ and $C_{\ell}^{g\kappa_{\rm{cmb}}}$ in  \citetalias{krolewski_2020} was validated on a set of 100 simulated Gaussian lensing and galaxy maps, with the actual masks used in the real data analysis applied.  In order to compare the recovered spectra to the input spectra, the input theory spectra are multiplied by the \verb|Namaster| band-power window
functions $W_i(\ell)$ (weights describing the binning scheme of the power spectrum, i.e., the band-power $C_{i \ \rm{binned}}$ is defined as $C_{i \ \rm{binned}} = \sum_{\ell}  W_{i}(\ell) C_{\ell}$), and then compared to the mean bandpowers of the simulated maps.
We detect moderately significant deviations from the input binned theory curves, particularly in the galaxy auto-correlation at $\ell < 400$ and $\ell > 800$, although in practice these deviations are $<1\%$ and at largest $< 0.5\sigma$ in units of the error bars on the real data auto-correlation measurements. These deviations are due to the sharp mask that we keep around stars and pixels with $>20\%$ area lost. Apodizing this portion of the mask is difficult. A Gaussian smoothing will smear out the mask and cause us to use pixels that were originally masked out with some weight $< 1$. On the other hand, the default \verb|Namaster| apodization schemes, which preserve the fully masked region, do not work well due to the very large number of masked regions, leading to an unacceptably low sky fraction after apodization.  Given these difficulties, we choose to therefore use the apodization scheme described above and apply a correction to the theory curves (on top of the bandpower window binning from \verb|Namaster|), i.e., a transfer function determined from the 100 Gaussian mocks. For instance, the values of the transfer function deviating the most from unity for the $C_{\ell}^{gg}$ data are 0.99158, 0.99142, 0.99752 for the blue, green, and red samples, respectively (all for the first bin centered at $\ell=151.5$). For the $C_{\ell}^{g\kappa_{\rm{cmb}}}$ data points, they are  0.97089, 0.97743, 0.97750 for the blue, green, and red samples, respectively (in the first bin centered at $\ell=60.5$). We apply the transfer function in our maximum likelihood analysis (see Section~\ref{sec:HODfitting}), by multiplying the binned theory power spectra by their respective transfer functions.

Masking different fractions of the Galactic plane has been tested in \citetalias{krolewski2021cosmological}, who found no significant change in the $C_{\ell}^{g\kappa_{\rm{cmb}}}$ data. However, the authors found a mild, scale-independent trend in the amplitude of $C_{\ell}^{gg}$ as the Galactic latitude cut changes, which might be caused by small changes in the galaxy population selected due to differing foreground dust levels at different Galactic latitudes.  However, \citetalias{krolewski2021cosmological} also suggests that $\mathrm{d}N/\mathrm{d}z$ changing on the sky is not a major systematic, and should not affect the analysis, as long as the $\mathrm{d}N/\mathrm{d}z$ and auto- and cross-correlations are inferred over the same sky region.  In short, we note that the galaxies comprising the \emph{unWISE} samples could change slightly with Galactic mask, so the results in our work should be taken to be specific to the choice of Galactic mask used here (or at least the Galactic latitude cut).

\begin{figure}
    \centering
    \includegraphics[scale=0.7]{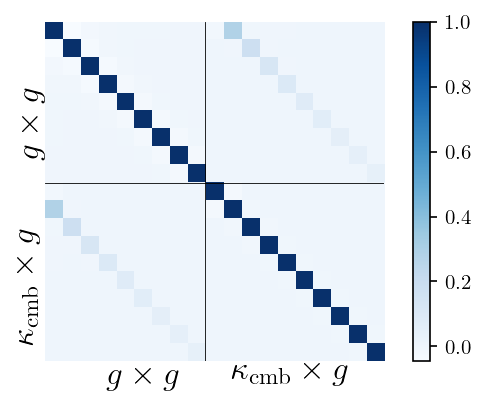}
    \includegraphics[scale=0.7]{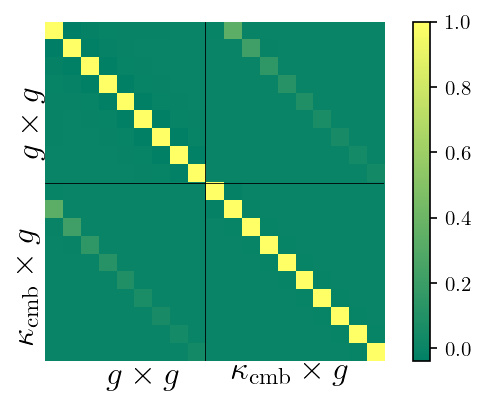}
    \includegraphics[scale=0.7]{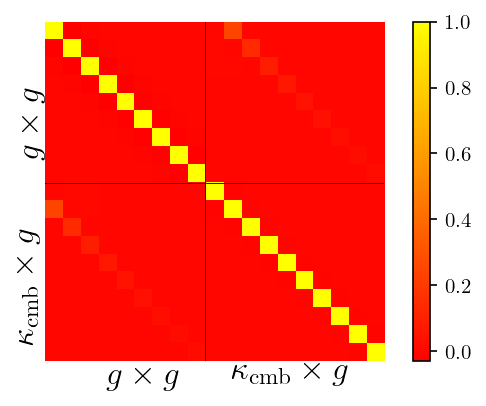}
    \caption{Correlation matrices for the nine galaxy-galaxy and ten CMB lensing-galaxy multipole bins used in the analysis for each \emph{unWISE} galaxy sample: \emph{unWISE} blue (left), green (middle), and red (right). Each matrix consists of 4 sub-matrices corresponding to nine $C_\ell^{gg}$ and ten $C_{\ell}^{g\kappa_{\rm{cmb}}}$ binned data points, as well as their cross-covariance.} 
    \label{fig:corr}
\end{figure}

The covariance matrices used in this analysis are recalculated for the exact mask used here (e.g., including the signal at the location of tSZ clusters), compared with the ones used in \citetalias{krolewski_2020, krolewski2021cosmological}. As in previous work, we also adopt the full covariance matrix from the analytic Gaussian approximations in \verb|Namaster| \cite{Efstathiou04,Couchot16,Garcia-Garcia:2019bku}, which are very close to the diagonal approximations given in Equations 2.1 and 2.2 of \citetalias{krolewski2021cosmological}. In Fig.~\ref{fig:corr} we show the correlation matrices (normalized covariance matrices) for each of the samples.

\begin{figure}
    \centering
    \includegraphics[scale=0.5]{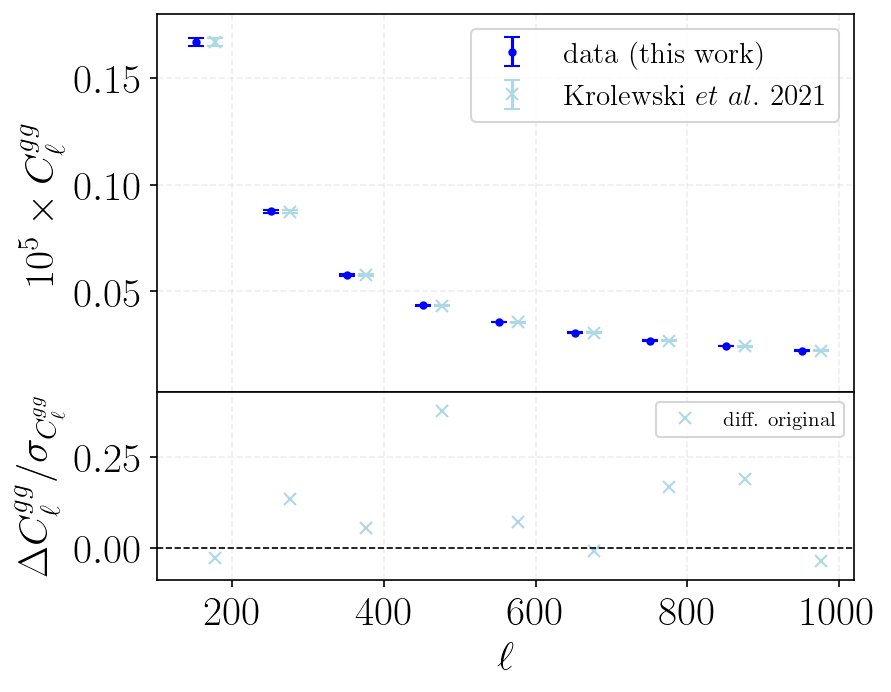}
    \includegraphics[scale=0.5]{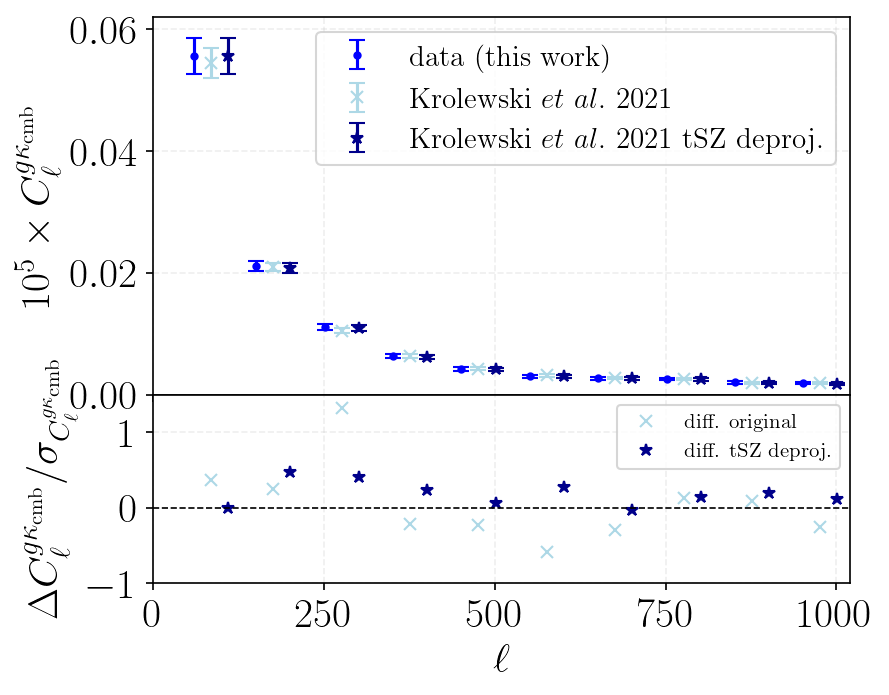}
    \includegraphics[scale=0.5]{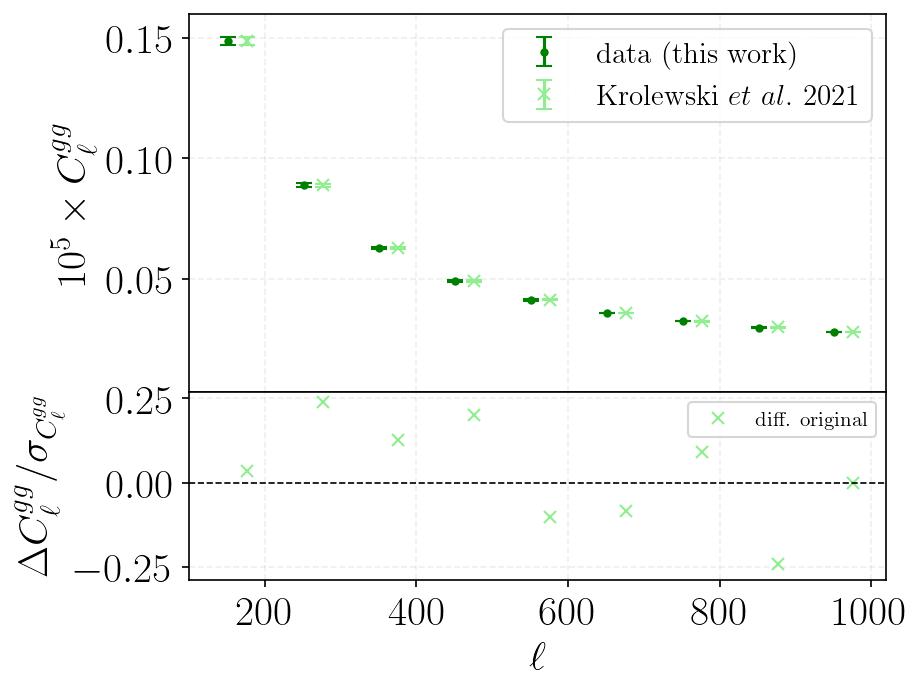}
    \includegraphics[scale=0.5]{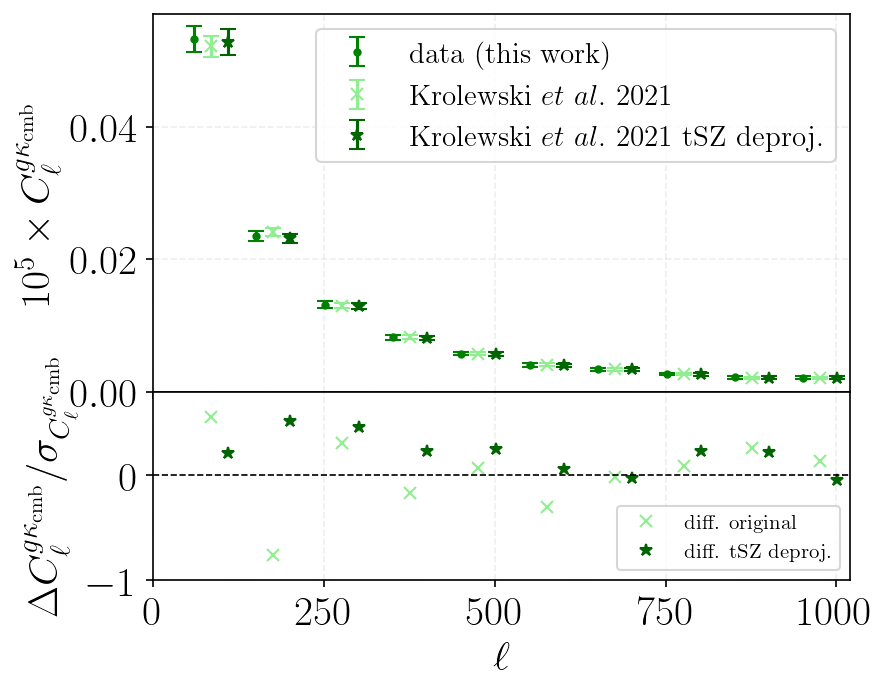}
    \includegraphics[scale=0.5]{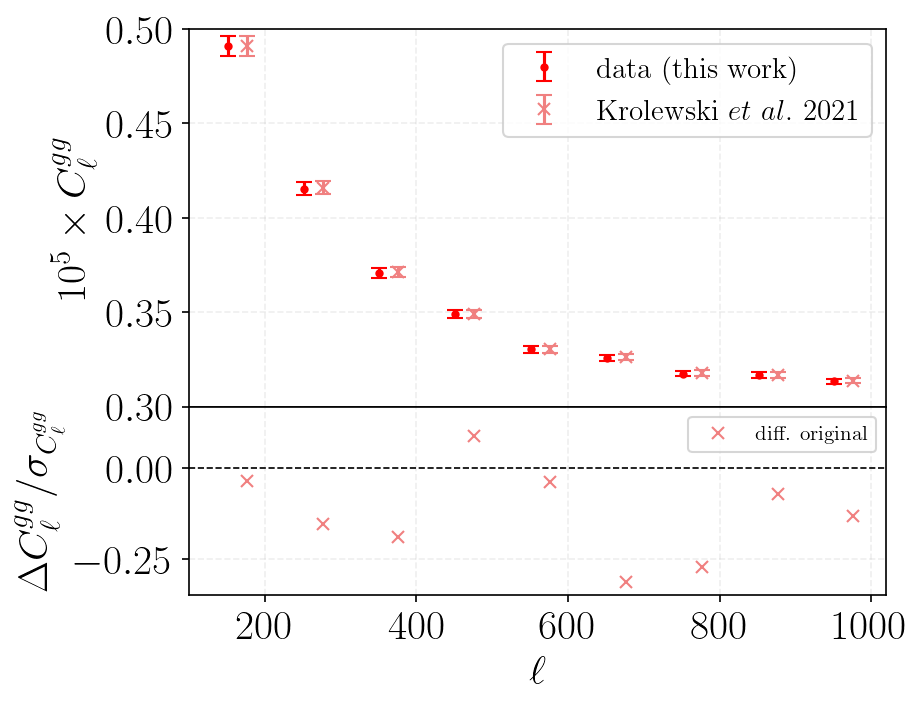}
    \includegraphics[scale=0.5]{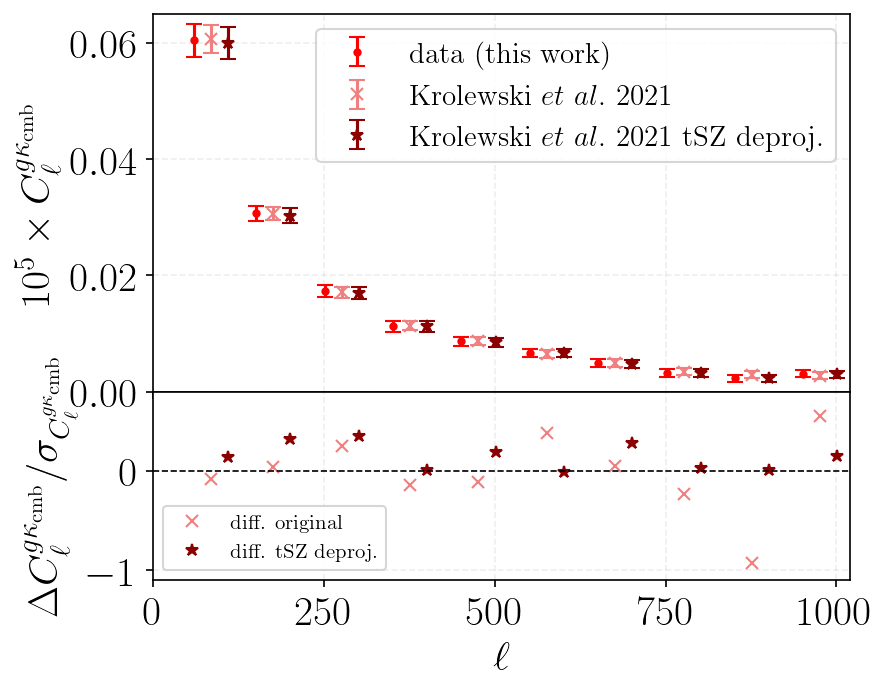}
    \caption{Comparison of the $C_{\ell}^{gg}$ (left) and $C_{\ell}^{g\kappa_{\rm{cmb}}}$ (right) data points used in this analysis (dots) with the original ones (crosses) and the tSZ-deprojected ones (stars) from \citetalias{krolewski_2020, krolewski2021cosmological} for the \emph{unWISE} blue (top panel), green (middle panel), and red (bottom panel) samples. Each plot also includes a bottom panel showing the difference (in units of the measurement error bars) between our measurements and those of \citetalias{krolewski_2020, krolewski2021cosmological} for the original data points (crosses) and the tSZ deprojected case (stars) for each sample; the differences are less than $1\sigma$ in all bins, except one bin in the blue $C_{\ell}^{g\kappa_{\rm{cmb}}}$ measurements. The $C_{\ell}^{gg}$ data points from \citetalias{krolewski_2020, krolewski2021cosmological} don't change between the original and tSZ-deprojected version (tSZ deprojection technique described in \citetalias{krolewski_2020, krolewski2021cosmological} only affects the CMB lensing data), so we only show one version for galaxy-galaxy. The original \citetalias{krolewski_2020, krolewski2021cosmological} data points (and their differences in the bottom panel) have been shifted by $\Delta \ell = +25$, and the tSZ-deprojected \citetalias{krolewski_2020, krolewski2021cosmological} ones by $\Delta \ell = +50$ for visual purposes.}
    \label{fig:data_SZdeproj_compare}
\end{figure}

In short, the  $C_{\ell}^{gg}$ and $C_{\ell}^{g\kappa_{\rm{cmb}}}$ data points used in this analysis are calculated using the pipeline described in \citetalias{krolewski_2020} and \citetalias{krolewski2021cosmological}, where the only differences are the use of a tSZ-deprojected temperature map in the CMB lensing reconstruction, allowing us to unmask the location of tSZ clusters and thereby avoid introducing a selection function to our theoretical model, and the use of slightly different mask apodization settings.  This yields a slightly noisier lensing reconstruction, because tSZ deprojection increases the noise on the temperature map and the reconstruction does not use polarization information. The final galaxy-galaxy $C_{\ell}^{gg}$ and CMB lensing-galaxy $C_{\ell}^{g\kappa_{\rm{cmb}}}$ measurements used in this analysis are presented in Fig.~\ref{fig:data_SZdeproj_compare} for each of the \emph{unWISE} samples. For comparison, we also show the $C_{\ell}^{gg}$ and $C_{\ell}^{g\kappa_{\rm{cmb}}}$ data points from \citetalias{krolewski_2020, krolewski2021cosmological}, which are very close to the measurements used here. The error bars shown in the plots are the square root of the diagonal elements of the covariance matrices obtained with \verb|Namaster|. 

\section{Likelihood analysis}
\label{sec:HODfitting}

In this section, we discuss how we perform the joint fit of the measured $C_\ell^{gg}$ and $C_\ell^{g \kappa_{\rm{cmb}}}$ data points (described in Section~\ref{sec:data}) to the halo model predictions implemented with \verb|class_sz| to constrain the model parameters. The final model of our measured auto- and cross-correlations, as described in Section~\ref{sec:HM}, includes the lensing magnification bias ($\mu_g$) contributions, as well as shot noise in the case of the galaxy auto-correlation.


A complete modeling of galaxy clustering power spectra, down to non-linear scales, should include the 1- and 2-halo terms of Eq.~\eqref{eq:clgg1h} and Eq.~\eqref{eq:clgg2h} as well as a shot-noise term. The shot noise is the random fluctuation inherent to the galaxy field, since  it is a discrete realization of the continuous matter field. It is therefore Poissonian in nature, and so it  has a constant  power spectrum.  In principle, its angular power spectrum is $C_\ell^{\mathrm{shot-noise}}=1/\bar{n}$ where $\bar{n}$ is the galaxy density of the sample in sr$^{-1}$. Ideally, the HOD should predict $\bar{n}$ as the comoving volume-integrated $\bar{n}_g$, weighted by the normalized redshift distribution of the sample, $\varphi^\prime_g$ (Eq.~\ref{eq:varphig}). But two main difficulties arise for this prediction to be accurate. First, due to the restricted range of scales in the non-linear regime, the scale-dependent part of the 1-halo term is generally not fully probed. This makes the 1-halo term and shot-noise term difficult to distinguish: they can be completely degenerate. Second, because of the complications due to the masks and halo exclusion, it is difficult to predict the galaxy abundance extremely precisely using the HOD formalism.
Thus, we include a free Poisson template in our model of the galaxy-galaxy auto-correlation determined by a free amplitude $A_{\mathrm{SN}}$  (multiplied by $10^7$ to match the order of magnitude of the power spectrum data). Nonetheless, for the reasons explained above, we do not expect the amplitude of this term to be a faithful representation of the actual shot-noise of the samples. Furthermore, to be conservative, we do not place specific priors on $A_{\mathrm{SN}}$ (such as Gaussian priors around the $1/\bar{n}$ values from Table \ref{table:unwise}).

Our model is thus:
\begin{equation}
    C_\ell^{gg, \mathrm{model}} = C_\ell^{gg} + 2C_\ell^{g \mu_g} + C_\ell^{\mu_g \mu_g} + 10^7 A_{\mathrm{SN}}
    \label{eq:clgg_model}
\end{equation}
\begin{equation}
    C_\ell^{\kappa_{\rm{cmb}} g, \mathrm{model}} = C_\ell^{\kappa_{\rm{cmb}} g} + C_\ell^{\kappa_{\rm{cmb}} \mu_g}
    \label{eq:clkg_model}
\end{equation}


We consider four HOD parameters, which calibrate the expectation value of the number of central and satellite galaxies, $N_c$ and $N_s$ (Eqs.~\ref{eq:N_c} and \ref{eq:N_s}), the parameter determining the truncation radius of the NFW profile  $\lambda$ (see Eq.~\ref{eq:ulm}), as well as the amplitude of the shot noise. Following the convention introduced in the DES-Y3 HOD analysis \cite{Zacharegkas2021}, we fix the $M_0$ parameter (the characteristic mass scale determining the expected number of satellites) to zero.  To summarize, the free parameters in our model are $\{ \alpha_{\mathrm{s}}$, $\sigma_{\mathrm{log} M}$, $M_\mathrm{min}^\mathrm{HOD}$, $M^{\prime}_1$, $\lambda$, $A_{\mathrm{SN}} \}$. 

We perform a joint fit of the nine $C_\ell^{gg}$ and ten $C_\ell^{ g \kappa_{\rm{cmb}} }$ observed, binned data points to the \verb|class_sz| halo model (as in Eqs.~\ref{eq:clgg_model} and \ref{eq:clkg_model}) to constrain these five HOD parameters and $A_{\mathrm{SN}}$ with a Markov Chain Monte Carlo (MCMC) analysis for each of the \emph{unWISE} samples. We assume a Gaussian log-likelihood:
\begin{equation}
    \ln \mathcal{L} (\vec{\theta})= -\frac{1}{2} (\textbf{d} -\textbf{t}(\vec{\theta}))^{T} \mathcal{C}^{-1} (\textbf{d} -\textbf{t}(\vec{\theta}))
\end{equation}
where $\vec{\theta}$ is the parameter vector, \textbf{d} is the data vector (consisting of nine $C_\ell^{gg}$ and ten $C_\ell^{g \kappa_{\rm{cmb}}}$ binned data points), and \textbf{t} is the model prediction vector of the same length, while $\mathcal{C}$ is the joint covariance matrix described in Section~\ref{sec:data} (in Fig.~\ref{fig:corr} we show the correlation matrices, i.e., the normalized covariance matrices, for each of the samples).  

The mass parameters are sampled on a logarithmic scale.  We put uniform priors on the model parameters, which are motivated by the DES-Y3 HOD analysis \cite{Zacharegkas2021}, and adjusted as needed for the different samples by determining how the change in parameters impacts our theory curves computed with \verb|class_sz| (see an example of the fractional change in the model when varying $\alpha_{\mathrm{s}}$, $\sigma_{\mathrm{log} M}$ by 10\% in Fig.~\ref{fig:varyHOD}). In most cases the priors are sufficiently wide to not be informative. They are summarized in Table~\ref{table:hod_results}. We fix the cosmological parameters to  the \emph{Planck} 2018 best-fit values (last column of Table~II of Ref.~\cite{Planck2018}), as quoted in Section~\ref{sec:intro}. We implement our likelihood in a modified version of the \verb|SOLikeT|\footnote{\href{https://github.com/simonsobs/SOLikeT}{https://github.com/simonsobs/SOLikeT}} package. To perform the fit, we run MCMC analyses with \verb|Cobaya| \cite{2019ascl.soft10019T,cobaya2021} separately for each of the \emph{unWISE} samples. The convergence criterion for the MCMC chains is that the generalized Gelman-Rubin statistic $R-1$ (as described in Ref.~\cite{Lewis2013}) satisfies $R-1 < 0.1$.


\section{Results}
\label{sec:results}
\begin{figure}
    \centering
    \includegraphics[scale=0.35]{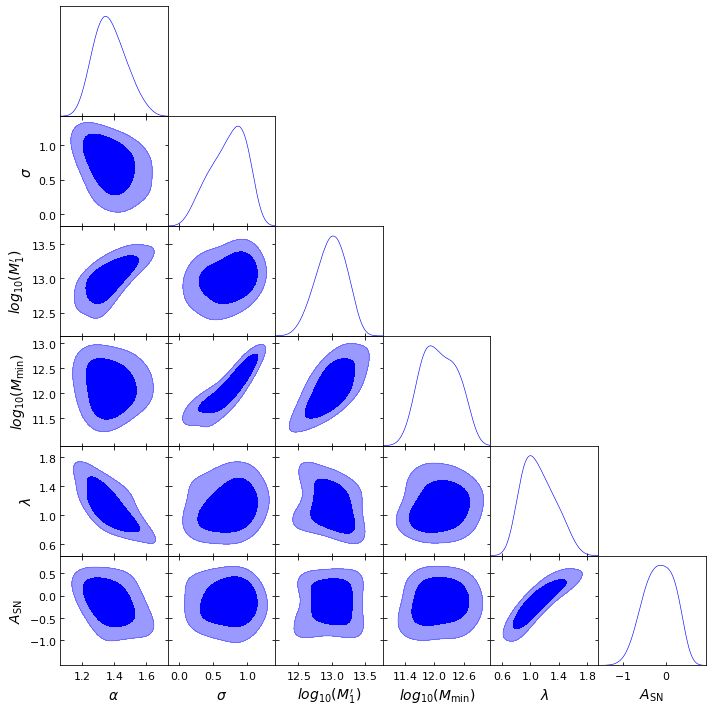}
    \includegraphics[scale=0.35]{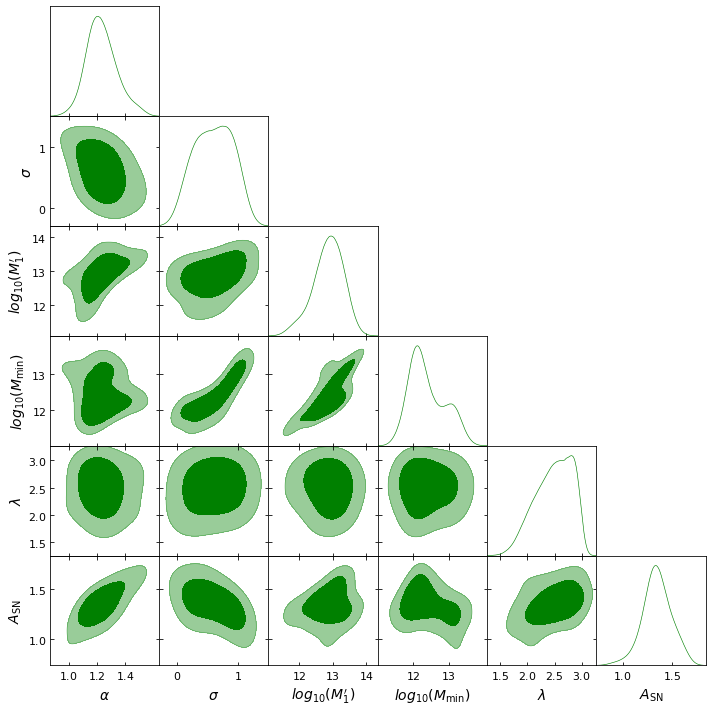}
    \includegraphics[scale=0.35]{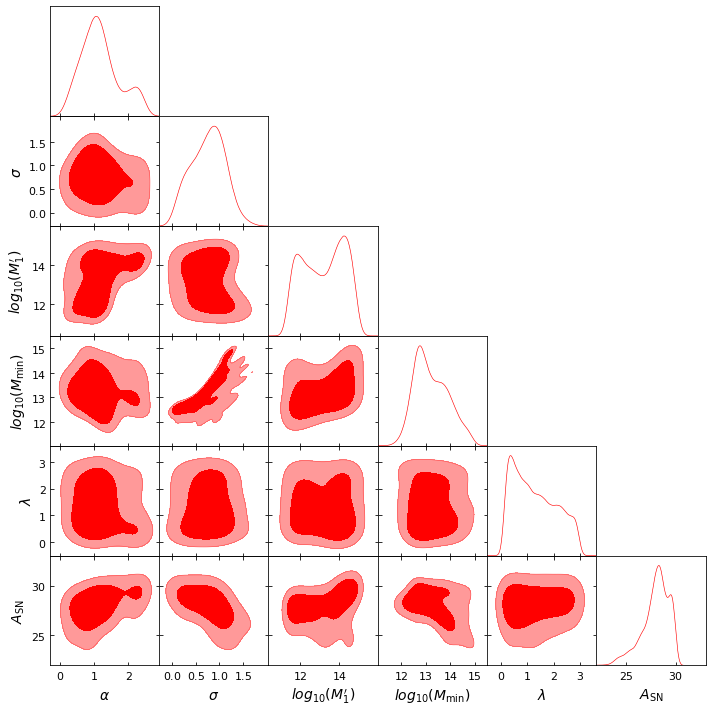}
    \caption{The 1D and 2D marginalized posterior distributions for the six model parameters $\sigma_{\mathrm{log} M}$, $\alpha_s$  $M_\mathrm{min}^\mathrm{HOD}$, $M_{1}'$, $\lambda$, and $A_{\mathrm{SN}}$ obtained as a result of fitting the measured \emph{unWISE} galaxy auto-power spectra and \emph{Planck} CMB lensing-galaxy cross-spectra (see Section~\ref{sec:data}) to our halo model  (Section~\ref{sec:HM}). Details of the fitting procedure can be found in Section~\ref{sec:HODfitting}. The priors imposed on the parameters are summarized in Table~\ref{table:hod_results}. Clockwise from top-left: posterior distributions for blue, green, and red \emph{unWISE} galaxy samples (also color-coded). The dark (light) shaded regions indicate 68\% (95\%) confidence intervals.}
    \label{fig:posteriors}
\end{figure}

In this section, we present the results of fitting the measured power spectra to the halo model predictions. As described above, these results are obtained by jointly fitting the \verb|class_sz| galaxy-galaxy and CMB lensing-galaxy halo model power spectra to the $C_\ell^{gg}$ and $C_\ell^{g \kappa_{\rm{cmb}}}$ measurements for \emph{unWISE} and \emph{Planck} CMB lensing, separately for each of the \emph{unWISE} samples (blue, green, and red). The obtained best-fit values for the six model parameters $\{ \alpha_{\mathrm{s}}$, $\sigma_{\mathrm{log} M}$, $M_\mathrm{min}^\mathrm{HOD}$, $M^{\prime}_1$, $\lambda$, $A_{\mathrm{SN}} \}$ for each of the \emph{unWISE} samples are shown in Table~\ref{table:hod_results}, along with the 1D and 2D marginalized posteriors in Fig.~\ref{fig:posteriors}. In Table~\ref{table:hod_results_mean}, we present a summary of the 1D marginalized parameter constraints for each of the model parameters.

        \begin{table}[t!]
        \setlength{\tabcolsep}{14pt}
        \renewcommand{\arraystretch}{1.3}
        \begin{tabular}{ |c|c|c|c|} 
        \hline
        \textbf{Parameter}  & \textbf{Best-fit Blue} &  \textbf{Best-fit Green} &  \textbf{Best-fit Red}  \\ 
        \hline\hline
        $\sigma_{\mathrm{log} M}$ & 0.687 & 0.973 & 0.403\\ 
        $\alpha_s$ & 1.304 & 1.302 & 1.629  \\ 
        $\mathrm{log}(M_\mathrm{min}^\mathrm{HOD})$  & 11.796 & 13.128 & 12.707\\
        $\mathrm{log}(M_{1}^\prime)$ &  12.701  & 13.441 & 13.519\\ 
        $\lambda$ & 1.087 &   2.746 & 0.184 \\
        $10^{7}A_{\mathrm{SN}}$ & -0.255 & 1.379 & 28.748 \\
        \hline
        $M_\mathrm{min}^\mathrm{HOD}$ [$M_{\odot}/h$]   & $6.251 \times 10^{11}$ & $1.342 \times 10^{13}$ & $5.096 \times 10^{12}$\\ 
        $M_{1}^\prime$ [$M_{\odot}/h$]  & $5.027 \times 10^{12}$ & $2.760 \times 10^{13}$ & $3.301 \times 10^{13}$\\ 
        \hline
        $M_h$ $[M_\odot /h]$ & $1.88 \times 10^{13}$ & $1.66 \times 10^{13}$ & $1.55 \times 10^{13}$ \\
        $b_g$ &  1.49 & 2.01 & 2.98\\
        $\alpha_{\mathrm{sat}}$   &  0.30 & 0.16 & 0.14\\
        \hline
        $\chi^2$ & 11.8 & 7.9 & 15.3 \\ 
        PTE & 0.544 & 0.850 & 0.289 \\ 
        \hline
        \end{tabular}
        \caption{Best-fit values for the six model parameters obtained by jointly fitting the measured \emph{unWISE} and \emph{Planck} galaxy-galaxy auto- and galaxy-CMB lensing cross-correlation to the halo model predictions, along with the $\chi^2$ and PTE for the best-fit (for 19 data points, i.e., 13 degrees of freedom), for each of the three \emph{unWISE} galaxy samples. We also include results for five derived parameters: $M_{1}^\prime$ and $M_\mathrm{min}^\mathrm{HOD}$ in units of $M_{\odot}/h$, the fraction of satellite galaxies $\alpha_{\mathrm{sat}}$ (see Eq.~\ref{eq:alpha_sat}), the mean galaxy bias $b_g$ (see Fig.~\ref{fig:bias}), as well as the average host halo mass $M_h$ (see also Fig.~\ref{fig:Mh}). The latter three are computed with the best-fit values of the HOD parameters from this table. 
        }
        \label{table:hod_results}
        \end{table}

        \begin{table}[t!]
        \setlength{\tabcolsep}{10pt}
        \renewcommand{\arraystretch}{1.7}
        \begin{tabular}{ |c|c|c|c|} 
        \hline
        \textbf{Parameter}  & \textbf{Blue}  & \textbf{Green} & \textbf{Red}  \\ 
        \hline\hline
        $\sigma_{\mathrm{log} M}$ & $0.73^{+0.33}_{-0.22}$ & $0.61^{+0.32}_{-0.32}$ & $0.75^{+0.42}_{-0.35}$\\ 
        $\alpha_s$ & $1.38^{+0.09}_{-0.12}$ & $1.23^{+0.10}_{-0.12}$ & $1.18^{+0.35}_{-0.72}$\\ 
        $\mathrm{log}(M_\mathrm{min}^\mathrm{HOD})$  & $12.11^{+0.37}_{-0.37}$ &  $12.39^{+0.52}_{-0.52}$ & $13.23^{+0.58}_{-0.84}$ \\
        $\mathrm{log}(M_{1}^\prime)$ &  $ 13.00^{+0.25}_{-0.21}$  &  $ 12.87^{+0.51}_{-0.38}$ & $13.20^{+1.10}_{-1.10}$\\ 
        $\lambda$ &  $1.11^{+0.20}_{-0.29}$ &  $2.50^{+0.45}_{-0.24}$ & $1.30^{+0.51}_{-1.10}$\\
        $10^{7}A_{\mathrm{SN}}$ & $-0.16^{+0.40}_{-0.34}$ &  $1.35^{+0.15}_{-0.15}$ & $27.95^{+1.90}_{-0.62}$ \\
        \hline
        $M_\mathrm{min}^\mathrm{HOD}$ [$M_{\odot}/h$] &  $1.83^{+0.41}_{-1.63} \times 10^{12}$ & $5.22^{+0.34}_{-4.80} \times 10^{12}$ & $6.60 ^{+0.30}_{-1.11} \times 10^{13}$\\ 
        $M_{1}^\prime$ [$M_{\odot}/h$]  &  $1.13 ^{+0.32}_{-0.70} \times 10^{13}$ & $1.18 ^{+0.30}_{-1.11} \times 10^{13}$  & $1.23 ^{+0.14}_{-1.17} \times 10^{14}$ \\ 
        \hline
        $M_h$ $[M_\odot /h]$  & $1.81 ^{+0.09}_{-0.08} \times 10^{13}$ & $1.64 ^{+0.08}_{-0.09} \times 10^{13}$  &  $1.52 ^{+0.31}_{-0.32} \times 10^{13}$\\
        $b_g$ &   $1.50^{+0.01}_{-0.02}$ &   $2.05^{+0.02}_{-0.03}$&   $2.92^{+0.06}_{-0.05}$\\
        $\alpha_{\mathrm{sat}}$  &  $0.23^{+0.06}_{-0.07}$ & $0.38^{+0.41}_{-0.36}$ & $0.36^{+0.47}_{-0.31}$ \\
        \hline
        \end{tabular}
        \caption{Statistical summary of the posteriors (mean and 68\% marginalized constraints) for the six model parameters obtained by jointly fitting the measured \emph{unWISE} and \emph{Planck} galaxy-galaxy auto- and galaxy-CMB lensing cross-correlations to the halo model predictions, separately for each of the three \emph{unWISE} galaxy samples. The 1D and 2D marginalized posterior distributions are shown in Fig.~\ref{fig:posteriors}. We also provide results for two derived parameters: $M_{1}^\prime$ and $M_\mathrm{min}^\mathrm{HOD}$ (in units of $M_{\odot}/h$), as well as the mean galaxy bias $b_g$ (see Fig.~\ref{fig:bias}), the average host halo mass $M_h$ (see also Fig.~\ref{fig:Mh} , and the fraction of satellite galaxies $\alpha_{\mathrm{sat}}$ (Eq.~\ref{eq:alpha_sat}). $\alpha_{\mathrm{sat}}$, $b_g$, and $M_h$ and their error bars (also corresponding to the 68\% CL) are obtained by computing $\alpha_{\mathrm{sat}}$ for the last 80,000 steps of the MCMC chains, which constitutes about half of the samples.  }
        \label{table:hod_results_mean}
        \end{table}

The model provides a good fit to the data; the best-fit $\chi^2$ values of the joint fit are $\chi^2 = 11.8$, $7.9$, $15.3$ for 19 data points for \emph{unWISE} blue, green, and red, respectively.  With 6 free parameters, the fit thus has $19-6 = 13$ degrees of freedom, and the $\chi^2$ values correspond to probability-to-exceed (PTE) values of 0.544, 0.850, and 0.289, respectively.  The $\chi^2$ values for the theory model computed with the best-fit parameter values when fitted to the $C_\ell^{gg}$ data points only are $\chi^2 = 4.2$, $2.4$, $8.4$, and when fitted to $C_\ell^{g \kappa_{\rm{cmb}}}$ only they are $\chi^2 = 7.2$, $5.9$, $7.1$ (these $\chi^2$ values do not add up to the $\chi^2$ of the joint fit because of the non-zero cross terms in the joint covariance matrix of the $C_\ell^{gg}$ and $C_\ell^{g \kappa_{\rm{cmb}}}$ measurements --- see Fig.~\ref{fig:corr} for the correlation matrices for the $C_\ell^{gg}$ and $C_\ell^{g \kappa_{\rm{cmb}}}$ measurements).  Note that the fit to $C_\ell^{gg}$ alone has $9-6 = 3$ degrees of freedom, while the fit to $C_\ell^{g \kappa_{\rm{cmb}}}$ alone has $10-5 = 5$ degrees of freedom (since there is no shot noise parameter in the $C_\ell^{g \kappa_{\rm{cmb}}}$ fit).  These $\chi^2$ and PTE values indicate that our model describes the data well, and our covariance estimates are reasonable.

For the central galaxy population, we constrain $M_\mathrm{min}^\mathrm{HOD}$, the characteristic minimum mass of halos that host a central galaxy, to be $M_\mathrm{min}^\mathrm{HOD} = 
1.83^{+0.41}_{-1.63} \times 10^{12} M_\odot/h$, $5.22^{+0.34}_{-4.80} \times 10^{12} M_\odot/h$, $6.60 ^{+0.30}_{-1.11} \times 10^{13} M_\odot/h$, and the width $\sigma_{\mathrm{log} M}  = 0.73^{+0.33}_{-0.22}$, $0.61^{+0.32}_{-0.32}$, and $0.75^{+0.42}_{-0.35}$ for \emph{unWISE} blue ($\bar{z}\approx 0.6$), green ($\bar{z} \approx 1.1$), and red ($\bar{z} \approx 1.5$), respectively. There seems to be an increasing trend in the $M_\mathrm{min}^\mathrm{HOD}$ parameter between the three \emph{unWISE} samples, with a higher value for the highest mean redshift red sample.  This aligns with expectations, since this sample is more highly biased than the blue or green samples~\citepalias{krolewski_2020}.

For the satellite galaxy population, the index of the power law $\alpha_s$ is constrained to be $\alpha_s = 1.38^{+0.09}_{-0.12}$, $1.23^{+0.10}_{-0.12}$, and $1.18^{+0.35}_{-0.72}$; and the mass scale at which one satellite galaxy per halo is found, $M^{\prime}_1$, is constrained to be $M^{\prime}_1 = 1.13 ^{+0.32}_{-0.70} \times 10^{13} M_\odot/h$,  $1.18 ^{+0.30}_{-1.11} \times 10^{13} M_\odot/h$,  $1.23 ^{+0.14}_{-1.17} \times 10^{14} M_\odot/h$ for \emph{unWISE} blue, green, and red, respectively. Again, we observe that $M^{\prime}_1$ is noticeably larger for the red sample. For $\alpha_s$ there is an inverse relationship between its value and the mean redshift of each sample, although the error bars are too large to draw a sharp conclusion.  The constrained shot noise amplitude values are $A_{\mathrm{SN}}= -0.16^{+0.40}_{-0.34}$,  $1.35^{+0.15}_{-0.15}$, and $27.95^{+1.90}_{-0.62}$ for each sample (blue, green, red). As described in Section~\ref{sec:HODfitting}, we allow the shot noise to be negative, as this parameter effectively absorbs any mismatch between the Poisson component of our 1-halo term and the Poisson level of the high-$\ell$ $C_{\ell}^{gg}$ data. The increasing value of the mass parameters $M_\mathrm{min}^\mathrm{HOD}$ and $M^{\prime}_1$ between samples seem to illustrate the redshift evolution of the \emph{unWISE} galaxies, particularly for the red sample.

From the 1D and 2D marginalized posterior distributions presented in Fig.~\ref{fig:posteriors}, we note that $\alpha_s$ is generally best-constrained, especially for the blue and green samples. This figure also shows that there are some degeneracies between parameters. As noted in the DES-Y3 HOD analysis~\citep{Zacharegkas2021}, a degeneracy between $M^{\prime}_1$ and $\alpha_s$ is expected based on the model for the expectation value of the number of satellites, $N_s$ (Eq.~\ref{eq:N_s}). There is also an expected degeneracy between the central parameters, $M_\mathrm{min}^\mathrm{HOD}$ and $\sigma_{\mathrm{log} M}$. We note that the $\lambda$ parameter is not very well constrained for the green and red samples. It is also reaching the upper prior boundary for the green sample, despite the fact that this prior is already very conservative.

\begin{figure}
    \centering
    \includegraphics[scale=0.4]{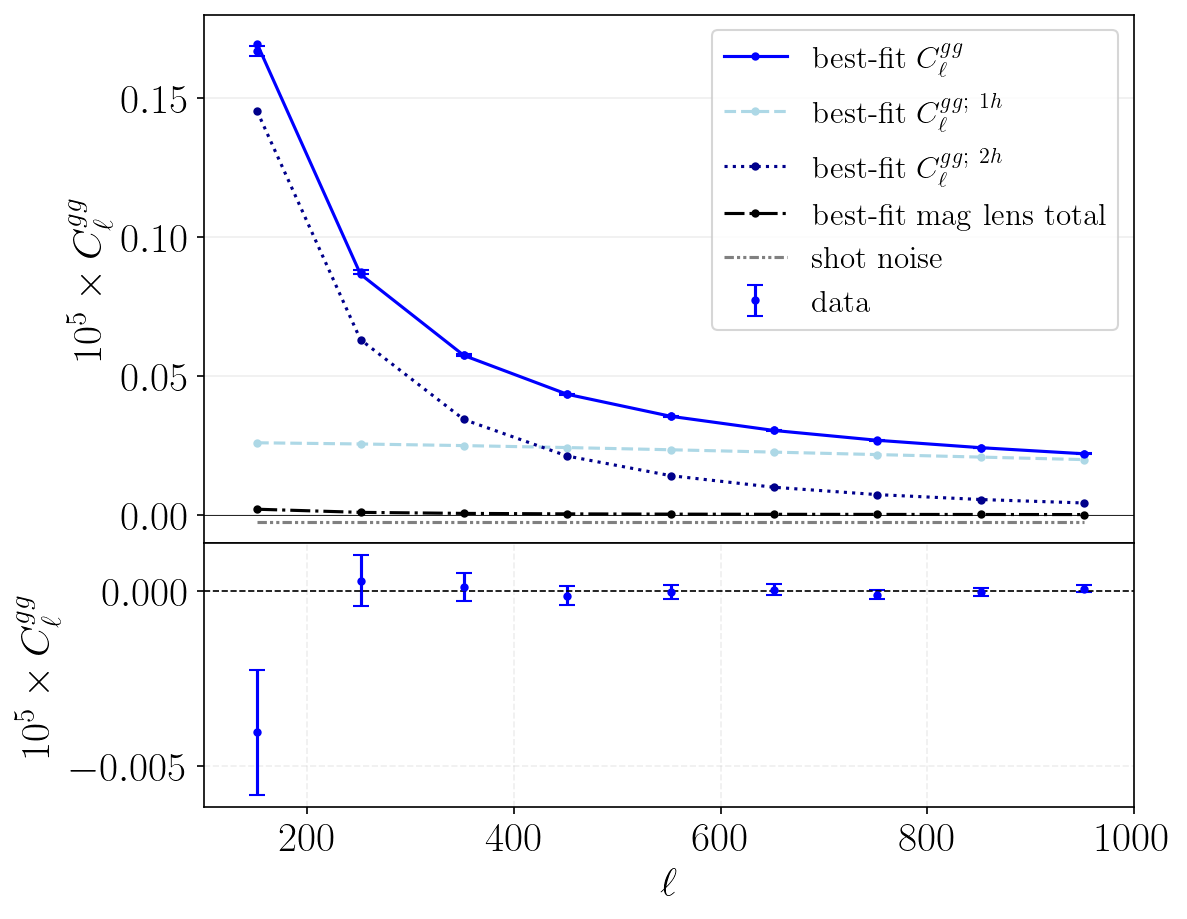}
    \includegraphics[scale=0.4]{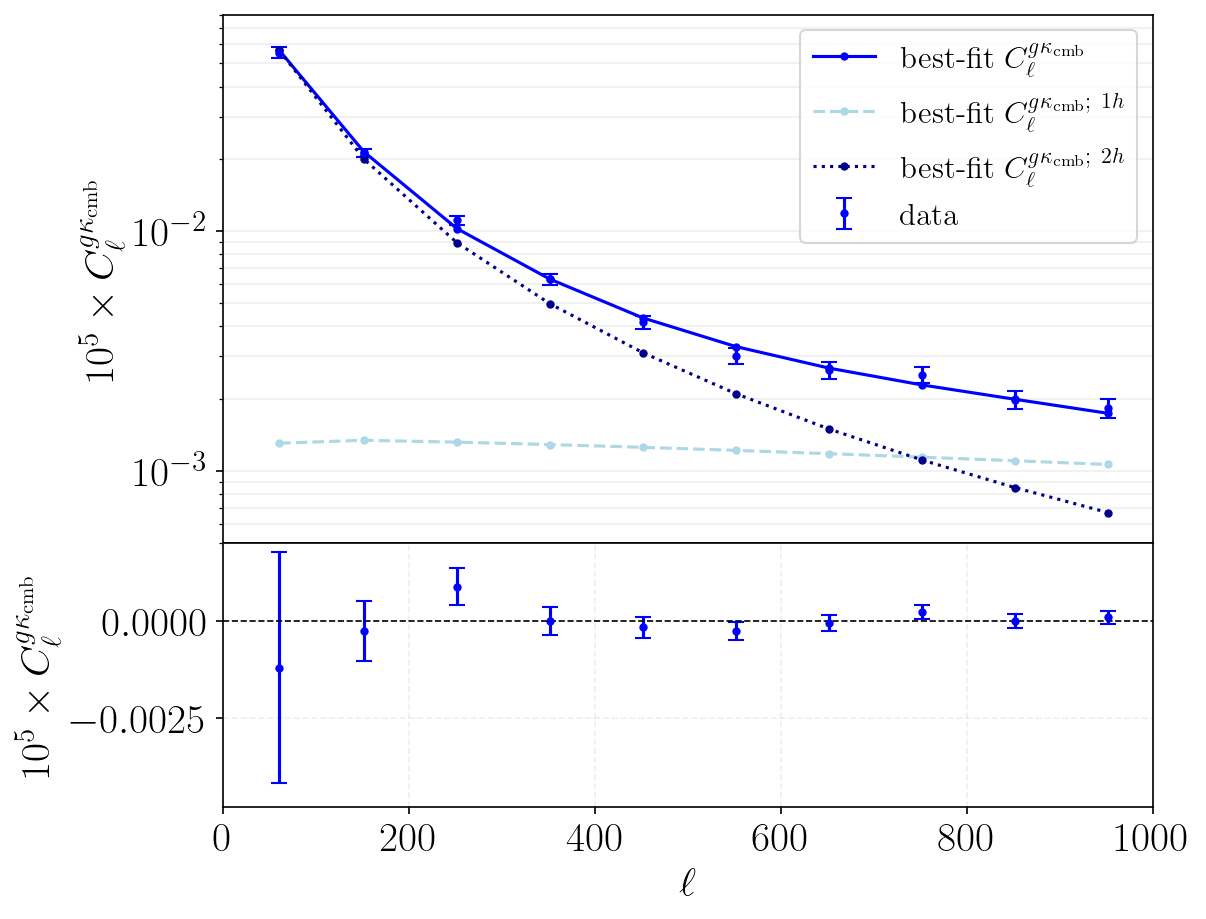}
    \includegraphics[scale=0.4]{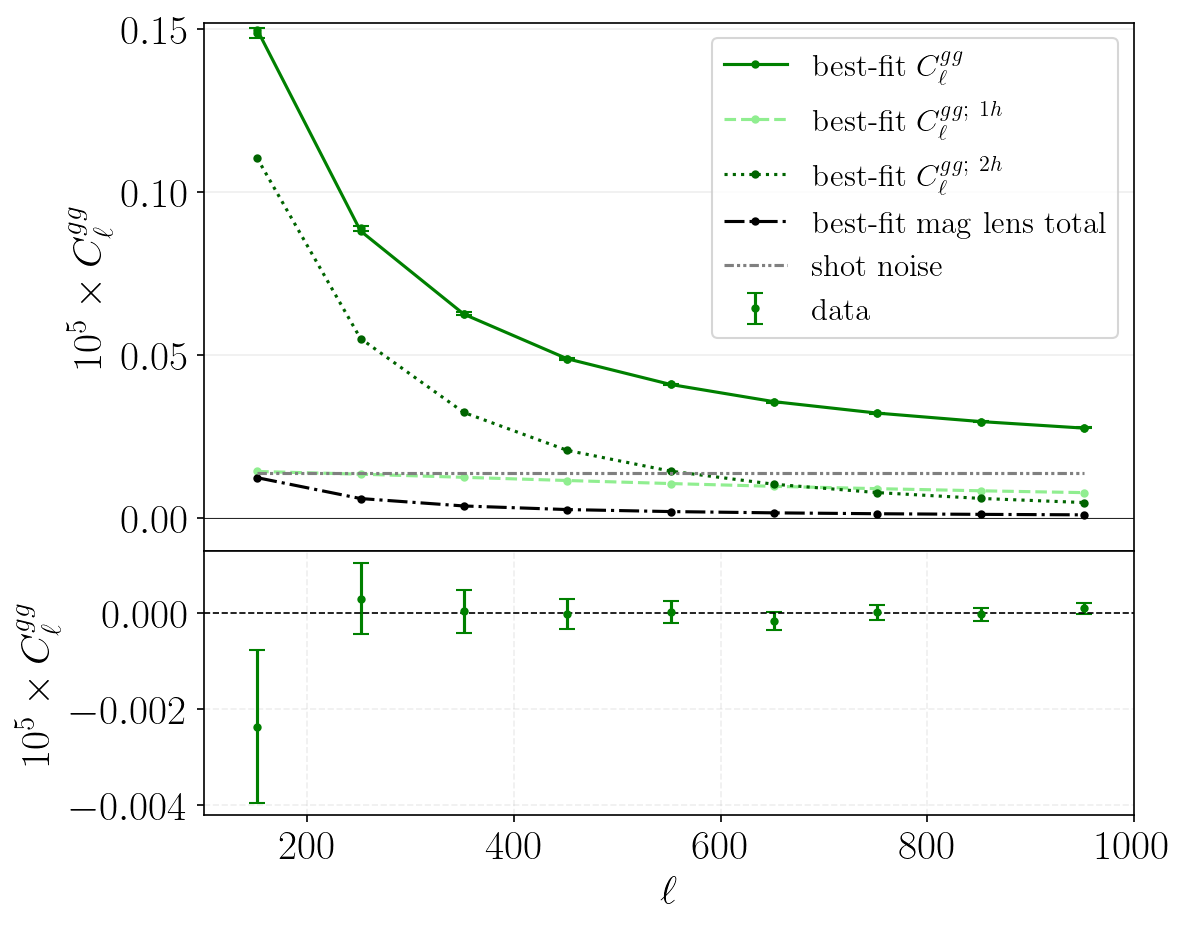}
    \includegraphics[scale=0.4]{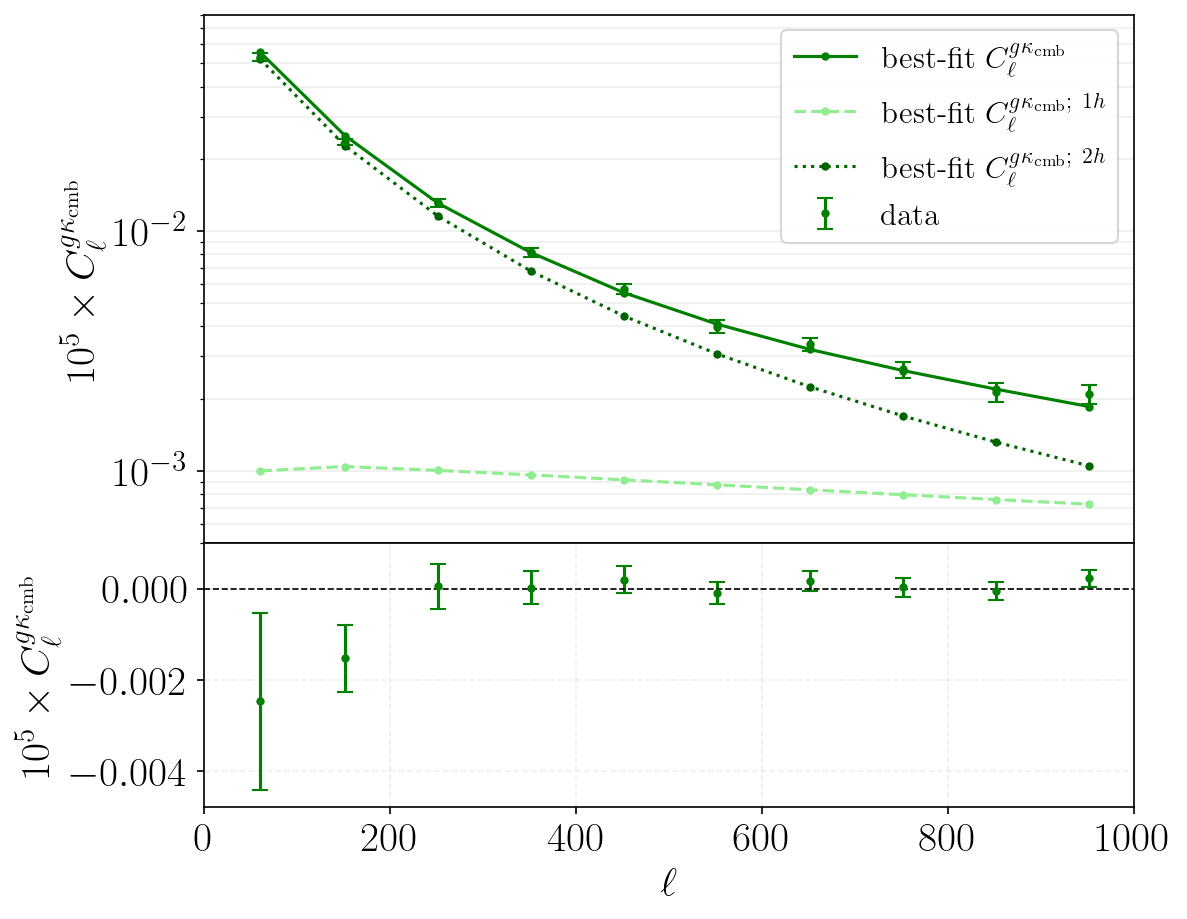}
    \includegraphics[scale=0.4]{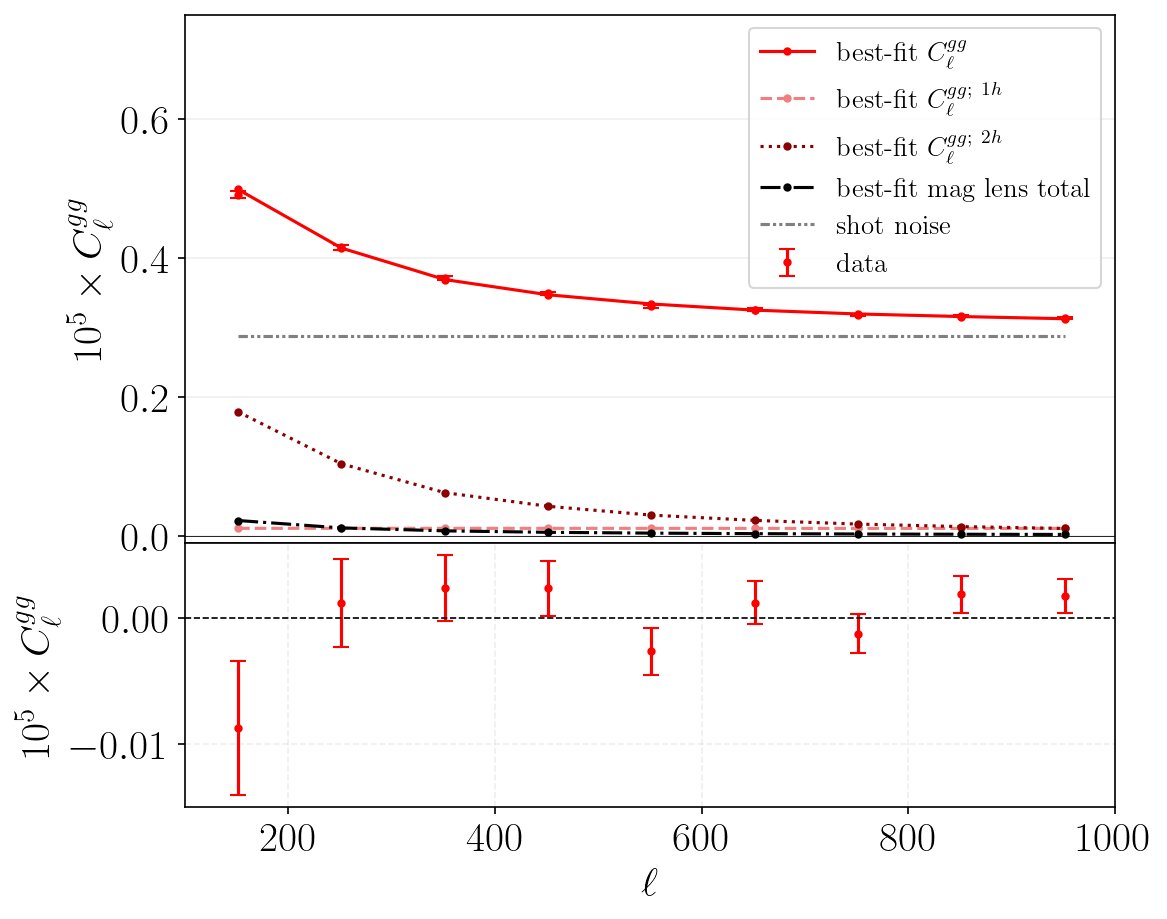}
    \includegraphics[scale=0.4]{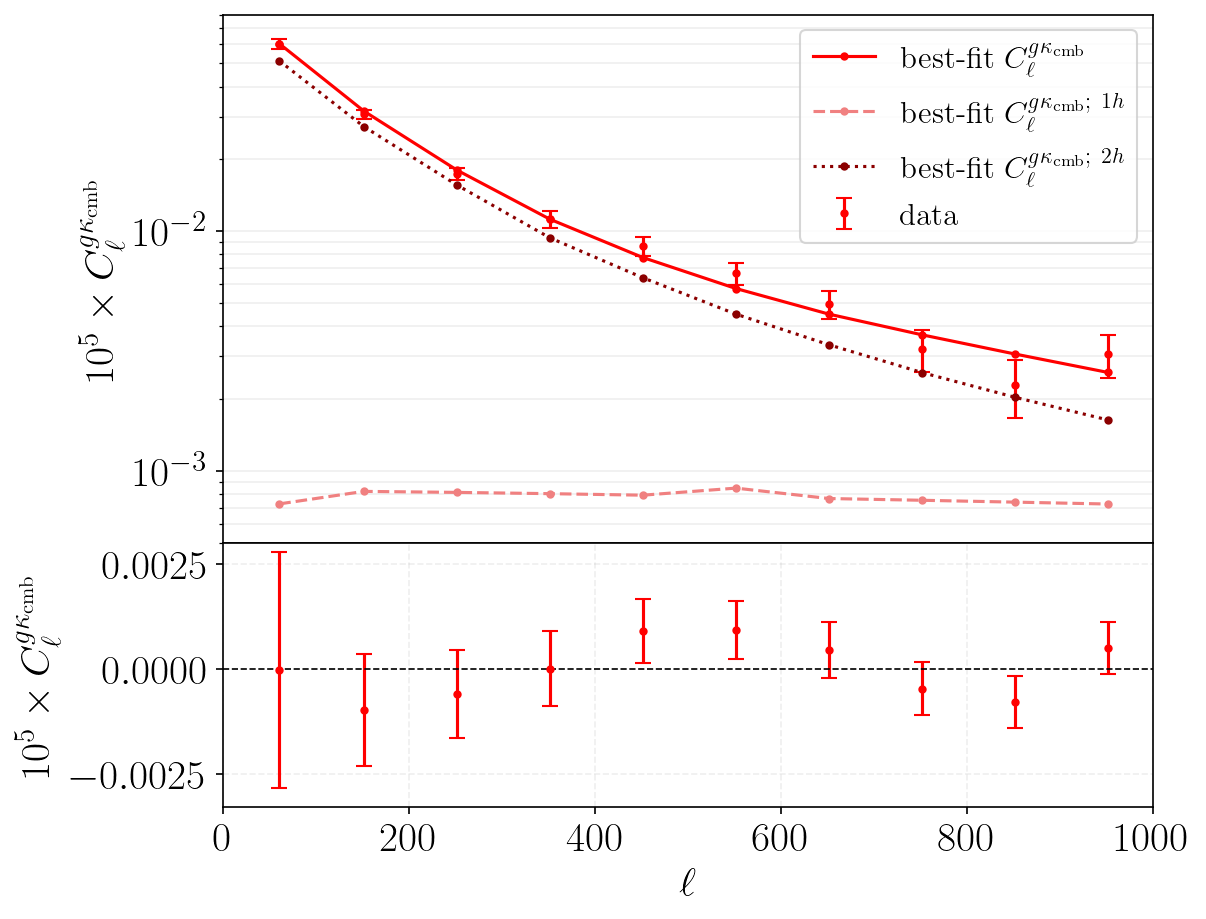}
    \caption{Measurements of the galaxy-galaxy auto-power spectrum,  $C_\ell^{gg}$, and the CMB lensing-galaxy cross-power spectrum, $C_\ell^{g \kappa_{\rm{cmb}}}$, for each of the \emph{unWISE} galaxy samples, along with our halo model theory curves for the best-fit model parameters (Table~\ref{table:hod_results}). The \emph{unWISE} galaxy samples are color-coded from top to bottom: blue, green, and red, with $C_\ell^{gg}$ on the left and  $C_\ell^{g \kappa_{\rm{cmb}}}$ on the right. On each galaxy auto-power spectrum plot, the solid curves are the best-fit total signal, the dotted curves show the best-fit 1-halo contribution to $C_\ell^{gg}$, the dashed show the best-fit 2-halo contribution to $C_\ell^{gg}$, the dash-dotted black show the total best-fit lensing magnification contribution, and the grey dash-dot-dotted show the best-fit shot noise contribution. On the CMB lensing-galaxy cross-spectra plots, the solid curves show the best-fit total signal, the dotted curves show the best-fit 1-halo contribution to $C_\ell^{g \kappa_{\rm{cmb}}}$, and the dashed show the best-fit 2-halo contribution to $C_\ell^{g \kappa_{\rm{cmb}}}$; the lensing magnification contributions are 3-4 orders of magnitude smaller than the presented curves and therefore not shown in the CMB lensing case.  Note that in the $C_\ell^{gg}$ plots the $y$-axis is shown on a linear scale, while for the $C_\ell^{g \kappa_{\rm{cmb}}}$ plots it is on a logarithmic scale. Each plot has a bottom panel that shows the residuals of the best-fit model for each bin. }
    \label{fig:bestfit}
\end{figure}

In Fig.~\ref{fig:bestfit} we present the best-fit \verb|class_sz| model and its different components, as described in Section~\ref{sec:HODfitting}, along with the data points (note that for the galaxy-galaxy auto-correlation, the y-axis is shown on a linear scale, while for the CMB lensing-galaxy cross-correlation it is shown on a logarithmic scale). For the galaxy-galaxy auto-correlations, we observe that the 1-halo term is nearly constant on the scales considered here (as expected since we do not resolve the satellite galaxy profiles well on 10 arcmin scales), becoming the leading term around $\ell \approx 400-700$.  Since the shot noise is also a constant term, it is particularly difficult to distinguish it from the 1-halo term. Thus, for $C_\ell^{gg}$, most of the constraining power on the HOD parameters comes from the 2-halo term, which also has a characteristic shape. The lensing magnification terms in the observed galaxy auto-power spectra are roughly two orders of magnitude smaller than the total prediction, yet they become more important for the higher-redshift samples with steeper luminosity functions, and thus are non-negligible for the green ($\bar{z} \approx 1.1$) and red ($\bar{z} \approx 1.5$) samples. For the CMB lensing-galaxy cross-correlations, the 2-halo term is again the leading term and the 1-halo term overtakes the 2-halo term only for the blue sample, at roughly $\ell \approx 750$. The lensing magnification term is not shown on the $C_\ell^{g \kappa_{\rm{cmb}} }$ plots, as it is too small to be relevant. In Fig.~\ref{fig:bestfit} we also present the residuals of the data with respect to the best-fit models.  As expected based on the good $\chi^2$ and PTE values, the residuals indicate that the best-fit models are consistent with the data.

Furthermore, in order to validate the obtained HOD parameter constraints, we also perform the analysis for the galaxy-galaxy and galaxy-CMB lensing data separately, instead of fitting them jointly. The results are presented in Appendix~\ref{app:gg_kg_separate}, and from the 1D and 2D marginalized posterior distribution in Fig.~\ref{fig:posteriors_kg_gg}, we note that all three analysis scenarios are consistent, which validates our main, joint analysis. 

\begin{figure}
    \centering
    \includegraphics[scale=0.5]{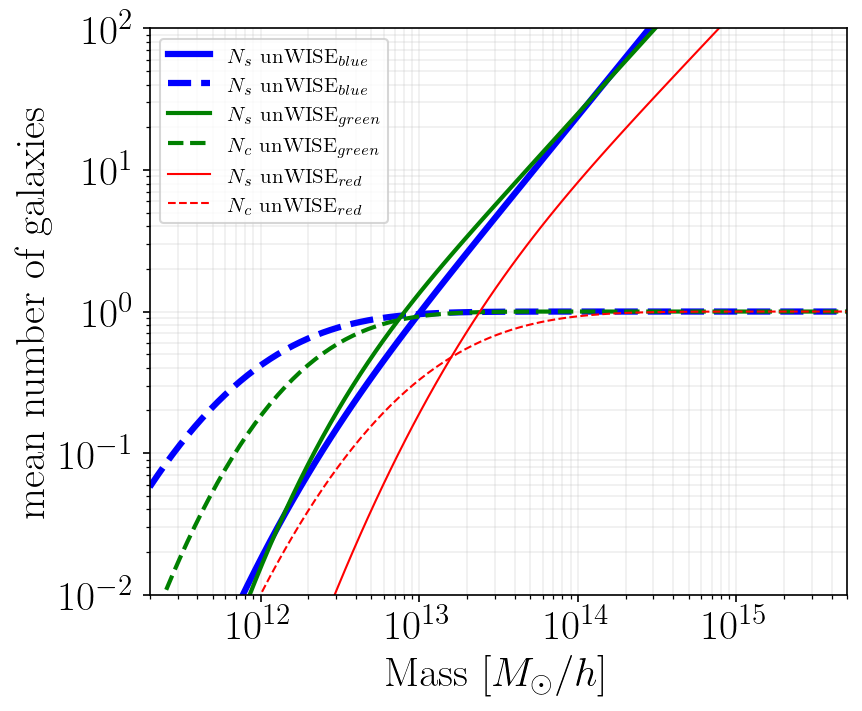}
    \includegraphics[scale=0.5]{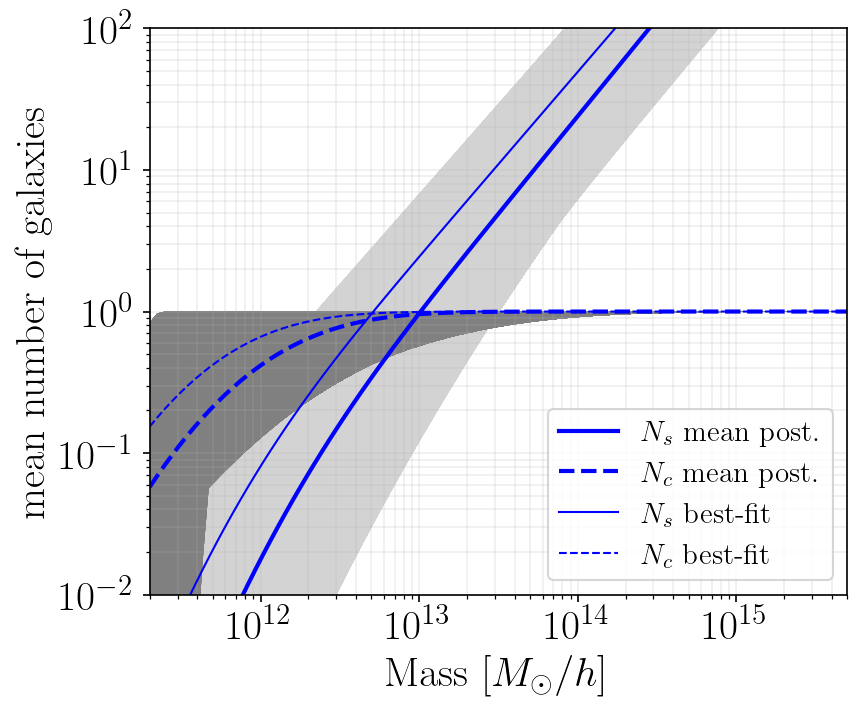}
    \includegraphics[scale=0.5]{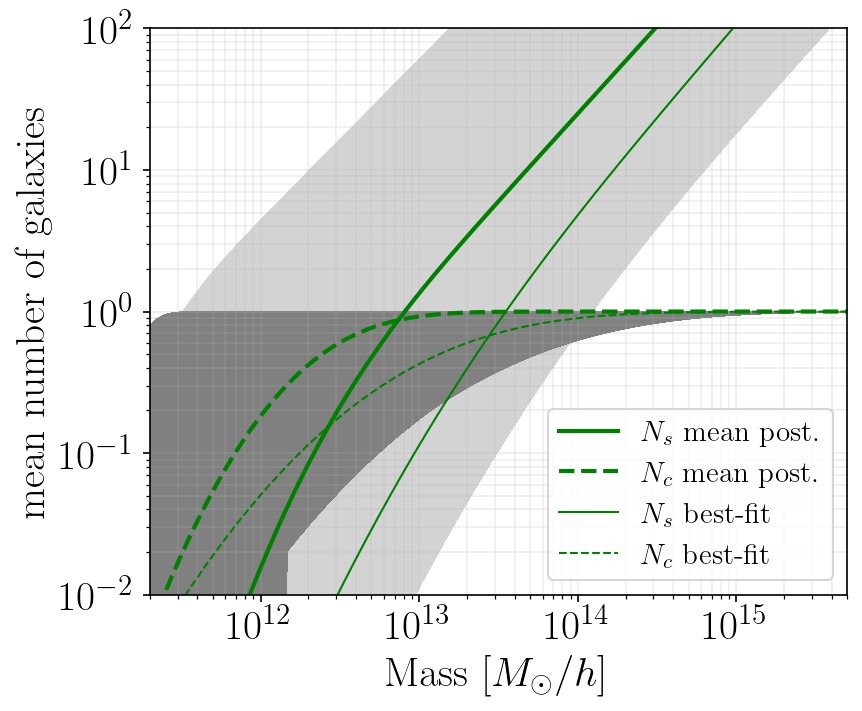}
    \includegraphics[scale=0.5]{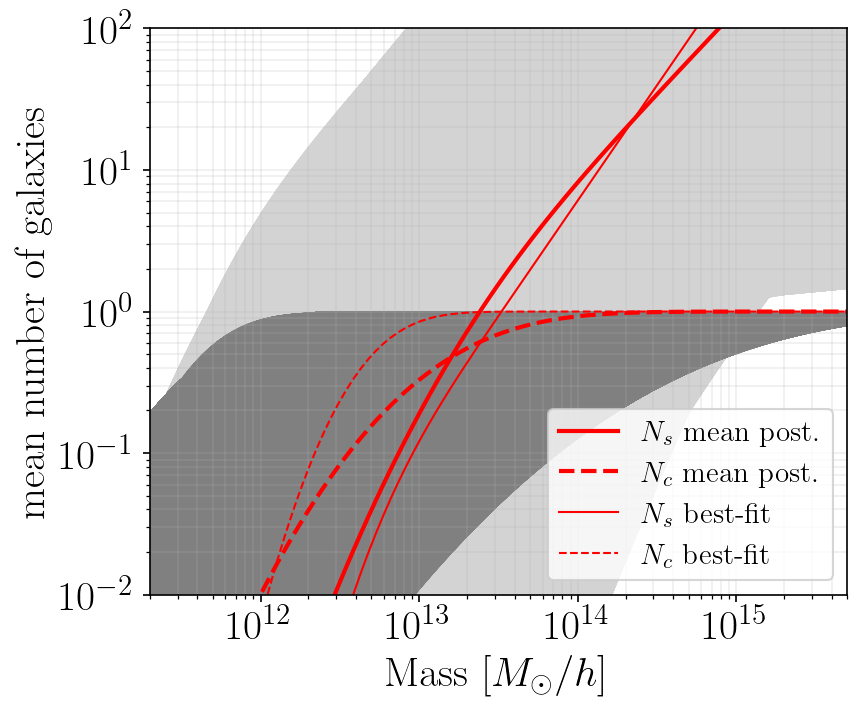}
    \caption{Mean number of central and satellite galaxies, $N_c$ and $N_s$, versus halo mass for the \emph{unWISE} samples, computed for the mean posterior values of the HOD parameters (Table~\ref{table:hod_results_mean}). The solid lines show $N_s$ and the dashed lines show $N_c$. Top left: all three \emph{unWISE} samples on one plot. Top right: blue sample.  Bottom left:  green sample. Bottom right:  red sample. For the individual plots, we also include the prediction computed for the best-fit values of the HOD parameters (Table~\ref{table:hod_results}) in thinner lines. The light grey (dark grey) regions show the $N_s$ ($N_c$) curves computed for the HOD parameter values from the last 80,000 steps of the MCMC chains to illustrate the uncertainty on the mean number of satellite (central) galaxies. }
    \label{fig:NcNs}
\end{figure}

Given our HOD model, we can derive various quantities with the obtained results. Firstly, we present the mean number of central and satellite galaxies, $N_c$ and $N_s$, for each of the \emph{unWISE} samples (Eqs.~\ref{eq:N_c} and \ref{eq:N_s}). Fig.~\ref{fig:NcNs} shows $N_c$ and $N_s$ as a function of halo mass, computed for the mean posterior values of the HOD parameters (Table~\ref{table:hod_results_mean}), for each of the \emph{unWISE} samples, along with shaded regions corresponding to the HOD values obtained from the last 80,000 steps of the MCMC chains to illustrate the uncertainty on the computed quantities. From Fig.~\ref{fig:NcNs}, we can see that the mean number of central galaxies is larger for lower halo masses for the blue sample than for the green and red ones. The satellite number $N_s(M)$ is very similar for the blue and green samples, and the mean number of satellites for these two samples is larger for lower halo masses than for the red one.  

\begin{figure}
    \centering
    \includegraphics[scale=0.5]{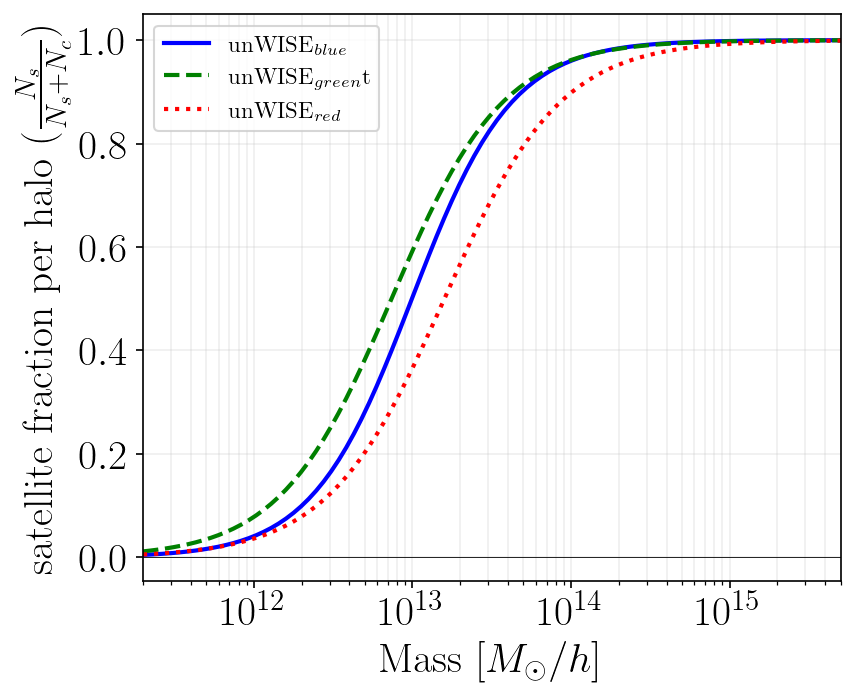}
    \includegraphics[scale=0.5]{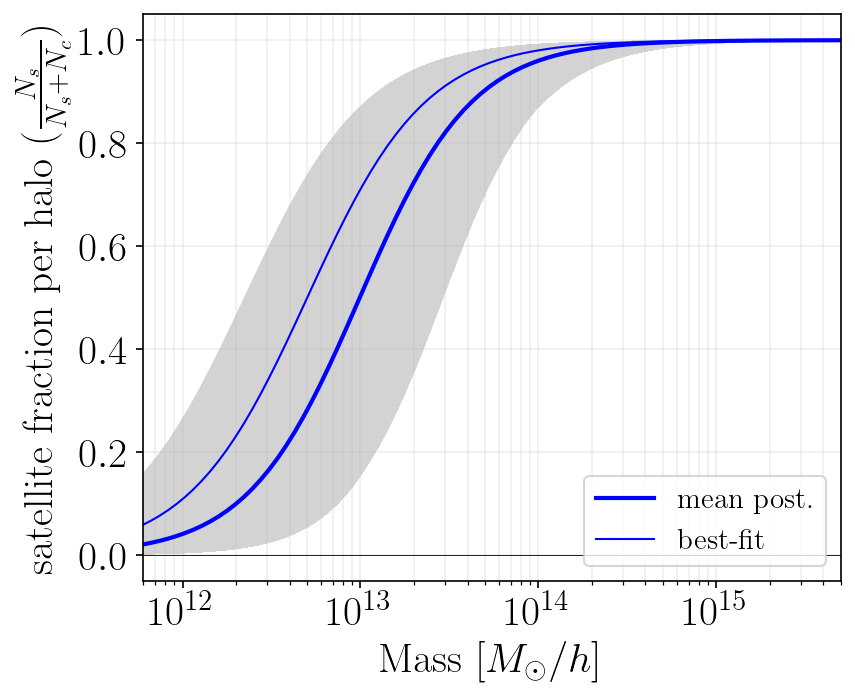}
    \includegraphics[scale=0.5]{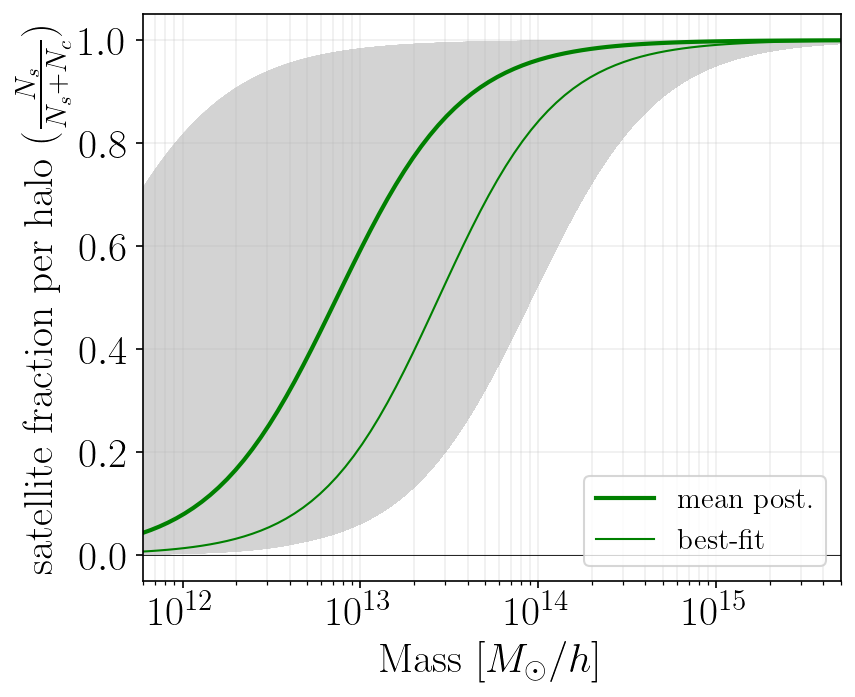}
    \includegraphics[scale=0.5]{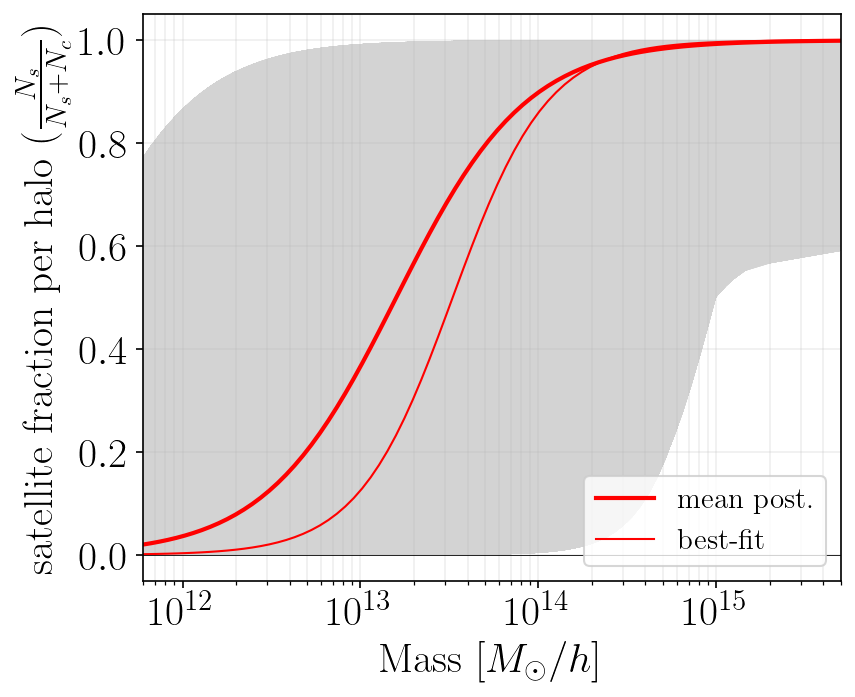}
    \caption{Fraction of the satellite galaxies per halo at a given halo mass, $\frac{N_s}{N_s+N_c}$ for the blue, green, and red sample, computed for the mean posterior values of the HOD parameters (Table~\ref{table:hod_results_mean}). Top left: all three \emph{unWISE} samples on one plot, blue (solid), green (dashed), and red (dotted). Top right: blue sample. Bottom left: green sample. Bottom right: red sample. For the individual plots, we also include the prediction computed for the best-fit values of the HOD parameters (Table~\ref{table:hod_results}) in thinner lines. The shaded regions correspond to the HOD values obtained from the last 80,000 steps of the MCMC chains to illustrate the uncertainty on the computed quantities. This figure is analogous to Fig.~\ref{fig:NcNs}}
    \label{fig:Ns_fract}
\end{figure}

From the mean number of centrals and satellites, we can also compute the satellite galaxy fraction per halo at a given halo mass, $\frac{N_s}{N_s+N_c}$. We show this quantity in Fig.~\ref{fig:Ns_fract} for each of \emph{unWISE} samples, computed for the mean posterior values of the HOD parameters (Table~\ref{table:hod_results_mean}). From this plot we note that at a given mass, there tends to be more satellites in the green sample than in blue and red, yet all three samples seem to have a similar fraction of satellites for a given mass, within the uncertainties. The computed $\frac{N_s}{N_s+N_c}$ also goes to one at high masses, which is an expected, physical result, as the sample is dominated by satellite galaxies at high halo masses. 

Similarly, we can define the total satellite fraction $\alpha_{\mathrm{sat}}$  in the entire sample
\begin{equation}
    \alpha_{\rm{sat}} = \int \mathrm{d} z \frac{1}{N_{\rm{g}}^{\rm{tot}}} \frac{\mathrm{d} N }{\mathrm{d} z} \int_{M_\mathrm{min}}^{M_\mathrm{max}}\mathrm{d}M  \frac{\mathrm{d}n}{\mathrm{d}M} \frac{N_s}{\bar{n}_g(z)}.
    \label{eq:alpha_sat}
\end{equation}
We find $\alpha_{\rm sat} = 0.24$, $0.28$, and $0.36$ for the blue, green, and red sample, respectively, using the mean values of the posteriors of the HOD parameters (Table~\ref{table:hod_results_mean}), and $\alpha_{\rm sat} = 0.30$, $0.16$, and $0.14$, when using the best-fit values of the HOD parameters (Table~\ref{table:hod_results}). To quantify the uncertainty, we calculate  $\alpha_{\rm sat}$ for the last 80,000 steps of the MCMC chains, and obtain $\alpha_{\rm sat} = 0.24^{+0.12}_{-0.11}$, $0.28^{+0.15}_{-0.11}$, and $0.36^{+0.47}_{-0.31}$, where the error bars denote the 68\% CL of the calculated $\alpha_{\rm sat}$ distribution. Given these results, we conclude that the majority of the galaxies in the \emph{unWISE} catalog are centrals, yet the number of satellites is non-negligible.

Secondly, we also present the effective linear galaxy bias as a function of redshift, for each of the \emph{unWISE} samples as predicted with our best-fit parameter values (Table~\ref{table:hod_results}). This quantity is just an integral over mass of the linear bias $b(M,z)$, the halo mass function $\frac{\mathrm{d}n}{\mathrm{d}M}$  (as noted in Section~\ref{subsec:HMframework}, we use the Tinker \textit{et al.} 2010 \cite{Tinker_2010} linear bias and Tinker \textit{et al.} 2008 HMF \cite{Tinker_2008}), and the mean number of galaxies, defined as 
\begin{equation}
    b_\mathrm{eff}(z) \equiv \frac{1}{\bar{n}_g(z)} \int_{M_\mathrm{min}}^{M_\mathrm{max}} \mathrm{d}M \frac{\mathrm{d}n}{\mathrm{d}M}  b(M, z)(N_c(M) + N_s(M)),
    \label{eq:eff_bias}
\end{equation}
where $\bar{n}_g (z)$ is defined in Eq.~\eqref{eq:n_g} and $N_c$ and $N_s$ are the HOD formulas of Eq.~\eqref{eq:N_c} and \eqref{eq:N_s}. By multiplying $b_\mathrm{eff}(z)$ by the normalized redshift distribution $\frac{1}{N_g^\mathrm{tot}}\frac{\mathrm{d}N_g}{\mathrm{d}z}$ (Eq.~\ref{eq:varphig}) of each sample and integrating over redshift we can also define the mean galaxy bias $b_g$ of each sample
\begin{equation}
        b_g = \int  \mathrm{d} z \frac{1}{N_{\rm{g}}^{\rm{tot}}} \frac{\mathrm{d} N }{\mathrm{d} z} b_\mathrm{eff}(z).
        \label{eq:b_g}
\end{equation}

\begin{figure}
    \centering
    \includegraphics[scale=0.4]{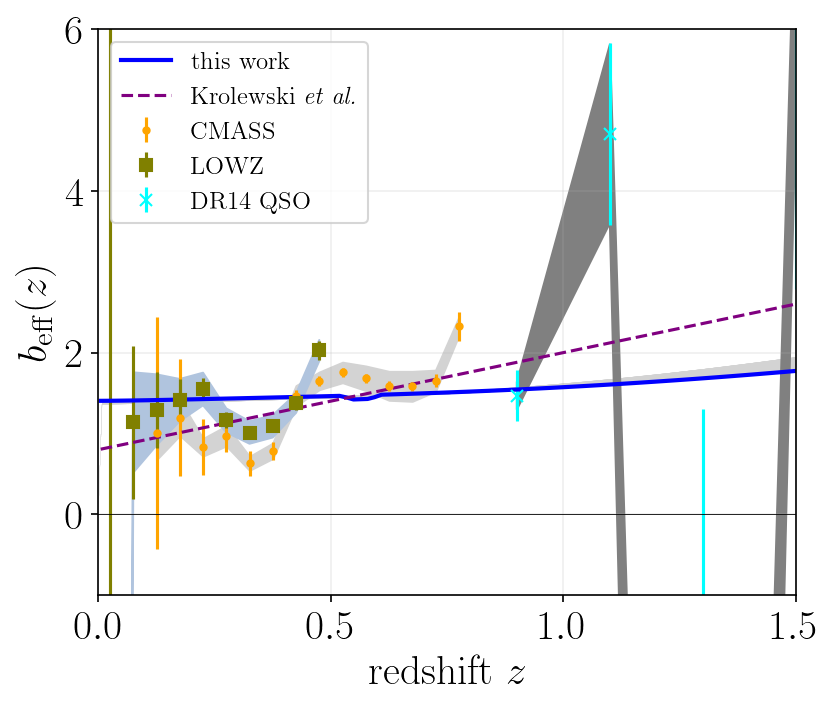}
    \includegraphics[scale=0.4]{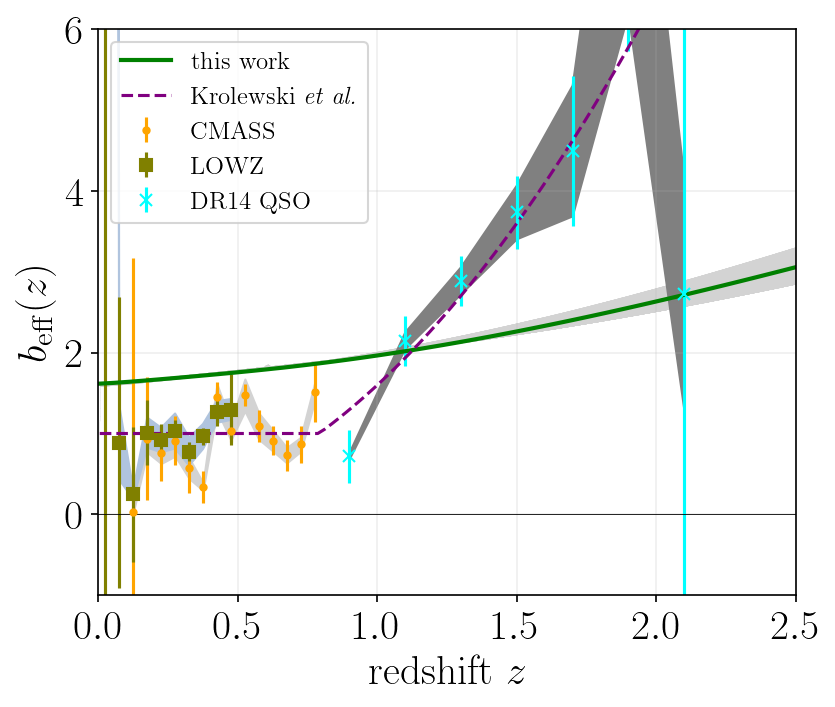}
    \includegraphics[scale=0.4]{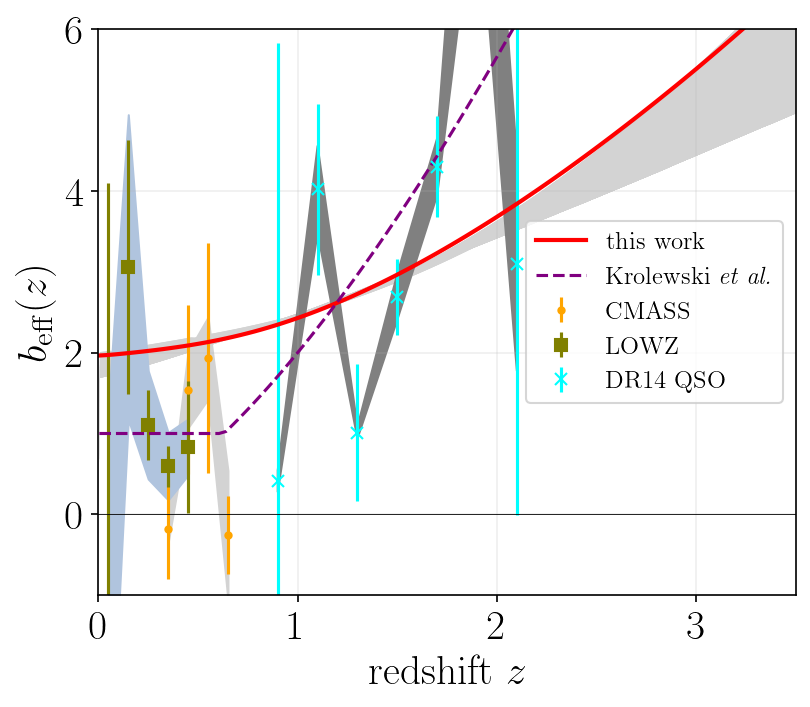}
    \caption{Effective linear bias $b_\mathrm{eff}(z)$ versus redshift $z$ for each of the \emph{unWISE} samples, from left: blue, green, and red. The solid curves show the effective bias calculated in this work (Eq.~\ref{eq:eff_bias}) for the Tinker \textit{et al.} \cite{Tinker_2010} linear bias for our best-fit model for each sample (Table~\ref{table:hod_results}), and the solid light grey lines show the bias curves computed for the HOD parameter values from the last 80,000 steps of the MCMC chains to illustrate the uncertainty on $b_\mathrm{eff}(z)$. The data points show the bias measurements from \citetalias{krolewski_2020, krolewski2021cosmological}, obtained by cross-correlating the \emph{unWISE} galaxies with spectroscopic galaxies from LOWZ (squares) and CMASS (dots) and quasars from DR14 (crosses). The dashed purple curves from  \citetalias{krolewski_2020, krolewski2021cosmological} were adjusted by hand to these data.  
    We also show additional uncertainty on these measurements as gray areas, which were obtained by propagating the uncertainty on the redshift distribution $\mathrm{d}N_g/\mathrm{d}z$, obtained in \citetalias{krolewski_2020, krolewski2021cosmological}.
   This plot is an independent check of our HOD model to measurements obtained with external data. }
    \label{fig:bias}
\end{figure}

In Fig.~\ref{fig:bias} we show the effective linear galaxy bias as a function of redshift computed with our best-fit parameter values for each of the \emph{unWISE} samples, from left: \emph{unWISE} blue, green, and red (also color-coded).  Again the grey curves are computed for the HOD parameter values from the last 5000 steps of the MCMC chains to illustrate the uncertainties on $b_{\rm eff}(z)$. As mentioned in Section~\ref{sec:intro}, \citetalias{krolewski_2020, krolewski2021cosmological} also investigated a simple HOD model for the \emph{unWISE} galaxies to test their cosmological inference pipeline, and measured $b(z)$ by cross-correlating the \emph{unWISE} photometric galaxies with spectroscopic quasars from BOSS DR12 \cite{SDSSquasars_eBOSS_DR122017} and eBOSS DR14 \cite{eBOSS_dr14_2017} and galaxies from BOSS CMASS and LOWZ \cite{BOSSgal_2015}. In Fig.~\ref{fig:bias}, we also show the bias measurements from \citetalias{krolewski_2020, krolewski2021cosmological}, obtained by cross-correlating with quasars from BOSS CMASS and LOWZ (squares), CMASS (dots), and DR14 (crosses), where the shaded grey areas correspond to additional uncertainty on these measurements from the uncertainty on the redshift distribution $\mathrm{d}N_g/\mathrm{d}z$, obtained in \citetalias{krolewski_2020, krolewski2021cosmological}.  The uncertainty on $dN_g/dz$ is important here, as the cross-correlation between \emph{unWISE} and the spectroscopic samples directly probes $b(z) dN_g/dz$, and thus to obtain $b(z)$ an independent estimate of the redshift distribution is required (as stated previously, this is determined from a cross-match to the \emph{COSMOS} data).  The purple dashed lines show an estimated (by eye) fit to the data from \citetalias{krolewski_2020, krolewski2021cosmological}. 
The redshift evolution of the effective linear bias obtained in this work is not as steep as that obtained in \citetalias{krolewski_2020, krolewski2021cosmological}, but it roughly agrees with the measured bias within the error bars and additional uncertainty of these measurements.  This cross-validation is non-trivial, as the data points shown in Fig.~\ref{fig:bias} are not used in our HOD fitting analysis.  One of the possible improvement strategies for our work would be to fully propagate the uncertainty in the redshift distribution $\mathrm{d}N_g/\mathrm{d}z$ into the HOD model results. As described in Sec.\ref{sec:data}, we use $\mathrm{d}N_g/\mathrm{d}z$ obtained by cross-matching \emph{unWISE} galaxies with \emph{COSMOS} objects, but we do not propagate any uncertainty on this quantity, due to computational expense. Furthermore, it might be interesting to include the \emph{unWISE} cross-correlation measurements with the spectroscopic galaxy and quasar samples from \citetalias{krolewski_2020, krolewski2021cosmological} directly in the HOD fitting analysis.  We leave exploration of these avenues to future work. For the values of the mean galaxy bias $b_g$ from Eq.~\ref{eq:b_g}, we obtain 1.49, 2.01, 2.98 for \emph{unWISE} blue, green, and red, respectively, which compare well with the values obtained in \citetalias{krolewski_2020, krolewski2021cosmological} (1.6, 2.2, 3.3).

\begin{figure}
    \centering
    \includegraphics[scale=0.5]{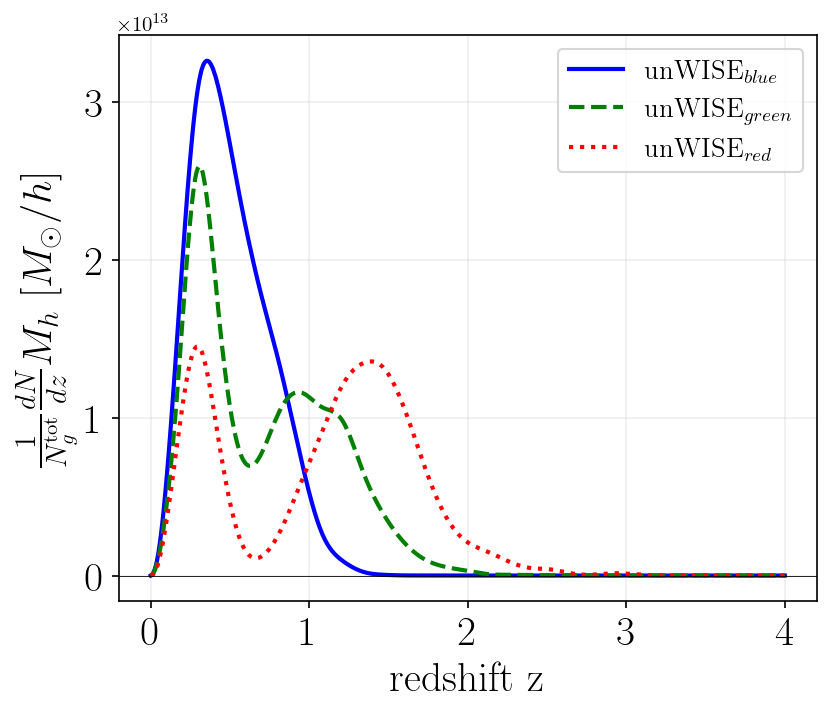}
    \includegraphics[scale=0.5]{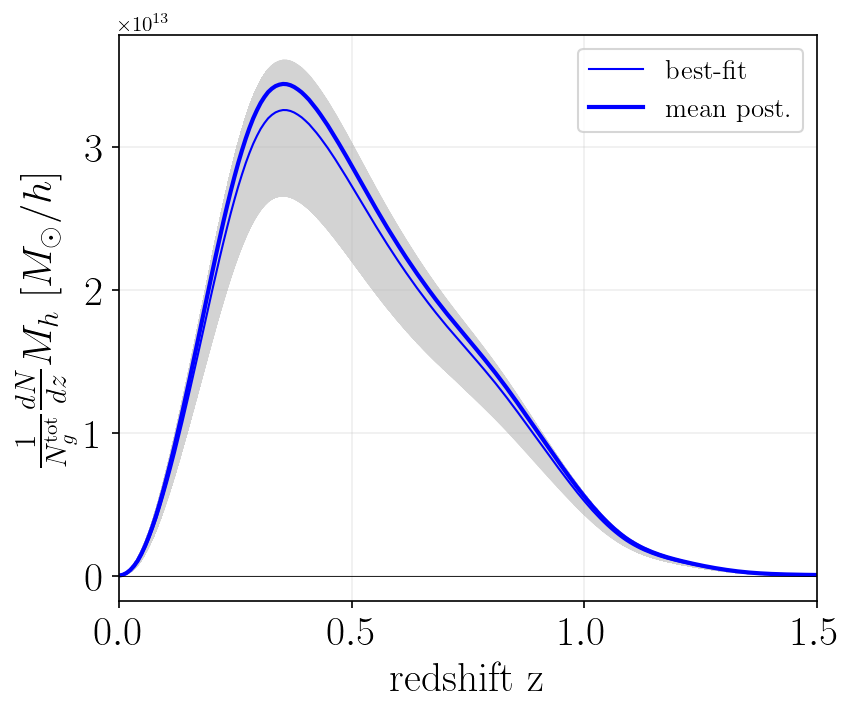}
    \includegraphics[scale=0.5]{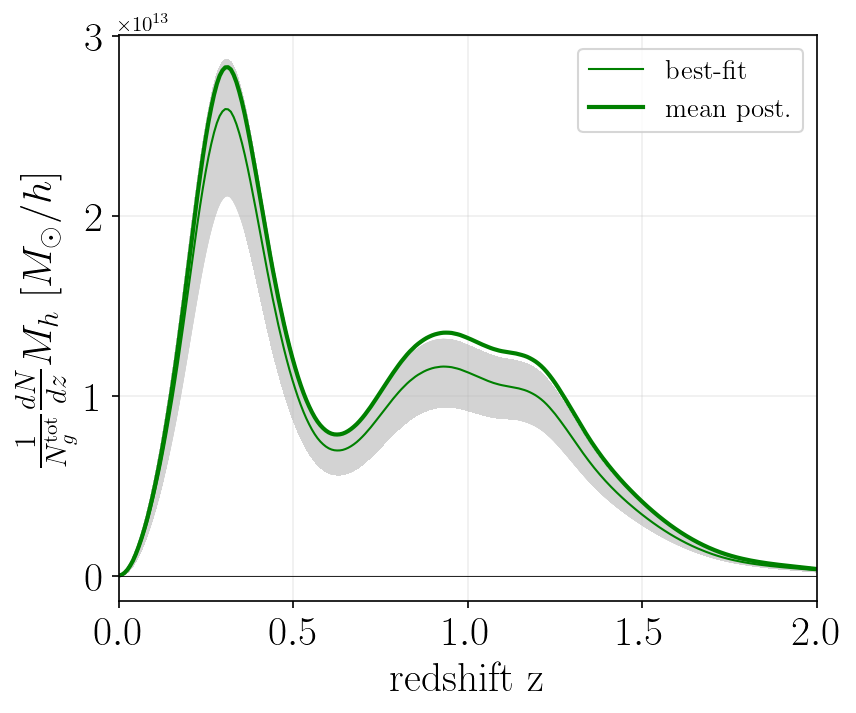}
    \includegraphics[scale=0.5]{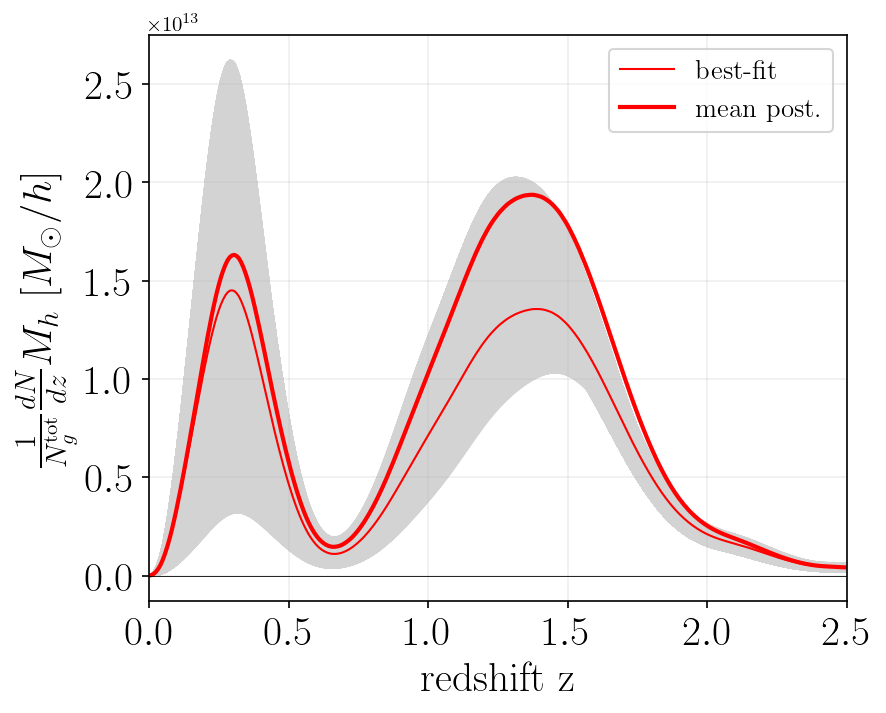}
    \caption{Top left: Average host halo mass $M_h (z)$ multiplied by the normalized redshift distribution $\frac{1}{N_g^\mathrm{tot}}\frac{\mathrm{d}N_g}{\mathrm{d}z}$ (Eq.~\ref{eq:varphig}) versus redshift for each of the \emph{unWISE} samples, blue (solid), green (dashed), and red (dotted), calculated for the mean posterior values of the HOD parameters (Table~\ref{table:hod_results_mean}). Top right: similar plot, only for the blue sample. Bottom left: for the green sample. Bottom right: for the red sample. For the latter three plots, we also include the prediction computed for the best-fit values of the HOD parameters (Table~\ref{table:hod_results}) in thinner lines. The shaded regions correspond to the HOD values obtained from the last 80,000 steps of the MCMC chains to illustrate the uncertainty on the computed quantities. }
    \label{fig:Mh}
\end{figure}

Finally, we can also derive the mean host halo mass as a function of redshift $M_h (z)$ for each of the \emph{unWISE} samples, which is defined as:
\begin{equation}
    M_h (z) = \frac{1}{\bar{n}_g (z)} \int_{M_\mathrm{min}}^{M_\mathrm{max}}\mathrm{d}M  \frac{\mathrm{d}n}{\mathrm{d}M}M\left(N_c+N_s\right).
\end{equation}
We plot this function multiplied by the normalized redshift distribution $\frac{1}{N_g^\mathrm{tot}}\frac{\mathrm{d}N_g}{\mathrm{d}z}$ (Eq.~\ref{eq:varphig}) in Fig.~\ref{fig:Mh} for each of the \emph{unWISE} samples.  Then by integrating this quantity over redshift, we calculate the mean host halo mass for each sample, which we define as $M_h$. We obtain $M_h = 1.99, 1.86, 2.04 \times 10^{13} M_\odot / h $ for \emph{unWISE} blue, green, and red, respectively, using the best-fit HOD parameter values (Table~\ref{table:hod_results}).  These results compare well with the mean halo mass estimates in \citetalias{krolewski_2020}, $1-5 \times10^{13} M_\odot / h $, which were inferred from the linear biases of the galaxy samples. 

\section{Discussion and Outlook}
\label{sec:discussion}

In this work, we have constrained the galaxy-halo connection for the \emph{unWISE} galaxies using the HOD and halo model approach. We fit the joint \emph{unWISE} galaxy-galaxy auto-correlation and galaxy-\emph{Planck} CMB lensing cross-correlation to the halo model predictions (see Section~\ref{sec:HM}) to constrain six model parameters $\{ \alpha_{\mathrm{s}}$, $\sigma_{\mathrm{log} M}$, $M_\mathrm{min}^\mathrm{HOD}$, $M^{\prime}_1$, $\lambda$, $A_{\mathrm{SN}} \}$, separately for each of the three \emph{unWISE} galaxy samples. The results are presented in Tables~\ref{table:hod_results} and~\ref{table:hod_results_mean} and the best-fit models are shown in Fig.~\ref{fig:bestfit}. This work is the first detailed HOD modeling of \emph{WISE}-selected galaxies. A basic HOD for \emph{unWISE} was considered in \citetalias{krolewski_2020} and \citetalias{krolewski2021cosmological}, where the authors investigated a simple model with redshift-dependent HOD parameters and fit it to the galaxy-galaxy, galaxy-CMB lensing, and also effective bias measurements by hand in order to test their mock pipeline. Our work provides a much more systematic and quantitative approach to constrain the HOD parameters in the \emph{unWISE} samples than \citetalias{krolewski_2020, krolewski2021cosmological} because of the detailed halo model description and quantitative fitting procedure. 

By performing the analysis for three different \emph{unWISE} galaxy subsamples with different redshift distributions (mean redshifts $\bar{z} \approx 0.6, 1.1, 1.5$) and characterized by different magnitude cuts (see Table~\ref{table:unwise_cuts}), we have probed the evolution of some of the HOD parameters between samples. We find a particularly strong sample-redshift trend in the mass parameters, i.e., $M_\mathrm{min}^\mathrm{HOD}$, the characteristic minimum mass of halos that can host a central galaxy and $M^{\prime}_1$, the mass scale of the satellite profile drop. One might be tempted to interpret this trend as a redshift evolution of the HOD parameters, however, this effect might be also caused by selection biases -- at higher redshifts, we can only observe brighter galaxies (due to surface brightness dimming at large distances) and therefore more massive halos. Thus, the mass parameters are the largest for the red sample.

We also note that comparing our constraints with the HOD descriptions of other galaxy samples (e.g., the DES-Y3 constraints in \cite{Zacharegkas2021}) is not necessarily straightforward, as those galaxies might have very different characteristics than the \emph{unWISE} catalog, related to how the objects are selected, even if they share similar redshifts. As a reminder, \emph{unWISE} is predominantly a rest-frame near-infrared catalog (observer-frame mid-infrared) up to redshift $z < 4$, thus the galaxy selection probes mainly the stellar emission directly, but is also sensitive to the thermal dust heated by starlight. 

There are several areas for future improvement in our analysis.  Firstly, as mentioned in Section~\ref{sec:results}, the uncertainty on the redshift distribution $\mathrm{d}N_g/\mathrm{d}z$ (obtained by cross-matching the \emph{unWISE} galaxies with \emph{COSMOS} objects, which have precise 30-band photometric redshifts) in this analysis has not been properly quantified.~(We performed an exploratory comparison of the impact of shifting the redshift distributions by 5\%-10\% in different directions, and the effect can be particularly large for the $\chi^2$ of the blue sample galaxy-galaxy data, given it has the smallest error bars). However, it would be a particularly difficult computational task to include this uncertainty properly in our MCMC fitting procedure, yet it could be done by parametrizing the error on $\mathrm{d}N_g/\mathrm{d}z$ (e.g., a constant shift to lower or higher $z$). Furthermore, an improvement in the redshift distribution estimation by cross-matching with ongoing and upcoming spectroscopic surveys, e.g., DESI, would also be very useful. We leave this for future work.
Secondly, the HOD, although extremely successful, is only an empirical model. Therefore a crucial step to validate our results will be to compare the HOD constraints with simulations. This could be done with dark matter simulations as a first step, populated with galaxies with a semi-analytic model (SAM), e.g., the Santa Cruz semi-analytic model \cite{SAM_1999, SAM_2017}, as they are less expensive than hydrodynamical simulations. There exist high-resolution hydrodynamical simulations, e.g., Illustris-TNG \cite{IllustrisTNG_2017}, but their volume is quite limited compared to our enormous galaxy sample, which spans redshifts up to $z \approx 4$ on the full sky. In both cases, the HOD modeling and constraints obtained in this analysis are, however, crucial to populate the simulations with galaxies in appropriate dark matter halos. This is also left for future work. 

We also intend to use the HOD constraints on \emph{unWISE} galaxies obtained in this work to carry on a joint analysis of the thermal and kinematic Sunyaev-Zel'dovich effects of \emph{unWISE} in the halo model framework (which constrain the electron pressure and density profiles, respectively), and then to further describe the thermodynamics of the electron gas. The kSZ measurement for \emph{unWISE} has been already performed in Ref.~\cite{Kusiak_2021} with \emph{Planck} data with the projected-fields method \cite{Dore2004, DeDeo, Hill2016, Ferraro2016}.  We plan to re-interpret this measurement in the halo model using the HOD results obtained here. 

It is also worth mentioning that, in principle, an HOD analysis like the one performed in this work could enable constraints on cosmological parameters (i.e., $\sigma_8$ and $\Omega_m$) that extend to higher $\ell_{\rm{max}}$ than the perturbation theory approach in \citetalias{krolewski2021cosmological}. We have made an initial exploration of this analysis and found that the degeneracies of the HOD and cosmological parameters are significant, thus hindering the constraints, but this could potentially be improved in future work by combining with weak lensing-galaxy cross-correlations using, e.g., DES weak lensing maps (as done for DES in Ref.~\cite{Kwan2017_DES}).

\section{Acknowledgements}
We thank Simone Ferraro for many helpful exchanges. We also thank Fiona McCarthy and Emmanuel Schaan for discussions about consistency of the halo model formalism, and the anonymous referee for useful comments. 
Some of the results in this paper have been derived
using the healpy and HEALPix packages \cite{healpy_paper1, healpy_paper2}. This research used resources of the National Energy Research Scientific Computing Center (NERSC), a U.S. Department of Energy Office of Science User Facility located at Lawrence Berkeley National Laboratory. AKK and JCH acknowledge support from NSF grant AST-2108536. The Flatiron Institute is supported by the Simons Foundation. AGK thanks the AMTD Foundation for support.

\appendix
\label{Appendix}

\section{Separate fitting of galaxy-galaxy and CMB lensing-galaxy data}
\label{app:gg_kg_separate}

\begin{figure}
    \centering
    \includegraphics[scale=0.25]{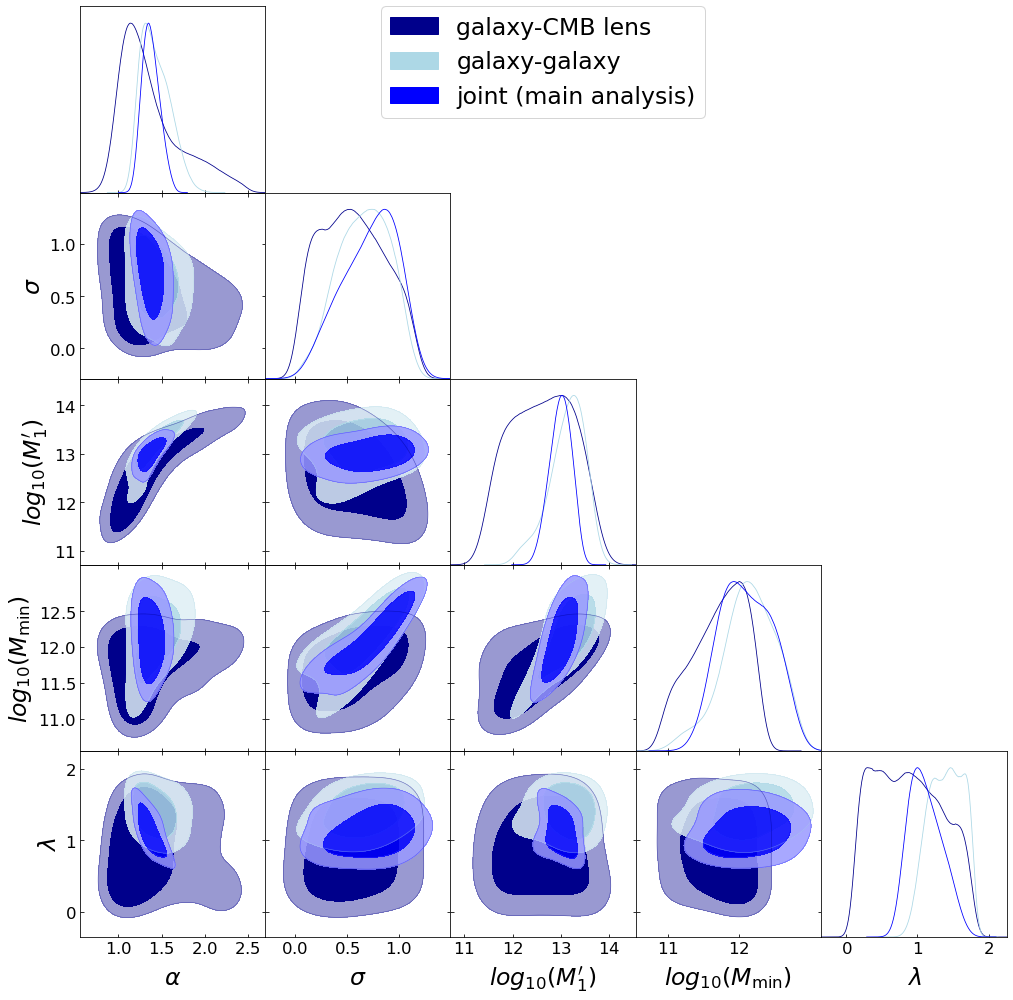}
    \includegraphics[scale=0.25]{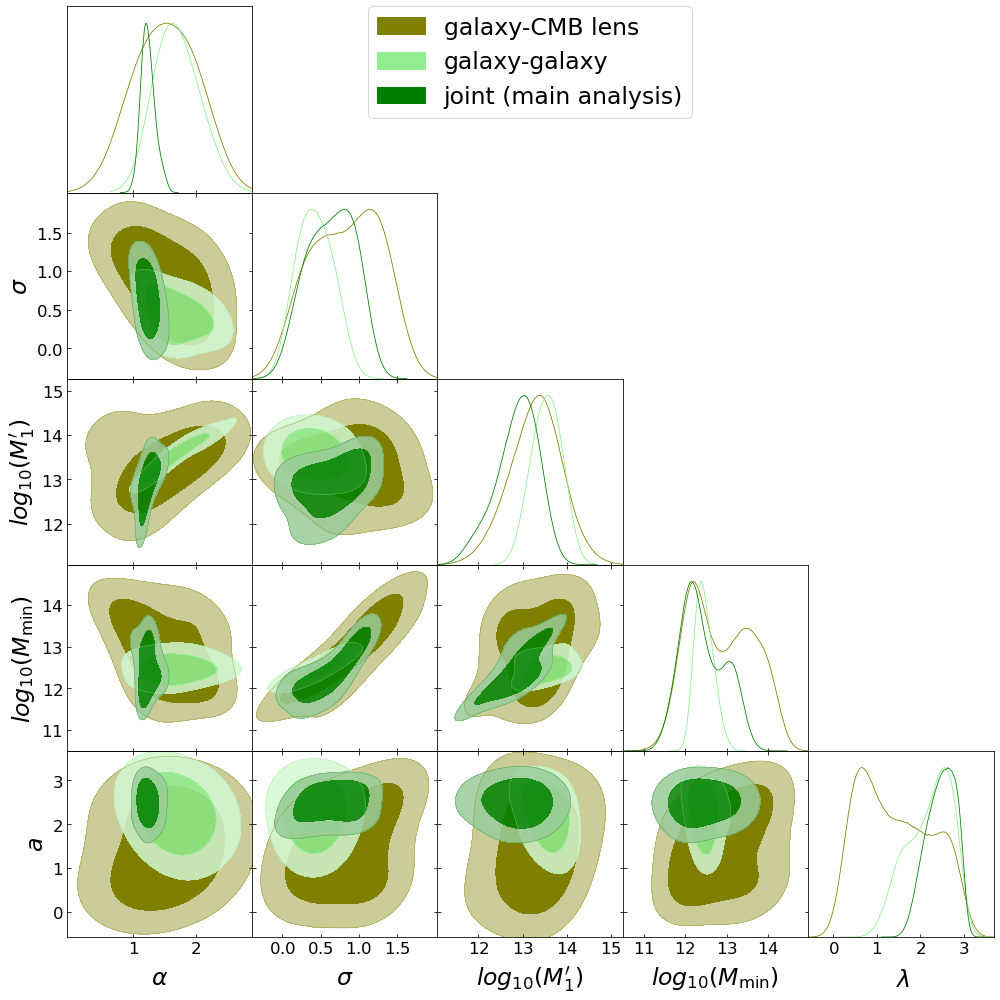}
    \includegraphics[scale=0.22]{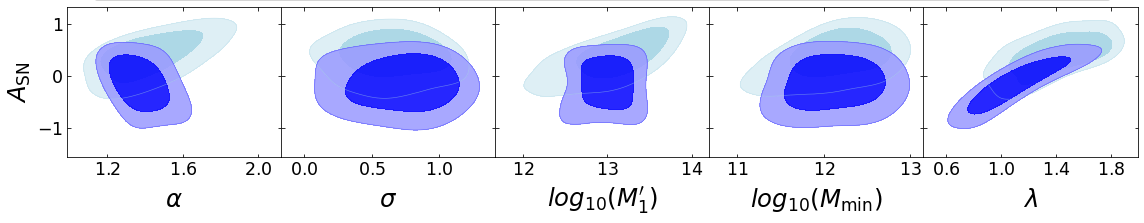}
    \includegraphics[scale=0.22]{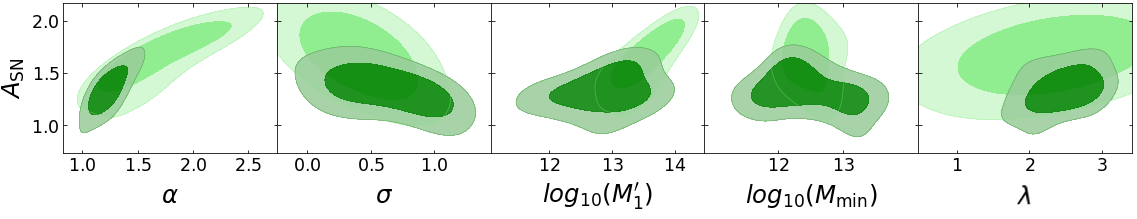}
    \caption{The 1D and 2D marginalized posterior distributions  of the model parameters $\{ \alpha_{\mathrm{s}}$, $\sigma_{\mathrm{log} M}$, $M_\mathrm{min}^\mathrm{HOD}$, $M^{\prime}_1$, $\lambda$ $A_{\mathrm{SN}} \}$ for three fitting scenarios: 1) fitting the $C_\ell^{gg}$-only data to our theory halo model predictions (Section~\ref{sec:theory}); 2) similar, but fitting $C_\ell^{g \kappa_{\rm{cmb}}}$-only; 3) fitting $C_\ell^{gg}$ and $C_\ell^{g \kappa_{\rm{cmb}}}$ jointly (our main analysis; see Section~\ref{sec:results} for the discussion of results). We show posterior distributions for two \emph{unWISE} samples: blue (left) and green (right).}
    \label{fig:posteriors_kg_gg}
\end{figure}

In this appendix, in order to validate the HOD constraints obtained in this work (Section~\ref{sec:results}), we present the results of fitting the galaxy-galaxy $C_\ell^{gg}$ and galaxy-CMB lensing $C_\ell^{g \kappa_{\rm{cmb}}}$ measurements to our halo model predictions separately, in contrast to fitting them jointly in the main analysis. Fig.~\ref{fig:posteriors_kg_gg} shows the 1D and 2D marginalized posterior distributions for the blue and green \emph{unWISE} samples, for three fitting scenarios to constrain the HOD parameters: 1) fitting $C_\ell^{gg}$-only data (9 data points) to the halo model galaxy-galaxy angular auto-power spectrum predictions; 2) fitting $C_\ell^{g \kappa_{\rm{cmb}}}$-only data (10 data points) to the halo model galaxy-CMB lensing angular cross-power spectrum predictions; 3) fitting $C_\ell^{gg}$ and $C_\ell^{g \kappa_{\rm{cmb}}}$ measurements (19 data points) jointly, as done and presented in the main analysis (see Section~\ref{sec:results}). In all scenarios, we follow the same fitting procedure as described in Section~\ref{sec:HODfitting}. In particular, we impose the same priors on the parameters of interest (Table~\ref{table:priors}). On the top plots we show the
5 model parameters that overlap for the three fitting scenarios $\{ \alpha_{\mathrm{s}}$, $\sigma_{\mathrm{log} M}$, $M_\mathrm{min}^\mathrm{HOD}$, $M^{\prime}_1$, $a \}$, while on the bottom one we show the $A_{\mathrm{SN}}$ parameter which is only relevant for the galaxy-galaxy auto and joint analyses.  For the galaxy-CMB lensing-only MCMC analysis, we use the same convergence criterion as used in our main analysis ($R-1 < 0.1$), while for the galaxy-galaxy-only MCMC analysis we use a slightly relaxed criterion, $R-1 < 0.3$.

From Fig.~\ref{fig:posteriors_kg_gg}, we note that in all cases the obtained constraints on the HOD parameters are consistent, which validates our main joint  analysis. We also point out that most of the constraining power in the joint analysis comes from the $C_\ell^{gg}$ data (which is expected due to the much higher signal-to-noise of the galaxy clustering measurements), yet the $C_\ell^{g \kappa_{\rm{cmb}}}$ data also provides some additional information on the parameters of interest.

\section{Choice of the halo mass function}
\label{app:HMF}

Although the Tinker \textit{et al.} (2010)  halo mass function (HMF) is normalized such that $\int \mathrm{d}\nu f(\nu)=1$ (see \cite{Tinker_2010}) while the Tinker \textit{et al.} (2008) version \cite{Tinker_2008} is not, we opt for the Tinker \textit{et al.} (2008) HMF in this work for two reasons:
 \begin{itemize}
     \item The normalization condition on the 2010 HMF is imposed via an analytical regularization, with a negative exponential of the form $e^{-g/\sigma^2}$ (see Appendix C of \cite{Tinker_2008}): this procedure is not based on simulation results (which do not extend to arbitrarily low halo masses). 
     \item Tinker \textit{et al.} (2010) \cite{Tinker_2010} do not provide a table with second derivatives with respect to $\Delta$ (for the overdensity mass definition) and therefore we cannot accurately interpolate the formula at arbitrary mass definition. For instance, the fitting parameter values are not available for $M_{200\mathrm{c}}$. Since we want to provide our results at $M_{200\mathrm{c}}$ (having in mind the comparison with HOD results from \cite{Zacharegkas2021} or future work involving gas pressure and density defined at this overdensity mass), by opting for the Tinker \textit{et al.} (2008) HMF we can avoid adding extra uncertainty associated with mass conversions. Indeed, the Tinker \textit{et al.} (2008) article \cite{Tinker_2008} provides a table of second derivatives which we use for a spline interpolation at any mass definition.
 \end{itemize}
Since the Tinker \textit{et al.} (2008) HMF is not normalized, we implement the prescriptions proposed by Schmidt to restore consistency \citep[see][for details]{fs2016}.
 
\begin{figure}
    \centering
    \includegraphics[scale=0.5]{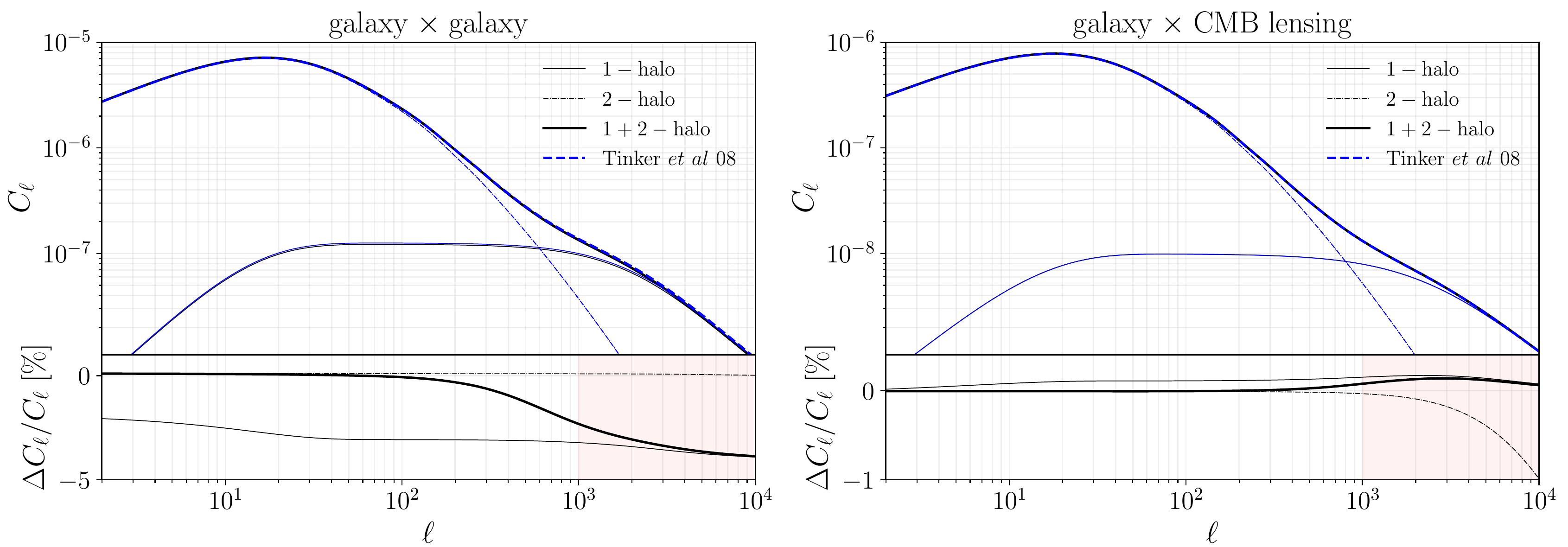}
    \caption{Galaxy-galaxy  (left) and galaxy-CMB lensing (right) power spectra with two different choices for the halo mass function: the Tinker \textit{et al.} (2008) formula \cite{Tinker_2008} (blue and blue-dashed) and the Tinker \textit{et al.} (2010) formula \cite{Tinker_2010} (black). The bottom panels show the fractional difference with respect to the Tinker \textit{et al.} (2008) formula. The thin lines are the 1-halo (solid) and 2-halo (dotted-dashed) contributions. Since the Tinker \textit{et al.} (2010) formula is not available at mass definition $M_\mathrm{200c}$, we compute both formulas for $M_\mathrm{200m}$ masses and convert to $M_\mathrm{200c}$ in the NFW profiles using the Bhattacharya \textit{et al.} (2013) concentration-mass relation \cite{Bhattacharya_2013}. The shaded areas in red show the multipole range that is not used in this analysis ($\ell >1000$). For other parameters and settings we assume our fiducial model (see Section~\ref{sec:intro}) and the best-fit HOD parameters of the \emph{unWISE} blue sample (see Table~\ref{table:hod_results}).}
    \label{fig:T08vsT10_gg_kg}
\end{figure}

For the sake of completeness, in Fig.~\ref{fig:T08vsT10_gg_kg}
we compare our power spectra predictions for the Tinker \textit{et al.} (2008) and Tinker \textit{et al.} (2010) HMFs. In the figure we see that  the differences between the 2008 and 2010 HMFs at the level of the angular power spectra are within $0-2.5\%$ in the multipole range of interest for our analysis, and are mostly driven by the 1-halo term.  For the galaxy-CMB lensing cross-power spectrum, the differences are much less than $1\%$ on all scales of interest, while for the galaxy auto-spectrum the differences reach $2.5\%$ at $\ell \approx 1000$.  Qualitatively, this illustrates the level of theoretical uncertainty in the modeling.
 
\section{Galaxy profile}
\label{app:profile}

In the DES-Y3 analysis \cite{Zacharegkas2021}, the relation between the satellite galaxies' radial distribution and the matter density profile was parameterized via $a\equiv c_\mathrm{sat}/c_\mathrm{dm}$. In this approach both satellite galaxies and matter follow the NFW profile, but with different concentrations. In our analysis we choose to parameterize this relation using the parameter $\lambda\equiv r_\mathrm{out}/r_\mathrm{200c}$, which sets the truncation radius of the profile $r_\mathrm{out}$ in terms of $r_\mathrm{200c}$ (see Section~\ref{sec:HM}).

\begin{figure}
    \centering
    \includegraphics[scale=0.5]{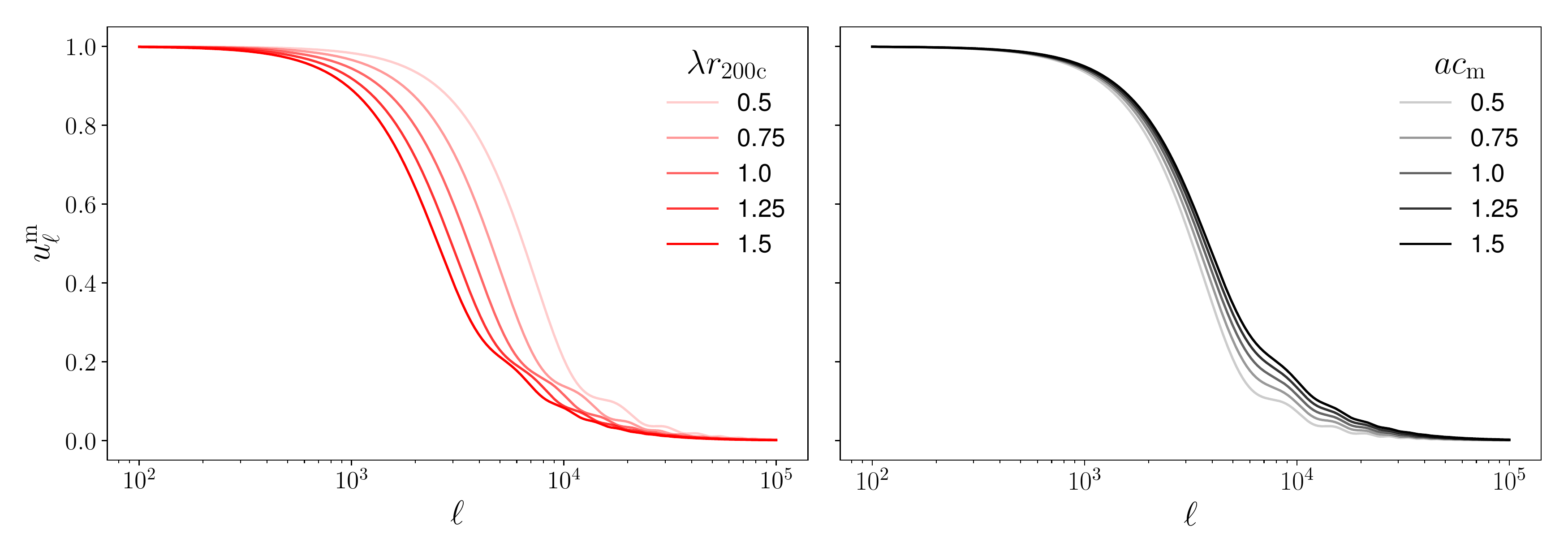}
    \caption{Impact of varying two different parametrizations $\lambda \equiv r_\mathrm{out}/r_\mathrm{200c}$ (left) and $a\equiv c_\mathrm{sat}/c_\mathrm{dm}$ (right) on the Fourier transform of the truncated NFW profile (Eq.~\ref{eq:ulm}).  Our analysis uses the parametrization in the left panel, while others (e.g.,~\cite{Zacharegkas2021}) in the literature have used the parametrization in the right panel.  For this plot, we use $\Delta=200$ with $M_\mathrm{200c}=3\times 10^{14}\, M_\odot/h$ at $z=1$. We also set $\chi=1317\,\mathrm{Mpc}/h$ (which can be used to convert between $\ell$ and $k=(\ell+0.5)/\chi$).  For this example halo, $c_\mathrm{200c}=3.4$ computed with the concentration-mass relation from \cite{Bhattacharya_2013}. Our fiducial truncation radius is $r_{200c}$.}
    

\label{fig:varying_cd_xout_in_ulm}
\end{figure}

In Fig.~\ref{fig:varying_cd_xout_in_ulm} we show how changes in $\lambda$ and $a$ impact the Fourier transform of the truncated NFW profile, $u_\ell^\mathrm{m}$, (Eq.~\ref{eq:ulm} in Section~\ref{sec:HM}), which enters the computation of the power spectra, e.g., Eq.~\ref{eq:ulkappacmb}. We note that the effects of $a$ and $\lambda$ are nearly equivalent, as these two parameters essentially determine the scale at which $u_\ell^\mathrm{m}$ goes from 1 to 0. Therefore, we conclude that choosing $a$ or $\lambda$ in the modeling does not affect the constraints on other HOD parameters.

\section{Varied-cosmology runs}
\label{app:cosmo}

The analysis is performed at fixed cosmology, namely \emph{Planck} 2018 best-fit parameters, as described in Sec.~\ref{sec:intro}. To assess the level of dependence on the \emph{unWISE} HOD constraints obtained in this analysis on the assumed cosmology, we run an exploratory MCMC varying the cosmological parameter $\ln(10^{10} A_s)$, the amplitude of the scalar power spectrum, which affects the clustering amplitude in the late-time universe (quantified by $\sigma_8$), that our galaxy-galaxy and galaxy-CMB lensing data is most sensitive to, out of all cosmological parameters. 

In practice, we apply a prior on the (derived) $S_8$ parameter, corresponding to the \emph{Planck} 2018 best-fit 1$\sigma$ error bar (as before, last column of Table II \cite{Planck2018}), keeping the matter density $\Omega_m$ constant. We start the MCMC with the covariance matrix from the main analysis, and keep all other values exactly the same as before. We do this exercise for the blue sample only, as these data points have the smallest error bars, so it will be the hardest to find a good fit. 

\begin{figure}
    \centering
    \includegraphics[scale=0.3]{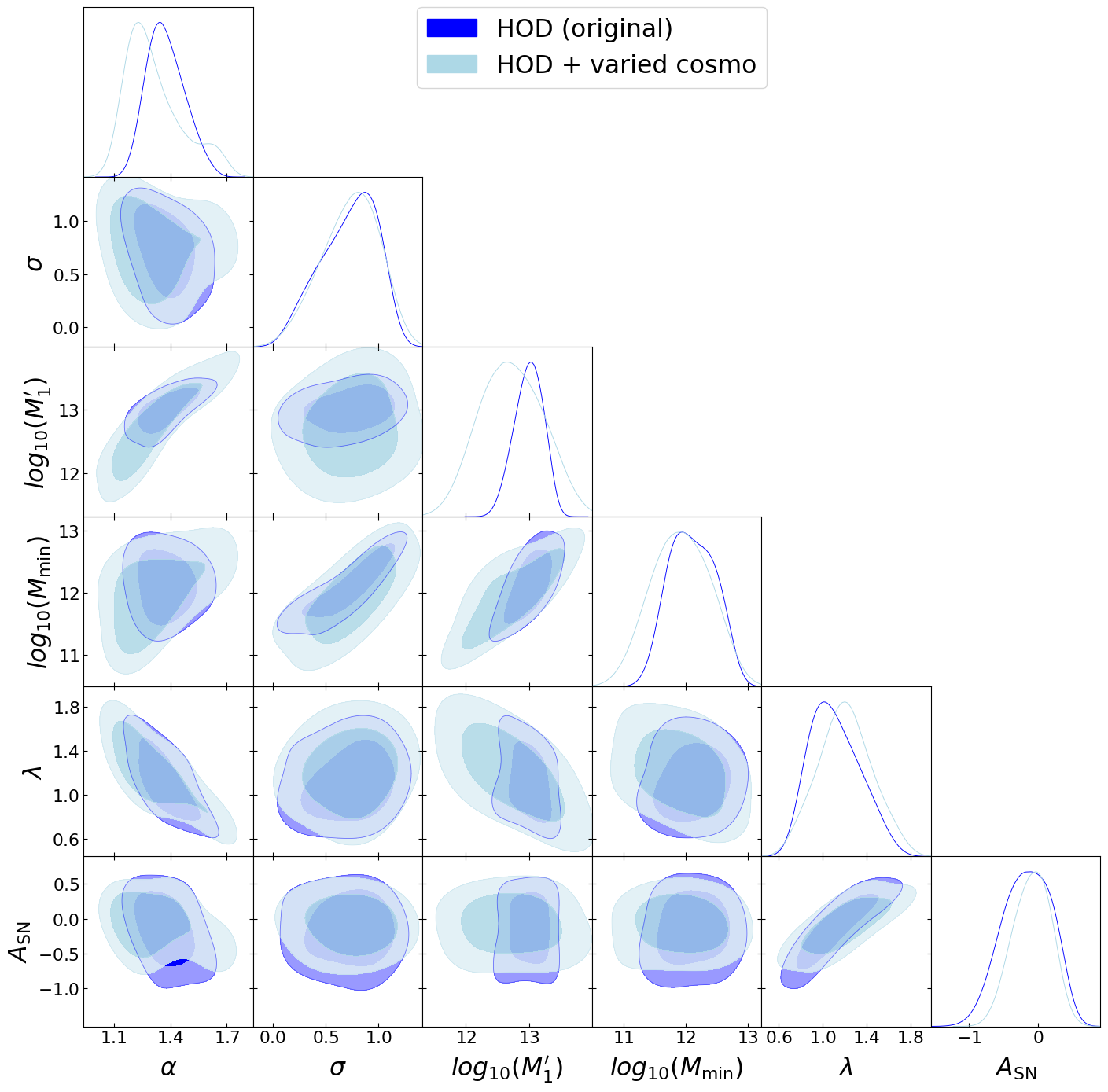}
    \caption{The 1D and 2D marginalized posterior distributions for two scenarios for the \emph{unWISE} blue sample: 1) the main analysis presented in this work (Fig.~\ref{fig:posteriors}), considering just the HOD parameters (blue contours); 2) same as 1), with addition of $\ln(10^{10} A_s)$ cosmological parameter, the amplitude of the scalar power spectrum 
    (light blue contours), thus illustrating the impact of varying the late-time clustering amplitude on our main analysis. We conclude that varying $\ln(10^{10} A_s)$ does not have a significant impact on our HOD constraints. See Appendix~\ref{app:cosmo} for more details. }

    \label{fig:posteriors_cosmo_hod}
\end{figure}

\begin{figure}
    \centering
    \includegraphics[scale=0.5]{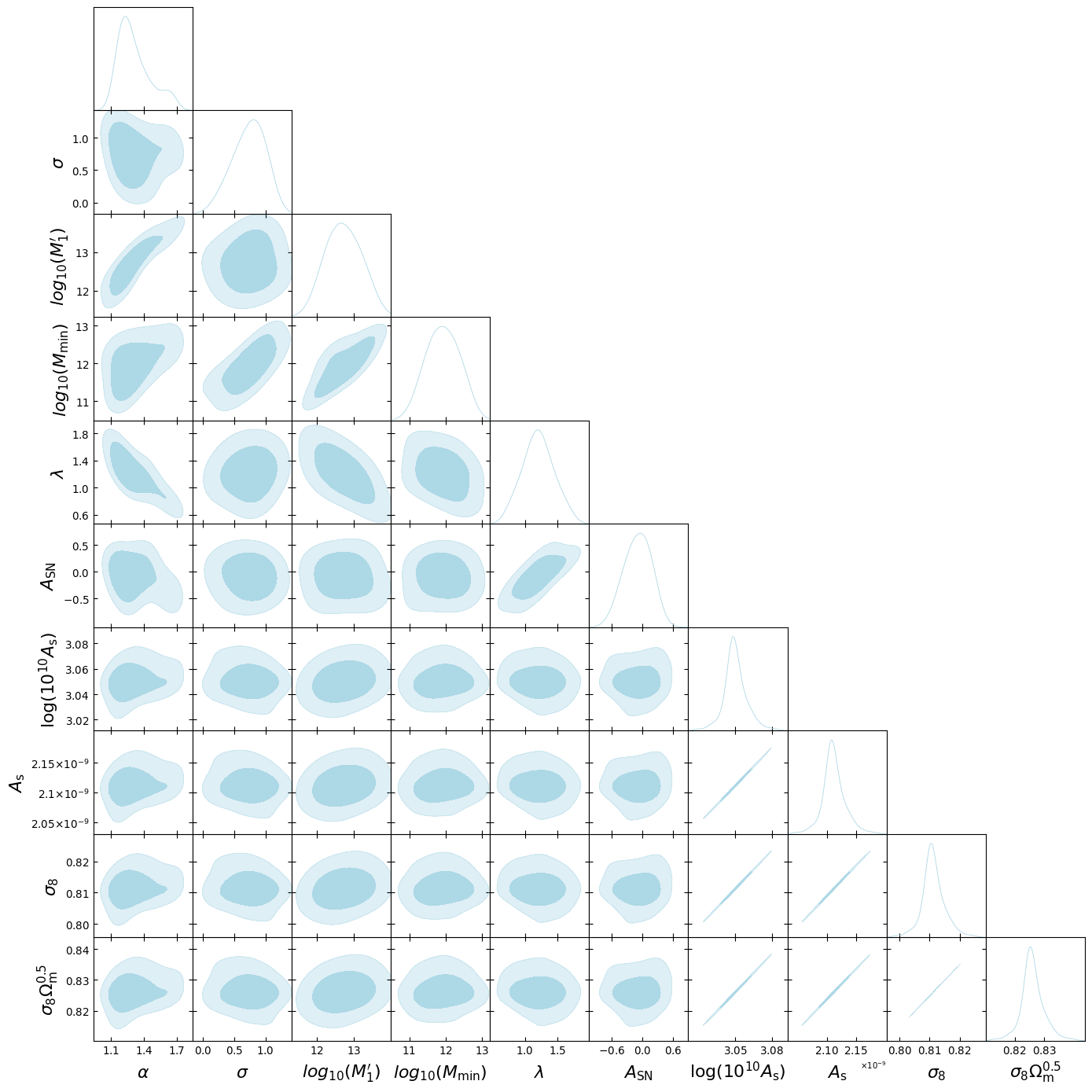}
    \caption{The 1D and 2D marginalized posterior distributions for the model HOD parameters, as well as the cosmological parameter $\ln(10^{10} A_s)$ (also showing the $\ln(10^{10} A_s)$-derived parameters: $A_s$, $\sigma_8$, $\sigma_8 \Omega_m^{0.5}$) for the blue sample. These contours are the same as the light blue posteriors in Fig~.\ref{fig:posteriors_cosmo_hod}, here just shown without the main analysis contours (in blue Fig~.\ref{fig:posteriors_cosmo_hod}), and including the derived parameters. }

    \label{fig:posteriors_cosmo}
\end{figure}

The results from this exercise are shown in Fig.~\ref{fig:posteriors_cosmo_hod} in comparison with the main analysis, and in Fig.~\ref{fig:posteriors_cosmo} individually. The chains are very slow to converge, resulting in Gelman-Rubin statistic $R-1 = 0.6$. From Fig.~\ref{fig:posteriors_cosmo_hod}, we conclude that the 1D and 2D marginalized posterior distributions for $\ln(10^{10} A_s)$ + HOD parameters (light blue) and HOD-only (blue, same as in Fig.~\ref{fig:posteriors}) are very similar, with the light blue contours being slightly larger than the original analysis (with the exception of the $M^{\prime}_1$ parameter, whose contours are noticeably larger), thus illustrating that our main results are not highly dependent on the exact value of $\sigma_8$ or $\ln(10^{10} A_s)$. 

There is an important caveat in regard to adding this cosmological parameter with a Gaussian prior, centered at the \emph{Planck} 2018 value. The 1D posterior of $\ln(10^{10} A_s)$ (as all derived parameters, that is $A_s$, $\sigma_8$, $\sigma_8 \Omega_m^{0.5}$) is not perfectly Gaussian, thus suggesting that the \emph{unWISE} galaxy-galaxy and galaxy-lensing late-time data prefers a lower value of $\sigma_8$.  However, because of the \emph{Planck} prior on this parameter, the MCMC cannot explore those regions (which results in convergence difficulties, as noted above).

\bibliographystyle{apsrev}
\bibliography{unWISE}

\end{document}